King Saud University
College of Computer & Information Sciences

# General-Purpose Visual Language and Information System with Case-Studies in Developing Business Applications

Mahmoud Samir Fayed

**Riyadh May 11, 2017**

**Master of Science Thesis**

King Saud University
College of Computer and Information Sciences
Department of Information Systems

# General-Purpose Visual Language and Information System with Case-Studies in Developing Business Applications

By

Mahmoud Samir Ibrahim Fayed

Approved by the Dissertation Committee, 11th of May 2017:

Dr. Atif M. Alamri, Associate Professor
Thesis Advisor
Signature:------------------

Dr. Ahmed Z. Emam, Associate Professor
(Examiner)
Signature:------------------

Dr. Djamal Ziani, Associate Professor
(Examiner)
Signature:------------------

# DEDICATION

*This Thesis is dictated to my Father; to my greatest Mather, to my lovely family.*

# DECLARATION

I hereby declare that I am the sole author of this thesis. I authorize King Saud University to lend this thesis to other institutions or individuals for the purpose of scholarly research.

## ACKNOWLEDGMENTS

I would like to express my gratitude to my supervisor Dr.Atif M. Alamri, for his inspired comments and his patience to lead this thesis to the best level. Also, I would like to say thanks to all those who helped me in this work.

Special Thanks to Dr. Ahmed Zayed Emam and Dr. Djamal Ziani for reviewing this thesis and providing very valuable advices and comments.

Thanks to Dr. Muhammad S. Al-Qurishi and Dr. Ahmad A. Al-Daraiseh from our research team for their great contributions.

Thanks to our Programming Without Coding Technology Team members:-

* Arcangelo Molinaro
* Stephen France
* Sameh Kamel
* Amine Debabsia
* Aboelsoud Abd Alraouf
* Ahmed Omar
* Abdulrazak Hussien
* Djardi Masoud
* Mouaddh Mebarki
* Ibrahim Abdelelah
* Dale Jones
* Rabeh Torchi
* Khalid Bentaleb
* Reem Mohamed
* Mohammed Fathy
* Shreen Hassan
* Sinan Qahtan
* Sozan Samy
* Vidlanovic Zlatko Aurel
* Waseem Salem
* Mohammed Al-Mahdi
* Antonio Fidel Perez Rubio
* Abobakar Ahmad Jallo
* Salama Mohammed Abo Mosslam
* Zerourou Bachir
* Navid Dezashibi
* Mahmoud Mohammed Hassan
* Belarir Abdelkader
* Mabrook AlRakhami
* Nabeel Hassanien

# ABSTRACT

Learning computer programming has been always challenging. Since the sixties of the last century, many researchers developed Visual Programming Languages (VPLs) to help in this regard. In this thesis, ten VPLs were specifically selected, studied, experimented with, and evaluated. A total of fifteen metrics were used to evaluate the tools. Comparisons, classification, and gap analysis were then presented. A list of requirements for a general-purpose VPL and a guide to help the novice programmer choose the right tool were generated and finally the PWCT (Programming Without Coding Technology, a novel general-purpose visual programming language) is developed and presented. The main objective behind developing this tool was to create not only a general purpose VPL but one that possesses textual language's capabilities. A language that can be used to develop similar programs to the ones developed in C++ or Java for example. As the name indicates, PWCT requires no coding at all. A person needs only to know basic programming concepts to be able to use the tool. PWCT has many attractive features such as: General Purpose, Open source, easy to use, Exports code to different languages, allows customizing visual components and creating new ones, uses keyboard shortcuts in addition to the mouse, and generates automatic documentation. PWCT has been launched as a Sourceforge project, which currently has more than 230,000 downloads for the language and more than 19,500,000 downloads for samples, tutorials and movies. Many business applications and projects are developed using PWCT, Also we developed the Supernova programming language and the Ring programming language using PWCT to prove that it can be used for advanced and large projects. Feedback from developers and results from the studies indicate that PWCT is a very appealing, competitive, and powerful language.

## الملخص

تعلم برمجة الكمبيوتر كان دائما تحدى ومنذ الستينات فى القرن الماضى اتجه الكثير من الباحثين نحو تطوير لغات البرمجة المرئية (الصورية) للمساعدة فى هذا الجانب. فى هذه الرسالة تم إختيارعشرة لغات برمجة مرئية ثم تم بعناية دراستهم وتجربتهم وعمل تقييم لهم وقد تم استخدام 15 معيار فى عملية التقييم وعمل المقارنات والتصنيف ثم تم عمل تحليل للفجوة الموجودة و تم تقديم قائمة من المتطلبات للغة برمجة مرئية متعددة الأغراض بالاضافة الى إرشاد لمساعدة المبتدئين فى تعلم البرمجة على إختيارالأداة المناسبة ثم تم تطوير و تقديم تقنية البرمجة بدون كود وهى لغة برمجة مرئية متعددة الاغراض و مبتكرة. الهدف من تطويرها ليس فقط انشاء لغة برمجة مرئية عامة ومتعددة الأغراض ولكن إنشاء لغة مرئية يمكن ان تقدم قوة وقدرات اللغات النصية على مستوى القدرة على تطوير التطبيقات والمشاريع التى تطور بلغات مثل السى++ والجافا على سبيل المثال. كما يشير الإسم فإن تقنية البرمجة بدون كود لا تتطلب كتابة كود والمستخدم يحتاج فقط لمعرفة أساسيات مفاهيم البرمجة ليتمكن من إستخدامها. تقنية البرمجة بدون كود لها الكثير من الخصائص مثل (عامة ومتعددة الأغراض – مفتوحة المصدر – سهلة الإستخدام – تنتج لنا الكود بلغات برمجة نصية مختلفة – قابلة لتخصيص المكونات وإنشاء مكونات جديدة – تتيح استخدام لوحة المفاتيح بجانب الفارة لانجاز المهام بمرونة – تنتج توثيق تلقائى لخطوات انشاء التطبيقات). تم نشر تقنية البرمجة بدون كود كمنتج حر مفتوح المصدر على موقع السورس فورج للمشاريع مفتوحة المصدر وتم تحميل تقنية البرمجة بدون كود أكثر من 230 الف مرة بالاضافة الى تحميل يتخطى 19 مليون و 500 ألف بالنسبة للامثلة والدروس التعليمية الخاصة بتعلم التقنية. أيضا تم تصميم و تطوير لغة البرمجة سوبرنوفا ولغة البرمجة رينج بإستخدام تقنية البرمجة بدون كود لإثبات أن التقنية يمكن ان تستخدم فى المشاريع المتقدمة والكبيرة. نتائج الدراسات وردود الفعل من المطورين تشير الى أن تقنية البرمجة بدون كود لغة برمجة مرئية قوية وقابلة للمنافسة.

# TABLE OF CONTENTS

## Table of Figures

# List of Tables

# List of Abbreviation

| | |
|---|---|
| VPL | Visual Programming Language |
| GUI | Graphical User Interface |
| GPVPL | General Purpose Visual Programming Language |
| VPE | Visual Programming Environment |
| VR | Visual Representation |
| SDE | Syntax Directed Editor |
| FE | Free Editor |
| VPLC | Visual Programming Language Compiler |
| VPLF | Visual Programming Language Framework |
| TPL | Textual Programming Language |
| PWCT | Programming Without Coding Technology |
| GCR | Graphical Code Replacement |
| GD | Goal/Module Designer |
| CB | Components Browser |
| IR | Interaction Designer |
| FD | Form Designer |
| VS | Visual Studio |

# Chapter 1: Introduction

## 1.1. Introduction to Visual Programming Languages

In the age of information technology, software development plays a vital role in responding to companies' and organizations' needs for high-quality information systems. Providing high-quality, reliable, scalable, efficient, easy to use, and cheap software is a nontrivial task. Meeting the requirements of such software requires proficient programmers and more productive software development tools. And so, there is an increasing need for software development tools that are easy to learn, general purpose, and highly productive [1].

As a result of the complexity in software requirements, many aspects of the software development process evolved and many tools were developed to help the programmers. Integrated development environments (IDEs) like Microsoft Visual Studio, NetBeans and Eclipse are essential for large projects. When using such tools, the programmers need to know not only the general programming paradigms but also the strict syntax of each programming language used. In these tools, the program representation in the source code is limited to text where photos and graphics can't be part of the source code. Moreover, the more expressive the programming language is the more complicated its syntax becomes and this makes it harder to understand and code with. This challenge made it more appealing to use VPLs, which attract more programmers, and can increase the productivity of developing software [2-5].

A VPL combines the features of integrated development environments and information systems to provide a visual tool for applications' development [5]. In addition to that, VPLs enable the development of applications and computer programs using more than one dimension, and provide a programming system based on interaction with graphical elements that mix between Text, Shapes, Colors, and Time, instead of typing text source code [6-10].

Currently, there are many VPLs in the market, but most of the successful and widely used ones are educational tools such as: Scratch, Alice & Kodu; or domain-specific such as: Max/MSP (Music and Multimedia) and LabView (Data Acquisition, Instrument Control and Industrial Automation). General-purpose VPLs like Limnor, Tersus, and Envision exist as well, but these languages are not widely used, According to TIOBE Index that measures the popularity of programming languages. The TIOBE Index is updated once a month and uses rating based on courses, skilled engineer and third party vendors. Also it uses the popular search engines to calculate the ratings.  [5, 11].

This survey is a step forward towards developing a new General-Purpose VPL that can be used by novice and mainstream-programmers to get more productivity without adding significant limitations that prevent the practical use of the proposed language. This research determines the limitations in previous implementations and defines the requirements of the new general-purpose VPL. In addition to that, a guide to help novice programming choose the right tool for their application is provided.

## 1.2. Examples of Visual Programming Languages

VPLs improve the methods in which programmers convey the display and processing of information. They allow the users to comprehend and then alter the results visually by utilizing graphics, animations, drawings and icons [20]. Experts in computer graphics, programming languages, and human-computer interaction need to collaborate to develop a VPL. The first VPL was put into practice in the 1960s, however, the success of visual programming was not achieved until the middle of the 1980s when graphics hardware became widely used [18, 19, 94].

Since 1960, more than 60 VPLs and visual programming environment have been developed in several domains. These tools can be classified into three main categories, namely, Forms/Reports based, Diagrammatic, and Iconic [8].

Most of the successful VPLs are domain-specific languages. For example, in the science and industrial domains, LabVIEW is an eminent VPL, which is designed for automating the usage of processing and measuring equipment in any laboratory setup [12]. Ladder Logic is also a VPL used to develop software for Programmable Logic Controllers (PLCs) in industrial control applications [13]. In the multimedia domain, OpenMusic is an object-oriented visual programming environment for musical composition based on Common Lisp [14]. In the educational domain, Alice [15] and Scratch [16, 17] are reputable VPLs that allow people of any age to be familiar with the programming world. Besides being domain specific, many of the existing solutions are unable to deal with large and complex programming problems let alone preserving a reasonable level of readability and maintainability. Moreover, many of these VPLs are still struggling to avoid frustrating programmers with strict limitations, unnecessary complexity, low productivity, and problems related to the quality of the final product.

Alice is a VPL where objects are manipulated in a 3D world (as shown in figure 1). It was developed at Carnegie Mellon University; it gives students the chance to learn about object-oriented programming concepts without the syntax frustrations imposed by text-based programming languages. With Alice, a programmer using a Graphical User Interface (GUI) environment selects program constructs and methods from lists of available choices [15, 21-22, 95].

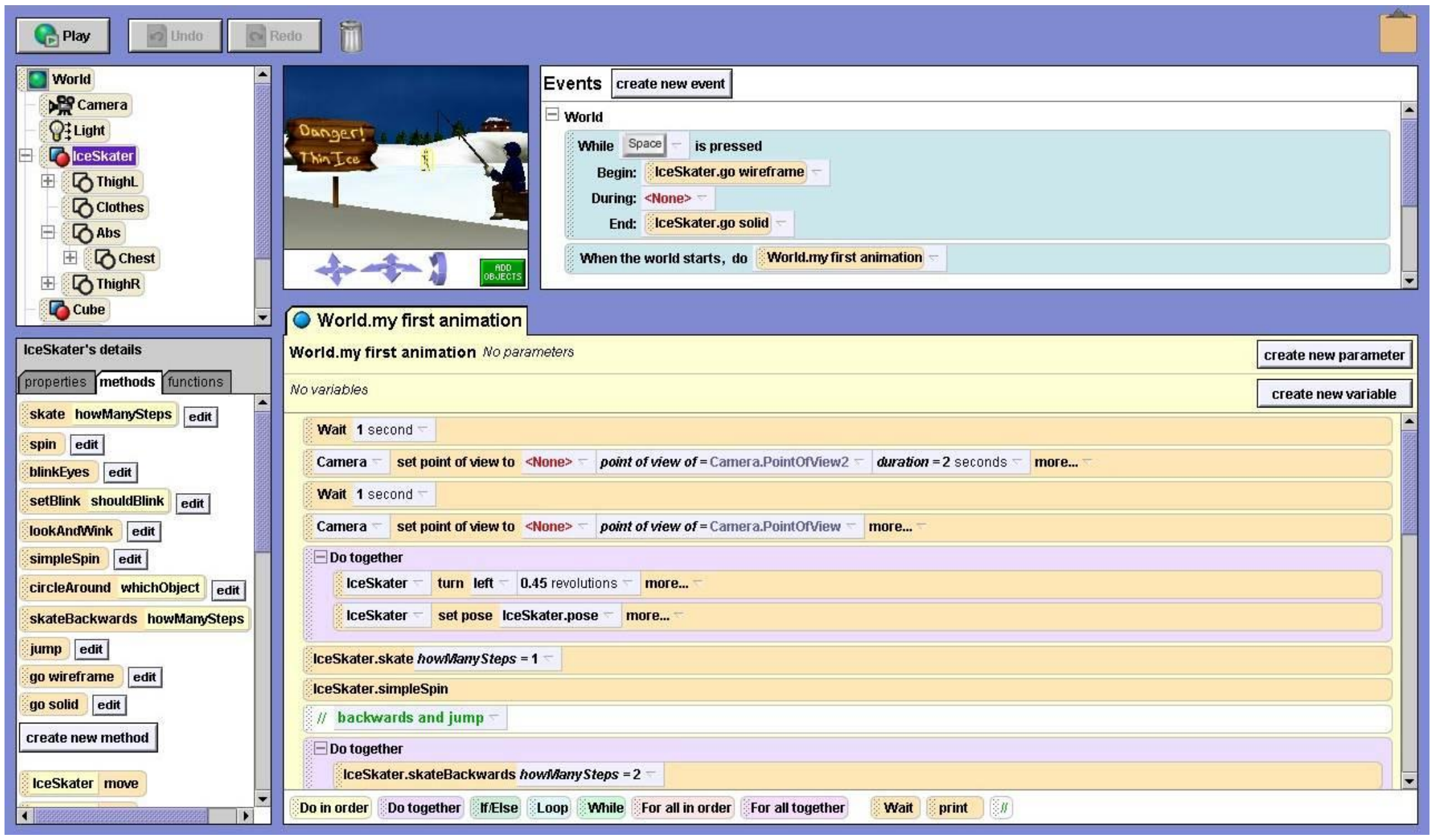

Figure 1: the Alice System [15]

Scratch is developed at the Massachusetts Institute of Technology Media Laboratory. It is a new VPL and environment that supports the creation of interactive stories, games, animations, music and art projects (as shown in figure 2). Scratch allows educators to reduce the cognitive load that learners experience when they are initially introduced to programming [23-24, 96].

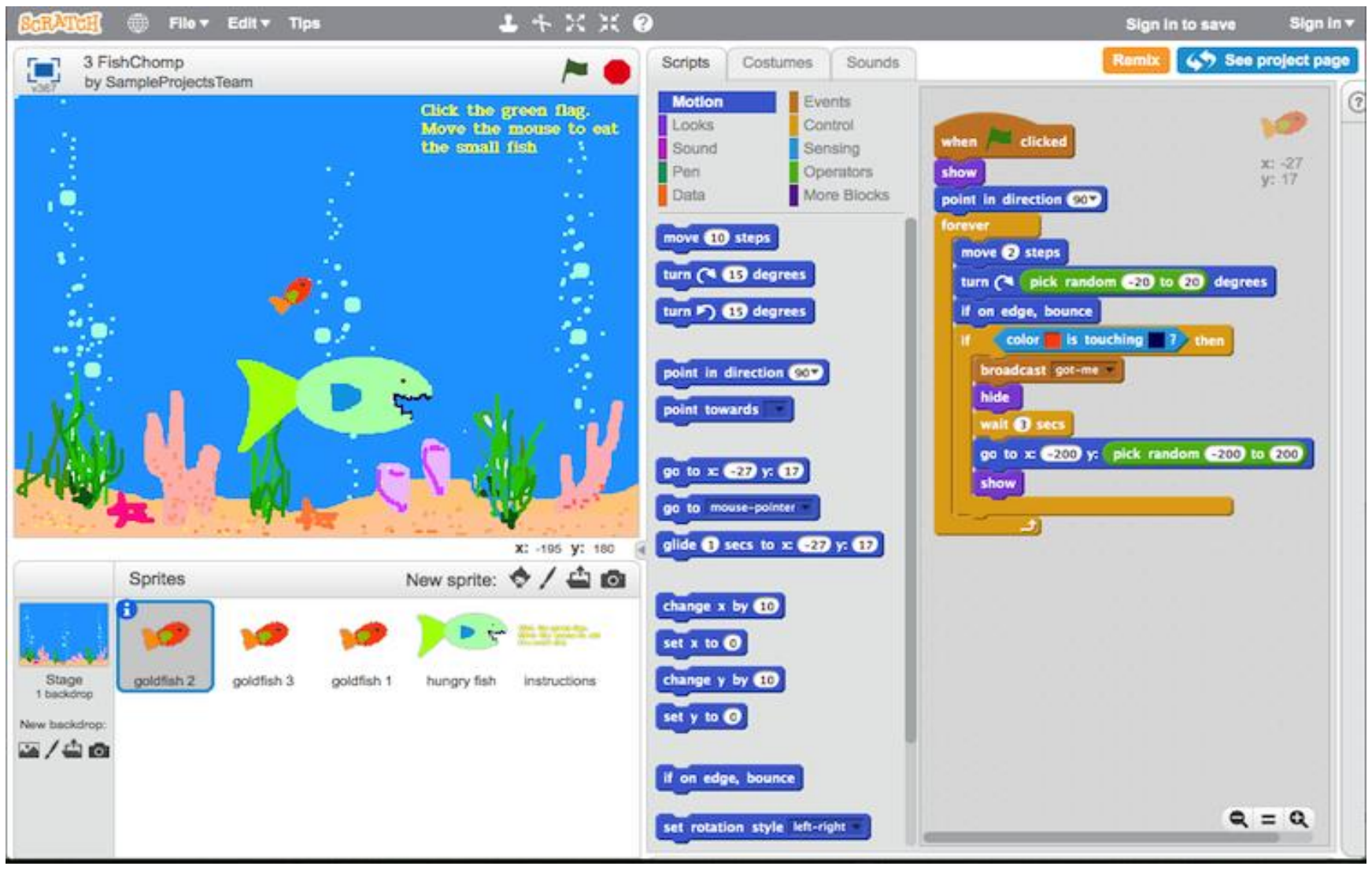

Figure 2: the Scratch System [23]

Kodu is an educational tool that was developed by Microsoft Research in 2010 to teach the basics of game development. Kodu is a new icon-based visual language programming made especially for creating games. It was designed to be appealing to children and enjoyable by everyone. The Kodu language is entirely event driven, where each line of programming is in the form of a condition and an action [25, 97].

LabView is a licensed software and it targets scientists and engineers. It is a graphical development environment with built-in functionality for data acquisition, instrument control, measurement analysis, and data presentation (Li et al., 2011). In LabView, the execution is determined by the structure of a graphical block diagram in which the programmer connects different function-nodes by drawing wires [13].

Tersus is an open source VPL developed by Tersus Software Ltd. It is used to build rich web and mobile applications by visually defining user interface, client side behavior, and server side processing (as shown in figure 3). It is a general purpose language that utilizes flow diagrams [93, 98].

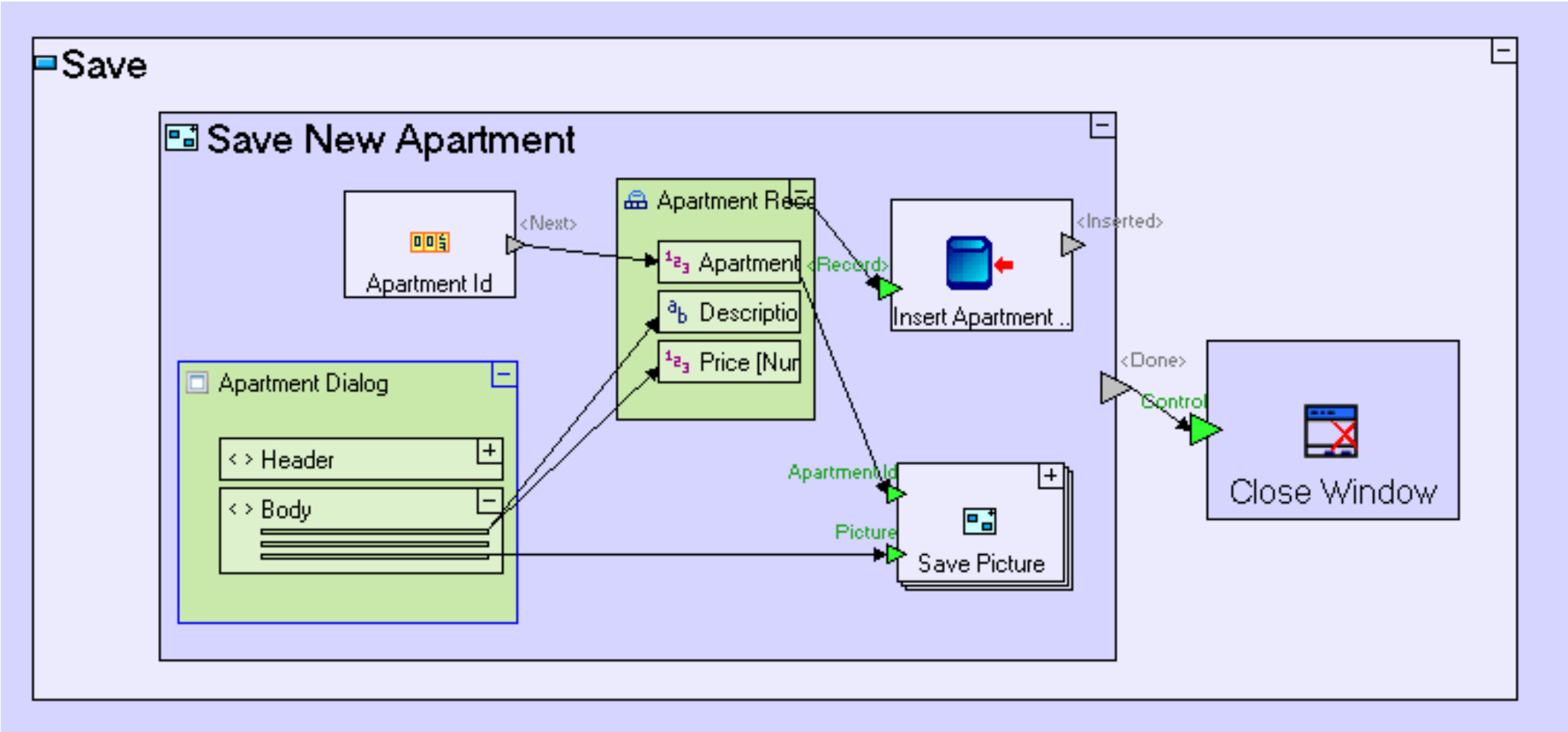

Figure 3: Application Modeling in the Tersus Platform [93]

Limnor (shown in figure 4) developed by Longflow Enterprises Limited is general-purpose VPL. It is based on visual studio .NET. It can be used to create most applications, interactive multimedia kiosks, sales presentations, database applications with interactive query and search, business management systems, internet payphone kiosks, etc. Non-technically oriented users can use it [26]. Limnor supports code generation only in C#.

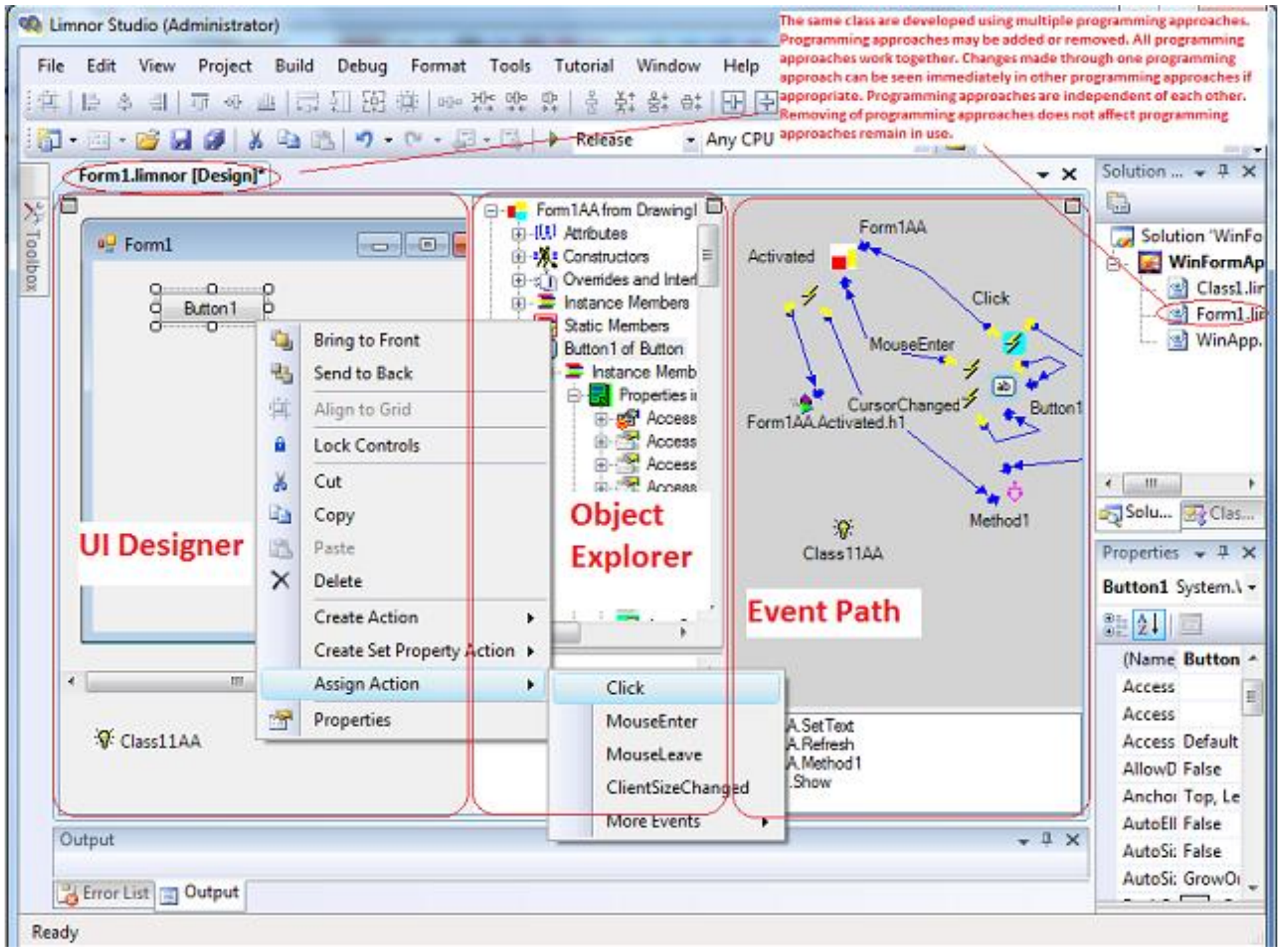

Figure 4: the Limnor System [26]

Prograph is a popular general purpose VPL licensed under Prograph CPX and Marten. The Object-Oriented, Data-flow & multi-paradigm language has long been known to reduce development time through drag and drop of iconic symbols to create programs[38].

Envision is a general purpose VPL that integrates imperative and object oriented programming paradigm in its design. It is an open source language that can be used in small and large scale projects as it can adopt both a High Level and Low Level programming. It partially supports Import/Export of Code from Java & C++. It's still under development [5].

VIPR is visual programming language that can be leveraged in small to large scale projects depending on the needs of a user. The language depicts both high level and low level functionalities on the basis of TCL and C++/TCL extensions respectively [90].

Lava is a general purpose VPL language but it is not used for production in the sense that it is only used for experimental purposes. As an open source language. The language is designed for Ease of Learning, Use, and Program Comprehension [92].

Forms/3 is a Declarative and spreadsheet-based high-level language that has long been used for general small and large projects. It is one of the language created to escalate development time in project realization[37].

Scratch, Alice, and Kodu are limited to the educational environment. Moreover, Scratch and Kodu were constricted at student level, while Alice can be used also at faculty level and other novice programmers to explore learning concepts of algorithmic thinking.

## 1.3. Motivation

The requirements for software applications are increasing because computers are now very large part of our everyday life. Software programs are more complex because of many factors. Nowadays programs run on a wide variety of hardware like high-performance clusters, personal computers, embedded devices and distributed systems. Applications are developed for different fields and the cost can vary significantly as free open-source software compete with proprietary software. Decreasing cost, improving reliability and increasing scalability are some of the requirements facing software developers. In the age of information technology, software development plays a vital role to respond to companies and organizations need of high quality information systems. This leads to the need to more programmers and more productive software development tools to be able to respond quickly to companies need with high quality [5].

After the success of many domain specific VPLs like Scratch, Alice by reaching millions of users worldwide, it's expected to find more interest in creating new VPLs that help novice programmers to learn programming and help mainstream programmers to create high-quality programs faster. Such languages must be designed carefully to solve current VPLs' issues without creating new problems and this is an important factor for new VPLs to gain popularity. Also, to increase the usage of these new VPLs, they need to compete and/or integrate with mature development tools that are based on popular programming languages such as C++, Java, C#, Python and Ruby [5-7, 13, 23].

## 1.4. Research Questions

The questions that we aim to answer during this thesis are:

1. What are the problems that prevent mainstream programmers from using the current visual programming languages?

2. What is the design of a general-purpose visual programming language that can solve mainstream programmers problems and encourage them to use the proposed visual programming language in developing practical software?

## 1.5. Research Objectives

The objectives of this research subject are:

1. Creating a Framework for creating visual programming languages to help us in creating the visual programming language components faster. The framework must contains the designers and domain-specific language to quickly create the visual programming language components. The implementation of each component should be done in little minutes to ensure the high-level of productivity.
2. Designing a General-Purpose visual programming language to make programming easier to understand for audiences other than programmers and to help expert programmers to quickly create programs and applications in less time.
3. Implementing the General-Purpose visual programming language through an information system that support programs and applications creation and manipulation, to reduce errors and to help users program faster.
4. Evaluating the system in developing business applications and comparing this to using other traditional programming languages that are text based.

## 1.6. Research Publications

The author had participated in two International Conferences. These conferences provided opportunities for the research to be shared with researchers. This first paper present the PWCT software while the second paper present a novel algorithm that we implemented using PWCT.

(1) Mahmoud S. Fayed, Muhammad Al-Qurishi, Atif Alamri, Ahmad A. Al-Daraiseh , PWCT: Visual Language for IoT and Cloud Computing Applications and Systems , ICC '17, March 22 2017, Cambridge, United Kingdom

(2) M Imran, MA Alnuem, Mahmoud S. Fayed, Atif Alamri, "Localized algorithm for segregation of critical/non-critical nodes in mobile ad hoc and sensor networks", Procedia Computer Science 19, 1167-1172

## Chapter 2: Background

### 2.1 Visual Programming Environment (VPE) vs Visual Programming Language (VPL)

It is worth explaining the difference between VPLs and VPEs. Programming languages like Visual Basic, Visual FoxPro, Visual C# and Visual J# are not VPLs. All of these languages are textual languages. The programmer must write textual code using the language's syntax. Other environments such as Microsoft Visual Studio are Visual Programming Environments (VPE) not VPLs [27]. Visual Studio (shown in figure 5), NetBeans (shown in figure 6) and Qt Creator ( shown in figure 7) provide designers and tools to create some parts of the application using visual components, but the textual code is necessary to complete useful and real applications. On the other hand, VPLs use different visual components instead of writing code.

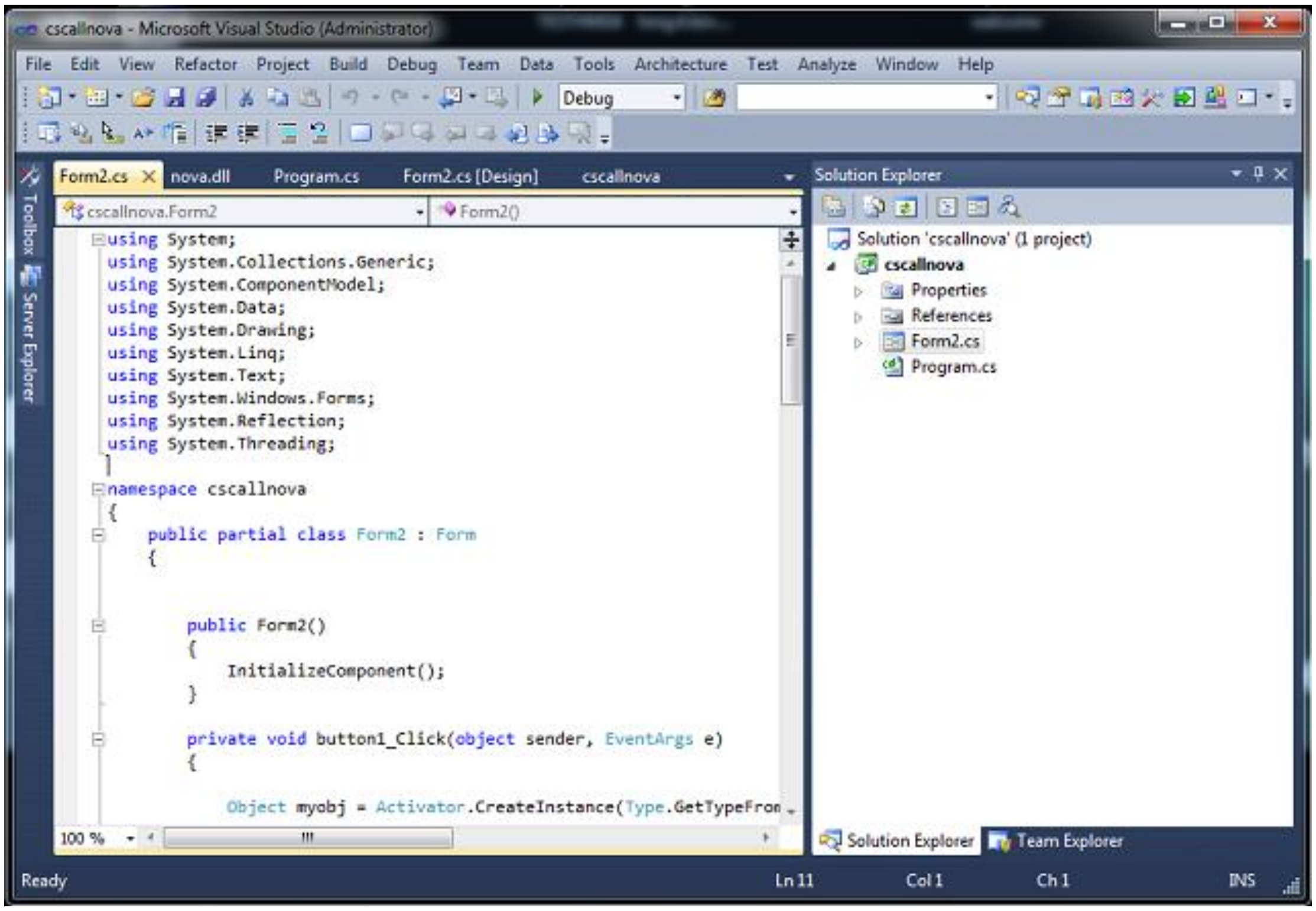

Figure 5: Visual Studio – Writing code using C# [106]

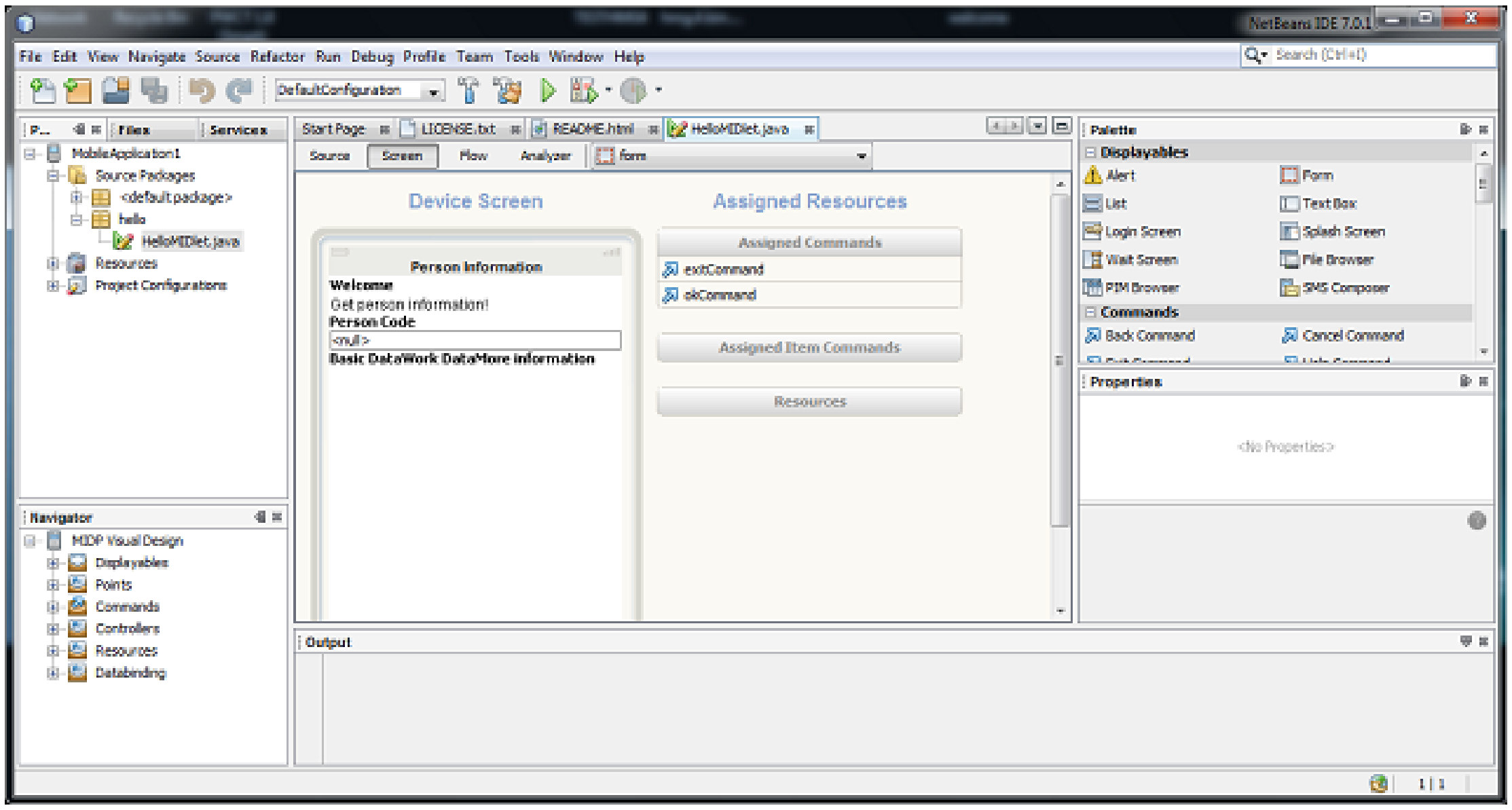

Figure 6: Netbeans – Using Visual Programming + Java textual code to create Mobile Application [106]

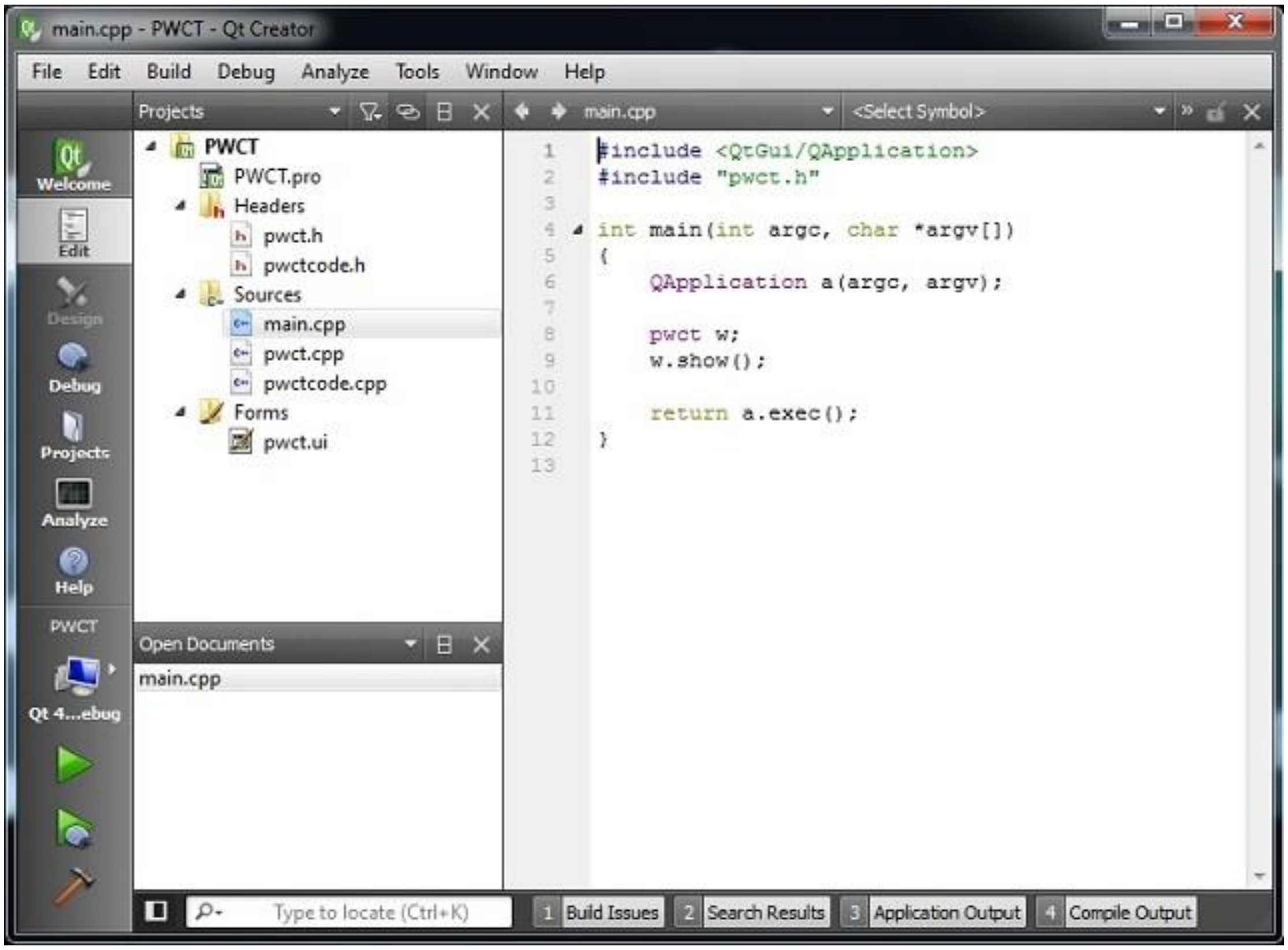

Figure 7: Qt Creator – using C++ and the Qt framework to create multi-platform GUI Applications [106]

## 2.2 Visual Programming Dimensions

In textual programming languages like C, C++ and Java, the code is text based and is unidimensional. In VPLs, the graphical code uses more than one dimension. Each graphical object has its place in 2D or 3D worlds. Each object may have its shape, colors and images. There are many relationships that can appear between objects such as: Inside, Outside, Touch, next-to, etc. Some languages also use the time as another dimension in the graphical code [27].

## 2.3 Visual Representations

A VPL can have one of the following four representations [28]:

1. Diagrammatic: uses components of shapes and text and use links to connect between shapes and represent the control or data flow.
2. Iconic: uses icons from the domain of the problem.
3. Form based: uses forms such as spreadsheet or data-entry forms.
4. Hybrid: uses a mixture of any of the above three.

## 2.4 Programming Paradigms

Many programming paradigms exist. Imperative, Declarative, Procedural, Object-Oriented, Functional, Concurrent, Data-Flow, and Event-Driven Programming paradigms are common. All of these paradigms can be used in VPLs beside the visual programming approach. Some VPLs use one paradigm only, while others are considered multi-paradigm [29].

## 2.5 Syntax Directed Editor vs Free Editor and VPL Compiler

Some VPLs use a syntax directed editor which is syntax-aware and prevents the programmer from making syntax errors. When used, the user can't make syntax errors, and after each step, the visual source or the graphical code will be correct, and can be executed directly after compiling without problems. Some VPLs compile the graphical code directly after each step during the design process. Other VPLs provide a free editor where the programmer have freedom to make mistakes during the development process and the VPL compiler can discover the errors during the compiling process. While the syntax directed editor is more suitable for novice programmers, the free editor provides more flexibility to the advanced ones [30].

## 2.6 Visual Programming Languages' Frameworks

A VPL's framework is a collection of tools that enable the developers to create new VPLs in less time with less efforts and better quality by utilizing ready to use and well tested tools [31].

In this thesis we created a visual programming languages' framework to use for developing the general-purpose visual programming language. In the Appendix C. we shows the designers in our visual programming languages' framework.

## 2.7 Visual Programming Languages' Issues

In general, four main issues need to be addressed when talking about the limitations of VPLs, namely, space, static representation, domain, and comments. In terms of space, almost all visual representations are physically larger than the text they generate, so the space used to show a program in a VPL is greater when compared with a text based program. Regarding the static representation, many programs created by VPLs look like a maze of wires that are hard to understand. Concerning the domain, a VPL can be classified as General or Specific domain. The most successful VPLs are designed for specific applications because general purpose languages are difficult to design and implement. In regard to comments, many VPLs don't provide a place for writing comments [32-35].

# Chapter 3: Literature Review

In a study about LabVIEW [36], the authors designed a survey to capture the cognitive effects of visual programming representations. The research was centered on a well-designed questionnaire to capture users' perceptions on visual effects of LabVIEW. After collecting and analyzing the data, it was found that most of the respondents felt that LabVIEW is an effective tool. Moreover, it was clearly established that the users found the visual aspects of LabVIEW to be more appealing than the other features rated during the survey.

In another study about Forms/3 and the Spreadsheet Paradigm [37], The main goal of the authors in this article is to show that procedural abstraction, graphics, and data abstraction are possible in the spreadsheet model. The authors presented a review based on forms/3 in a bid to show how limitations in a spreadsheets paradigm of programming can be alleviated for better end-user functionality. The author utilizes forms/3 in this research as a prototype from a language design perspective and from a human oriented perspective. It was found that it was actually possible to improve spreadsheets beyond the state they were at the time of documentation of this survey. Some of the features that were seen to have potency for improvement in spreadsheets were graphical types, gestural programming, dynamically-sized grids, generalized abstractions, graphical I/O, time travel, and steering.

In a Survey about Data Flow VPLs [18], The Authors did an in-depth review of fifteen data flow VPLs. The tools were first categorized depending on their application domain. Then, they were studied for characteristics and strengths. The authors argue that for a language to be considered powerful it should have predefined functions and programming constructs. Lastly, they gave an overview of unsolved problems in the topic such as, the large interfaces of such VPLs.

The article concluded by giving a general overview of the characteristics of data flow VPLs, the strengths and weaknesses at the time of research.

It's worth mentioning here that the authors stated that the ***creation of visual languages that suit a wide array of application domains can be a good area of research, not to mention that users should also be allowed to choose their level of expertise rather than being presented with rigid functionalities by languages***.

In a Pleminary Report about PROGRAPH [38], the authors presented a review of PROGRAPH which is a VPL based on data flow computational model. The main goal was to provide the reader with a detailed analysis of the features of PROGRAPH plus its built-in operations. And hence, the article was structured to make it easy for novice PROGRAPH enthusiast to get an understanding of the core concepts of the language. The authors integrated many examples in the discussion in a bid to cover the whole language and to ensure that the reader understands this VPL better.

In a study about the Users of Scratch [39]**,** the authors analyzed Scratch to identify the problems faced by skilled programmers when using the software. The authors created four teams of programmers and asked them to provide a comparison between Scratch and traditional integrated development environments. They discussed the usability of the tool at the fundamental programming level as well as at the graphical user interface level. Later, the experience of the programmers on each of the two interfaces was categorized based on the Nielsen usability framework. The authors found that there is a broader field for possible improvements in the operation of Scratch based on the logical errors that were encountered by some of the users. Furthermore, the developers of the tool need to do a better job in explaining some of the tool's aspects.

In a study about LabVIEW and Simulink [40]**,** the authors evaluated both LabVIEW and Simulink and did a comparison between them. Although there are many perceived differences between these two tools, the authors tried to find the relationship between LabVIEW and Simulink based on the criteria of their functionality as well as ease of use. The results showed that the majority of the users preferred using LabVIEW over Simulink due to its capabilities and ease of use. According to the authors, there is need to carry out a proportional study on LabVIEW and Simulink based on other criteria apart from the ease of use and functionality. Furthermore, even though LabVIEW has an advantage over Simulink based on graphical display and DSP hardware integration tools, it is not necessarily the best deal among the two. Extensive research is needed to develop a comprehensive analysis between the two languages.

In another study about Teaching Programming Concepts to High School Students with Alice [22]**,** the authors examined the possibility of using Alice to teach programming concepts to high school students. In this study, the authors divided 166 students into two groups and taught

C++ to one group and Alice to the other. The results of the achievement test showed that the Alice group scored higher than the C++ group. The results suggest that students who were taught Alice had a better understanding of the concepts of programming and were more interested in exploring the different aspects of programming on their own.

In a study about Student Opinions of Alice in CSI [42], using a questioner, the authors studied the responses of eighty-four college students who were taught Alice first and then Java. The authors discovered that the majority (59.5%) of the students felt that their experience with Alice helped them learn Java, and 66.7% of the students felt that the department should keep using Alice in the curricula.

In another study about Alice and Pair – Programming [43], the authors used 6 non-computing major sections with a total of 89 students to determine to which extent the students enjoyed the programming environment provided by Alice. In two sections the students were asked to work separately, however, in the other four, the students worked in pairs. The authors gave the students few assignments and later asked them to write an essay about their experience. The results showed that the student enjoyed working with Alice, became more confident, and had a better understanding of programming concepts.

In another study about Children Learning Computer Science Concepts via Alice Game Programming [44], the authors studied 325 middle school students to see if they can gain knowledge on the complex concepts of computer science. The students were asked to develop simple games using Alice. The authors reported that the students were able to learn many concepts such as, abstraction, modeling, control structure, and events' handling.

In a study about the effects of the media to promote the scratch programming capabilities creativity of elementary school students [87], the authors tried to see the effect of using Scratch on elementary school student's creativity. They developed a simple program using Scratch to provide guidelines to students and used a sample of 60 students in their first grade to use it. The authors reported that using such application had a positive impact on the students' creativity.

In a study about Prograph 2.0 [88], the author expressed his opinion of a specific version of Prograph that includes a compiler. The compiler facilitates the production of executable applications that operate as standalone. The author stated that this tool is well thought out, well integrated with Macintosh environment, and has good documentation.

In a Review on Teaching and Learning of Computational Thinking through Programming [89], the authors reviewed 27 articles that focused on introducing computational thinking to K-12 curriculum. The authors found that the majority of the studies focused on introducing computational thinking concepts and few considered practice and perspective. And hence, they encouraged researchers to do more studies focusing on the latter two aspects.

In a study about Visual Programming Languages [90], the authors studied five VPL tools namely, Ark, Vipr, Prograpgh, Forms/3, and Cube. They presented different classifications, and touched on the theory of VPLs in general. Later, they showed where the five tools fall. Lastly, they commented on VPLs related issues such as, control flow; and procedural and data abstraction.

In a study about programming environments and languages for novice programmers [91], the authors in this article provided a taxonomy of a large number of tools. Some of the tools are VPLs and others are not. They categorized the tools in two major groups, tools to help teaching programming and others to support the use of programming (empowering) to achieve other goals. Then they divided the teaching category into multiple categories based “mechanics of programming”, “social learning” and “providing reason to program”. The empowering category was divided into two groups “mechanics of programming” and “activities enhanced by programming”. Each of the group was later divided into other groups based on different aspects. Each of the tools was classified according to the different categories mentioned above. Lastly, the authors concluded that the majority of the tools focused on the mechanics of programming and there is a need for tools to provide reasons so more people start learning programming.

### 3.2. Evaluation and Comparative Analysis for Selected Characteristics

In this section we will compare between different VPLs, We have three stages, In the first stage we will do the selection process and In the second stage we will do the comparative analysis process. And in the third stage we will provide the implications as demonstrated in figure 8.

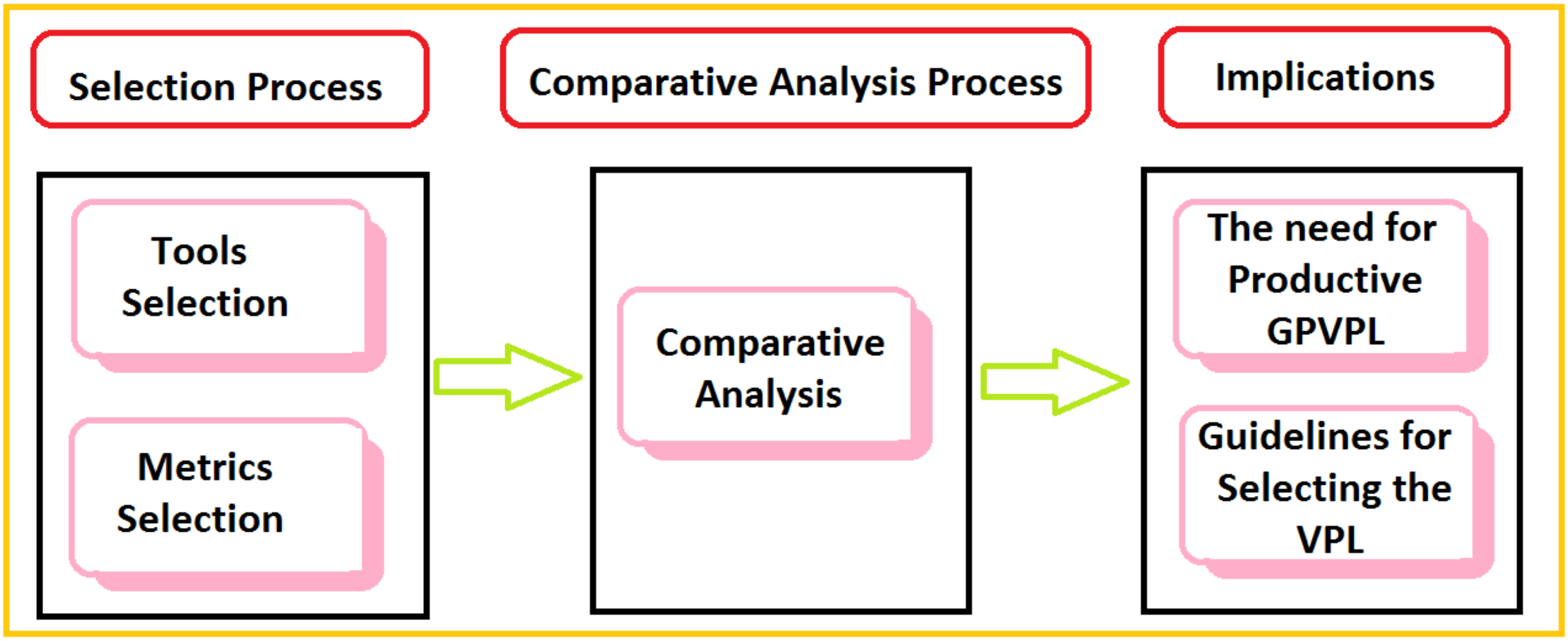

Figure 8. Evaluation and Comparative Analysis

#### 3.2.1. Selection Process

The selection process is divided into two stages, In the first stage we will select the tools and in the second stage we will select the metrics.

##### 3.2.1.1 Tools Selection

In this chapter, ten different VPLs were studied and analyzed to show their strengths and weaknesses. The main goal of this study is to come up with a set of requirements for a general-purpose VPL that is easy to learn, fast, can be appreciated by both novice and expert programmers, used to develop small and large applications, and produces output code in different languages. A minor goal of this study was to help a person decide which of the ten tools suits his needs.

There exist many visual programming languages. In this research, ten tools that cover different types of visual representations and design goals were selected. The tools were selected for different reasons, for example, Scratch, Alice and LabVIEW were selected because they are the most popular VPLs according to TIOBE index.  LabView, Prograph, VIPR and FORMS/3 were selected because they are very popular in the literature. Tersus and Limnor were selected because they are general purpose and used in practical and different types of applications. Finally, Lava and Envision were selected because they look familiar to programmers with knowledge of textual code and paradigms like object-oriented.

### 3.2.1.2 Metrics Selection

To start with, we had to decide on the criteria for assessment/evaluation. A team of researchers backed with few experts in the field [107] suggested fifteen different criteria as shown in Table 1. Most of the criteria are based on the common non-functional requirements in every software [108] and the common criteria of programming languages in general [109].

| Criteria | Code | SubCr | Description | SubCr | Description |
|---|---|---|---|---|---|
| Domain & Scope | Cr1 | A | Domain Specific | B | General-Purpose |
| | | C | High Level Programming | D | Low Level Programming |
| | | E | Can Create Large Programs | F | Creates only Small Programs |
| Compatibility (Exporting/ Importing text code) | Cr2 | A | Export to many languages. | B | Import from many languages. |
| | | C | Export to one languages. | D | Import from one languages. |
| | | E | Export not  supported | F | Import not supported |
| | | G | Allows scripts and code. | | |
| Programming Paradigm | Cr3 | A | Multi-paradigms | B | One programming paradigm |
| | | C | Imperative Programming | D | Declarative Programming |
| | | E | Event-Driven Programming | F | Object-Oriented Programming |
| | | G | Data-Flow Programming | H | Procedural Programming |
| | | I | Spread-sheet based | J | Control-Flow |
| License & Cost | Cr4 | A | Published Software | B | Not published |
| | | C | Free Software | D | Commercial |
| | | E | Open Source | F | Closed Source |
| Development Time (Decreased:Dec Increased: Inc) | Cr5 | A | Dec by Drag-And-Drop | B | Dec by High Level components. |
| | | C | Dec by allowing writing input such as equations. | D | Dec by using other dimensions such as the Time dimension |
| | | E | Dec by Keyboard shortcuts | F | Inc: Mouse only is used. |
| | | G | Inc: many ops to do simple tasks such as assignment. | H | Inc: missing editor features like (Search, Replace, etc.). |
| | | I | Dec by providing tools such as GUI builder to create UI. | J | Dec using Liveness |
| Size of programs | Cr6 | A | Small (uses small blocks or icons). | B | Large (uses large components and many connections). |
| | | C | Small (hides optional details | D | The size is dynamic |
| Readability (Decreased:Dec Increased: Inc) | Cr7 | A | Inc: using (text, colors, shapes & icons) | B | Dec due to missing features like changing the font size |
| | | C | Dec due to the use of many connections between blocks. | D | Dec by allowing text similar to textual code. |
| Performance | Cr8 | A | Fast enough for applications | B | Inc: compiles to languages like C |
| | | C | Inc: allows C extensions | D | Application dependent |
| Memory Usage | Cr9 | A | No critical problems found | B | Some Critical Problems |
| | | C | Needs Large RAM (>1 GB) | D | Needs small RAM (256MB) |
| | | E | Optimized by the programmer | F | Depends on the application. |
| Portability | Cr10 | A | Multiplatform(Win, Mac, etc.) | B | One Platform |
| | | C | Only MS-Windows | D | Only Web Based (Online) |
| | | E | Only Apple Mac OS X | F | Only Sun/Solaris |
| Security and Safety | Cr11 | A | No critical issues | B | Depends on the developed app' |
| | | C | The tool is a prototype (issues) | | |
| Scalability | Cr12 | A | Suitable for small applications | B | Suitable for large applications |
| | | C | Hard to manage large projects (missing OO) | D | OO is used to manage large projects. |
| | | E | Divides Apps into subroutines | F | App can run across multi-cores |
| | | G | Uses containment or nesting | H | Uses Zoom in and Zoom out |
| Extension | Cr13 | A | Support Extension | B | No support for extension |
| | | C | uses HTTP and JavaScript | D | Done by the Tool Developers |
| | | E | The Ch interpreter | | |
| Documentation | Cr14 | A | Reference | B | Getting Started |
| | | C | Tutorials | D | Book |
| | | E | Research Papers | F | No Documentation |
| | | G | Documents for Contributors | H | Online Documents |
| Support | Cr15 | A | There is support | B | No Support |
| | | C | Provided by a Company | D | Provided by (Forums) |
| | | E | No English Forum | | |

Table 1. Evaluation Metrics

Table 1 shows all fifteen criteria and 92 sub-criteria. The majority of them are self-explanatory, and hence, there is no need to define them. It is important to know that, general-purpose here means the language can be used to develop similar programs to those done with textual languages such as C++.

To evaluate the tools, each tool was evaluated by three different team members for all 15 criteria. The results of each of the evaluators were then gathered and compared. To claim whether a tool has a specific feature or not, two of the evaluators at least needed to agree. Few randomly chosen points were re-evaluated by the main author to make sure that the evaluation process is accurate and complete.

To gather the required information, the evaluators had to download, install, and experiment with the tools. They also visited the tools' websites, read published articles about the tools, searched the web, and sent questions to developers or to forums. Each of the evaluator did their evaluation independently to be as comprehensive and unbiased as possible. The results of the evaluation are discussed in the next section.

### 3.2.2. Comparative Analysis Process

Table 2 below shows the results of evaluating the ten tools. One of the main findings is that eight of the tools claim to be general purpose. According to our definition above, a VPL is general purpose if it can be in different domains and develop different types of applications. In other words, it has to be comparable to a textual language such as C++. Only the authors of Scratch and Alice state that they designed their VPLs to be used in education while the others claim to have developed general purpose VPLs and not for a particular domain.

In practice Lava, VIPR and Forms/3 are just research tools and are known to be used for serious development. Tersus is used in web and mobile database applications only. Prograph is used in desktop database applications and only mac os x. Limnor is used in database applications for windows – desktop and web but not for low level programming. LabView is used by engineers for automating the usage of processing and measuring equipment in any laboratory setup. Evaluating The other criteria doesn't provide a proof that the above VPLs are general purpose that can replace C++ and java for system programming and applications development.

| **Cr'** | **#** | **Description** | Scratch | Alice | ProGraph | L-VIEW | Forms/3 | VIPR | Limnor | Tersus | Envision | Lava |
|---|---|---|---|---|---|---|---|---|---|---|---|---|
| Cr1 | A | Domain Specific | T | T | | | | | | | | |
| | B | General-Purpose | | | T | T | T | T | T | T | T | T |
| | C | High Level Programming | T | T | T | T | T | T | T | T | T | T |
| | D | Low Level Programming | | | | | | T | | | T | |
| | E | Can Create Large Programs | | | T | T | T | T | T | T | T | T |
| | F | Creates only Small Programs | T | T | | | | | | | | |
| Cr2 | A | Export to textual code in many languages | | | T | | | T | T | | T | |
| | B | Import from textual code in many languages. | | | | | | | | | T | |
| | C | Export to textual code in one languages. | | T | T | | | | | T | | T |
| | D | Import from textual code in one languages. | T | | | | | | | | | |
| | E | No support to export to textual code. | T | | | T | T | | | | | |
| | F | No support to import from textual code. | | T | | T | T | T | T | T | | T |
| | G | You can write scripts and code too. | | | | T | | | | | | |
| Cr3 | A | Multi-paradigms | T | T | T | | | T | T | | T | |
| | B | One programming paradigm | | | | T | T | | | T | | T |
| | C | Imperative Programming | T | | | | | T | | | T | |
| | D | Declarative Programming | | | | | T | | | | | |
| | E | Event-Driven Programming | T | T | | | | | T | | | |
| | F | Object-Oriented Programming | | T | T | | | T | T | | T | T |
| | G | Data-Flow Programming | | | T | T | | | | T | | |
| | H | Procedural Programming | | | | | | T | | | | |
| | I | Spread-sheet based | | | | | T | | | | | |
| | J | Control-Flow | | | | | | | T | | | |
| Cr4 | A | Published Software | T | T | T | T | T | | T | T | T | T |
| | B | Not published | | | | | | T | | | | |
| | C | Free Software | T | T | T | | T | | T | T | T | T |
| | D | Commercial | | | T | T | | | | T | | |
| | E | Open Source | T | T | | | | | | T | T | T |
| | F | Closed Source | | | T | T | T | | T | | | |
| Cr5 | A | Increased using Drag-And-Drop | T | T | T | T | T | | T | T | | |
| | B | Increased using High Level components that hide the details. | | T | T | T | | | T | | | |
| | C | Increased because some input can be written directly like equations. | | | | T | T | | | | T | T |
| | D | Increased because of using other dimensions like the Time dimension | | | | | T | | | | | |
| | E | Increased because of using the Keyboard shortcuts to quickly create programs. | | | | | | | | | T | T |
| | F | Decreased because of using the Mouse only in the interaction. | T | T | T | T | | | T | T | | |
| | G | Decreased because of the need to connect between many components to create small tasks. | | | T | T | | | | | | |
| | H | Decreased due to missing editor features like (Search, Replace & Multi selection). | T | T | | | | T | | | | |
| | I | Increased using the right tool for the right task like using GUI builder to create UI Forms. | | | | | | | T | | | T |
| | J | Increased using Liveness | T | T | | | | | | | | |
| Cr6 | A | Small (for example: using small blocks or text and small icons). | T | T | | | | | | | T | T |
| | B | Large (for example: using large components and many connections). | | | T | T | T | T | T | T | | |
| | C | Small due to hiding details that are optional | T | T | | T | | | | | | |
| | D | The size is dynamic and can be changed by selecting more options | | T | | | | | | | | |
| Cr7 | A | Enhanced using visual representation (text, colors, shapes & icons) | T | T | T | T | T | | T | T | T | T |
| | B | Decreased due to some missing features like changing the font size | T | T | | | | | | | | |
| | C | Decreased due to the use of many lines and connections between blocks. | | | T | T | | T | T | T | | |
| | D | The text used in the visual representations looks like the textual code. | | | | | | T | | | T | |
| Cr8 | A | No critical problems (The speed is enough to run most of the applications | T | T | T | T | T | T | T | T | T | T |
| | B | Increase Performance by creating Native Applications | | | T | T | | | | | | |
| | C | Can be increased by better programmers (Like adding C extensions ) | | | | | | T | | | | |
| | D | Depend on the application (can be compared to textual code) | | | | | | T | T | T | T | T |
| Cr9 | A | No critical problems found that prevent using the software | T | T | T | T | T | T | T | T | T | T |
| | B | Critical Problems in memory usage | | | | | | | | | | |
| | C | Development tool need 1GByte RAM | | T | | | | | T | | | |
| | D | Development tool need 256Mbyte RAM | | | T | | | | | | | |
| | E | Memory usage can be optimized by the programmer | | | | T | | | | | | |
| | F | Memory usage depend on the application. | | | | | | T | | T | T | T |
| Cr10 | A | Multiplatform(Windows, Linux & Mac OS X) | T | T | | T | | | | T | T | T |
| | B | One Platform | | | T | | T | T | | | | |
| | C | MS-Windows | | | | | | | T | | | |
| | D | Web Based (Online) | T | | | | | | | | | |
| | E | Apple Mac OS X | | | T | | | T | | | | |
| | F | Sun/Solaris | | | | | T | | | | | |
| Cr11 | A | No critical problem the prevent using the software | T | T | T | T | | | T | T | T | T |
| | B | Depend on the application under development and the programmer skills | | | | T | | T | T | T | T | T |
| | C | The tool is just a prototype (bugs and problems are present) | | | | | T | | | | | T |
| Cr12 | A | Very suitable for small applications | T | T | | | | | | | | T |
| | B | Very suitable for large applications | | | T | T | T | T | T | T | T | |
| | C | The absence of object-orientation/functional programming makes it difficult to scale | T | | | | | | | | | |
| | D | The usage of Object-Orientation is good to manage large projects. | | T | T | | | | T | | T | T |
| | E | Dividing programs to subroutines | | | | T | | T | | | | |
| | F | Applications Automatically scales and run across multiple cores | | | | T | | | | | | |
| | G | Containment or nesting (avoid the graph edge problem) | | | | | | T | | | | |
| | H | Scalable using Zoom in and Zoom out | | | | | | | | T | T | |
| Cr13 | A | Support Extension | T | | | T | | | | | | |
| | B | No support for extension | | | | | | | | | | |
| | C | Can be done using HTTP requests and JavaScript | T | | | | | | | | | |
| | D | Done by the Tool Developers | | T | T | | T | T | T | T | T | T |
| | E | The Ch interpreter | | | | T | | | | | | |
| Cr14 | A | Reference | T | T | | T | | | T | | | |
| | B | Getting Started | T | T | | | | | T | T | | |
| | C | Tutorials | T | T | | | T | | T | T | | T |
| | D | Book | T | T | T | | | | | | | |
| | E | Research Papers | T | T | | | T | | | | T | T |
| | F | No Documentation | | | | | | T | | | | |
| | G | Documents for Contributors | | | | | | | | | T | |
| | H | Online Documents | T | T | | T | | | | T | | T |
| Cr15 | A | There is support | T | T | | T | | | T | T | | T |
| | B | No Support | | | | | T | T | | | | |
| | C | Provided by a Company | | | | T | | | T | T | | |
| | D | Provided by open source community (Forums) | T | T | | | | | | | T | |
| | E | No English Forum | | | T | | | | | | | T |

Table 2. Evaluation and a comparative analysis

Below we provide a brief description of the findings related to each tool.

### 3.2.2.1. Scratch

The first Language under the microscope is scratch-a domain specific high level language. This language is domain specific in the sense that it is tailored to create Games, Stories & Animations courtesy of an imperative and event driven programming paradigm. Written in Written in Squeak and ActionScript, Scratch has a good number of advantages that makes it attractive for use with one of them being the fact that it is open source under GPL v2. and Scratch Source Code Licenses. The development time is scaled down a lot with this language thanks to the use of components and drag-and-drop way of programming. Most of blocks depend on small number of parameters that can be modified directly while the block size is compact to hide the unnecessary detail. The readability in this platform is also enhanced due to use text, colors and shapes. Its performance in terms of speed is also good enough to run most of applications (presentations, stories & games) created by students. It is also very portable as it is supported on Mac, Linux, Windows and on the Web. It also has free support and documentation that is also quite palatable for both starters and advanced users-this is through Getting Started, Reference & Video Tutorials.

On the downside, it should be noted that some features are missing that might decrease the speed of maintaining programs. Some of these features are Search, Replace and multi-selection of block, expand/collapse of inner blocks. It is also suitable for small programs and applications given that the absence of object-orientation and functional programming makes it difficult to manage large projects. There is also No Standard support to export or Import from text based source code using languages like C/C++, C# & Java. However, this can be added by the aid of library for reading/writing MIT's Scratch file format.

### 3.2.2.2. Alice

This is another domain specific high level language that is popular for programming in 3D environment mainly for education purposes. Apart from the fact that this language is open source, it has a wide array of features to put the icing on the cake. One of the most pronounced features is the speed of creating programs using advanced components and drag-and-drop functionalities, with most of blocks in use also depending on small number of parameters that can be modified directly. The size of these blocks is also pretty compact to hide unnecessary details. It can also be used for big projects but it is highly recommended for small projects thanks to its adoption of object orientation. Furthermore, it is highly portable in the sense that it supports Mac, Linux and Windows not to mention that it works well with normal memory specs. In terms of compatibility, the language supports source code generation using Java. The presence of complete documentation and free support is also welcome among the community members. It is however important to note that for better performance, a 2 GB or more RAM, VGA graphics card capable of high (32 bit) color and at least 1024x768 resolution and a sound card are recommended.

**3.2.2.3. Prograph**

This is a popular general purpose VPL licensed under Prograph CPX and Marten. The Object-Oriented, Data-flow & multi-paradigm language has long been known to reduce development time through drag and drop of iconic symbols to create programs. Though the size of the programs may look larger, this platform design alleviates this through the use of text, colors and shapes that increase readability. The ability to create classes or simply the object oriented nature of the language makes it very ideal in dealing with small and large programs. When it comes to portability, the language is supported in both Mac and Windows while it exhibits awesome speeds performance wise. The other colossal aspect is that it comes with books that can be handy not only to beginners but advanced users as well. On a negative dimension is that it requires MacOS X 10.6.8 or better and it does not have an official support platform barring email support. Also, it is not open source and one can only get trial versions for testing purposes.

**3.2.2.4. LabView**

This is yet another Object-Oriented, Data-flow & multi-paradigm language that is popular among engineers and scientists. It is a high level general purpose language that has found great use in Data acquisition, instrument control and industrial automation among other scientific and engineering tasks. It was developed to accelerate development time among users through sophisticated elements coupled by drag and drop functionalities. This time is also reduced given the possibility to divide of programs into VI and the ability to run any written code to on multiple with no changes. Perhaps one of the most advanced feature of LabView is in its extensibility and Compatibility. In the realm of compatibility, the language can be integrated with graphical, text-based, and other programming approaches not to mention that users can write code and scripts too! This is made possible through a variety of possible extensions such as the Ch interpreter that can be added to Labview for scripting. Users are also afforded the possibility of buying add ons for specific applications. The language also boosts a high quality documentation plus lots of tutorials and forums for user support.
The only challenge is that large programs can grow complex especially with the presence of large blocks, lines as connections between blocks. The complexity has however been mitigated by enhanced readability through the use of text, colors, shapes and images in the visual representation.

### 3.2.2.5. Forms/3

This is Declarative, spreadsheet-based high-level language that has long been used for general small and large projects. It is one of the language created to escalate development time in project realization. This has been well orchestrated by a declarative and drag and drop programming capability it affords programmers. A good example of this is the ability to write or demonstrate equations and the ability to use time dimension in cells. It performs quite well and it does not consume much computer memory especially in research projects.
This open source language has however been pegged back by security issues from bugs and other problems that is compounded by the lack of proper support and documentation since 2004. This has also played a role in the difficulty of extension development. It is also not supported by common Operating systems as it is only supported by Sun/Solaris and Hewlett-Packard/HP-UXat.

### 3.2.2.6. VIPR

This is yet another powerful VPL language that can be leveraged in small to large scale projects depending on the needs of a user. The language depicts both high level and low level functionalities on the basis of TCL and C++/TCL extensions respectively. This way, the language is known to exhibit imperative, procedural and object oriented programming paradigms tailored for improved development time. Some admirable advantages of these is that it is to execute run-time animation of the code. Code debugging is also made easy through single stepping through program snapshots. The use of containment or nesting otherwise known as “Follow the Ring into the nesting” has also made the language scalable in that eliminates graph edge problem in programs. The language performance highly depends on the nature of program but it is worth saying that reduced latency between command issue and execution has increased performance. The readability of this platform rests on graphical primitives like rings, arrows, and parameter rings. The memory requirements will also depend on the program being executed while its security will also be influenced by the programmer.
The disadvantages of this platform is the lack of official support and documentation which has been a drawback has far as extensibility is concerned. It is also limited to Mac platform only.

### 3.2.2.7. Limnor

This is another open source control flow VPL that can be leveraged for a wide array of applications ranging from building Windows, Kiosk and Web Applications. One of the huge advantages with language is the compatibility with common object oriented programming languages like C#, ASP.NET, PHP & JavaScript conglomerated with a control-flow paradigm. The drag and drop mode of programming coupled by use of the mouse for simple tasks-instead of the keyboard, cuts development time by far. Programs are generally large as shapes and arrows are used but readability is enhanced through text and color.

The negative aspect of this platform is the lack of documentation and the memory requirements- it needs a CPU speed of over 2GHZ and 1GB RAM. Regardless, most of the support can be found in Limnor forum. The other disadvantage is the lack of third party extensions and plugins to scale functionality. It is also limited to windows platform only.

### 3.2.2.8. Tersus

This is an open source data flow VPL language that is designed for creating mobile and web applications by virtue of visual diagrams. Based on AJAX, the language generates java code from visual programs meaning that it performs pretty well in with normal computer specs. Its extensions can also be done using java code. The other positive aspect of this language is the fact that it is supported by every common computer platform be it Linux, Unix, Mac and Windows. Its interactive design is another strong point of the language as it deploys drag and drop functionality in development. This plus the use of natural language for description and plenty of well blended color makes it quite easy for a user to navigate around hence reducing development time. Starters do not have to hassle a lot to learn as there is a documentation and lots of good tutorials. For advanced users, there is a forum that helps solves problems in the realm of the platform.

### 3.2.2.9. Envision

This is yet another general purpose VPL that integrates imperative and object oriented programming paradigm in its design. It is an open source language that can be used in small and large scale projects as it can adopt both a High Level and Low Level programming. It partially supports Import/Export of Code from Java & C++.

It actually uses programs like that look like textual code but only this time the development time is reduced as there are no syntax errors. Just like most of the VPL languages discussed, Envision enhances interaction through a blend of color and texts. It also affords the user a chance to zoom in and out of large code bases and millions of objects on screen. It is supported by common operating systems namely: Windows, Linux & Mac OS X. Its security, performance and memory requirements are all dependent on the kind of application being designed.

The only glitch with this language is the lack of proper documentation, forums and tutorials but this is bound to change as a forum is under development.

### 3.2.2.10. Lava

This is one other general purpose VPL language but the only one in the bunch that is not used for production in the sense that it is only used for experimental purposes. As an open source language, it is exudes quality in building small and large scale projects courtesy of an object oriented programming paradigm. The development time of a user is boosted with this platform given that it uses point-and-click structure editor, a GUI builder etc. The program size is relatively smaller as it uses Treeview enhanced with natural language text to boost readability. The language is supported on Mac, Linux and in Mac. Its performance, scalability and memory usage will depend on the application being developed.

To the negatives now and it can be seen that, the platform cannot be extended by third party developers-it can only be done by the Lava team using C++ and Qt. There are lots of online documentations to learn from but no English forum to get support from other developers.

### 3.2.3. Implications

In this section we will present the implications from the comparative study

### 3.2.3.1 The need for productive general purpose visual programming language

From the aforementioned tables, we see that there is a wide array of VPL languages in the programming science to choose from depending on the needs and preferences. This paper presents some of the most common of these languages with an in-depth analysis of these languages based on a variety of metrics. The metrics of interests in this case are Domain & Scope, Compatibility, Programming Paradigm, License & Cost, Development Time, Size of programs, Readability, Performance, Memory Usage, Portability, Security and Safety, Scalability, Extension, Documentation and Support.

It can be seen that this VPL vary depending on the metrics of interest. Alice and Scratch are domain specific while the rest are general purpose. In terms of portability, Scratch, Alice, Prograph, LabView, Tersus Envision and Lava cover Mac, Linux and windows platforms. Forms/3 runs on Sun/Solaris and Hewlett-Packard/HP-UXat while Limnor and VIPR will only run on Mac and Windows respectively.

Most of these VPL are open source except Prograph and LabView. Turning attention to support and documentation, it is seen that VIPR, Envision and Lava lag behind the rest which have some sort of support. As mentioned earlier, these VPL languages reduce development time but their programs are generally large due to the use of visual graphics. It should however be distinguished that Lava and Envision have relatively small visual representations than the others. LabView is the most extensible of the bunch as it allows a variety of possible extensions such as the Ch interpreter to be added for scripting. Users are also afforded the possibility of buying add ons for specific applications.

In this section we conducted an evaluation between well-known general-purpose VPLs (Tersus and Limnor) and two reputable VPE (Visual Studio and Net Beans). One professional programmer for each language was hired to develop programs in that specific language. All programmers were asked to develop 20 simple to intermediate applications. The authors monitored the time and memory needed to complete each task. Table 3 shows the averages of all tasks. We can find that the time required to create a simple application using VPE is the smallest amount of time among the other languages. This criterion is very important because it tackles the productivity. In terms of time, Visual Studio comes in the first place with approximately 117 seconds, then Tersus 140 seconds, Limnor with 144 second and lastly NetBeans 165 seconds.

With respect to the number of user steps used to build the application. We can notice that using Visual Studio and NetBeans need 26, and 33 steps respectively, whereas Limnor, and Tersus need 41,56, and 46 steps respectively. Building applications in VPLs involves many steps; however, most of these steps are duplicated and simple, which make them easy to learn for the novice user. On the other hand, Visual Studio and NetBeans require less number of user steps but one of these steps is writing code, which is complex and advanced. Regarding to the professionals, the productivity is very important factor, and as we notice from the table, Visual Studio proof the minimum development time regardless the number of user steps. Finally, memory usage by Tersus is near to 12.12 MB which is the least size for the created application; however, Visual Studio has 58 MB.

| Criteria\ Alt. | General Purpose VPLs | | Visual Programming Environments | |
|---|---|---|---|---|
| | Limnor | Tersus | Visual Studio | Netbeans |
| Average time required to create and test the application (Seconds) | 144 | 140 | 115.66 | 165 |
| The steps to create and test the application (steps) | 54 | 46 | 26 | 33 |
| Average time required for each step (Seconds) | 2.66 | 3.04 | 4.45 | 5 |
| Average Time spent using keyboard in user steps (Seconds) | 13 | 18 | 31.1 | 30 |
| Memory used before application design starts (Kbyte) | 13460 | 50268 | 51276 | 156052 |
| Memory used after the application design (Kbyte) | 29292 | 62392 | 109420 | 173352 |
| Memory consumed in application representation during design time (Kbyte) | 15832 | 12124 | 58144 | 17300 |

Table. 3 the average results of a simple program

### 3.2.3.2 A guide to help the novice programmer choose the right tool

The next flowchart (figure 9) provides a guide to help the novice programmer choose the right tool based on the purpose (Educational, Development or Research) and age.

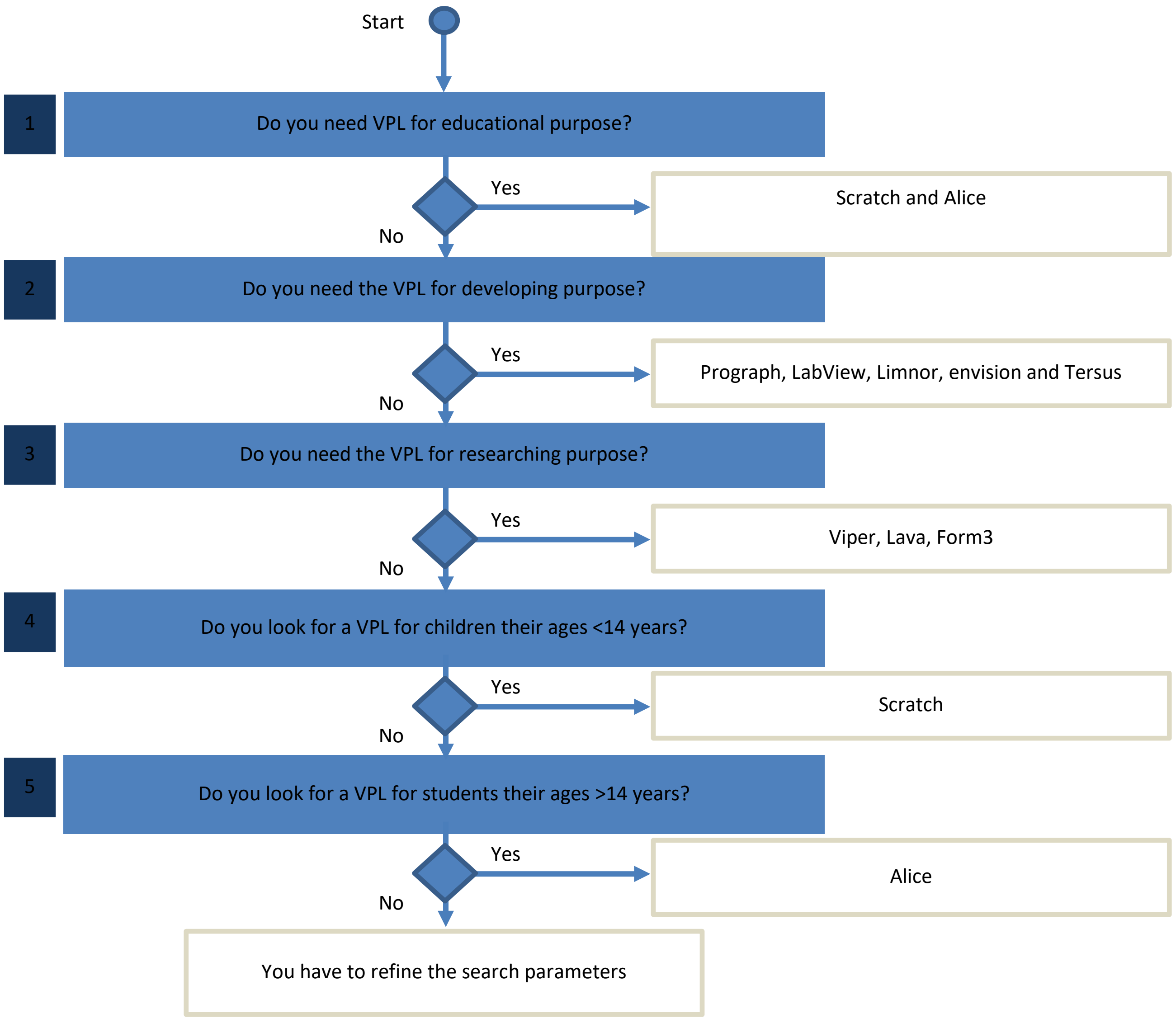

Figure 9: Flowchart to help the novice programmer choose the right tool

# Chapter 4: Main Design

## 4.1 The requirements for the new General Purpose Visual Programming Language

We talked about the progress in the visual programming languages, the success in different programming domain, and the features of the most popular visual programming languages. According to our literature review and comparative analysis we see a gap in attracting mainstream programmers to using visual programming languages in each new programming task. The current VPLs are not equivalent to the current textual programming languages at the practical side. Yes, some limitations are a result of missing features that can be added but this doesn't count only for "what we can do?" but for "how we can do things with respect to the productivity of the programmer during software development and the quality of the final software".

We will provide the requirements of a general purpose visual programming tool that can be used as replacement for the usage of popular programming languages like ( C,C++, C# and Java) + visual programming environments like Visual Studio and Netbeans). We expect that after the implementation and marketing of this new tool, many new software projects can be developed using Visual Programming Language in less time with high quality. Also we expect attracting more users to real programming industry after the existence of this tool.

The next table provides the requirements for the new general purpose VPL. These requirements are the result of our comparative study where our research team selected the best features that we need to have together in one visual programming language.

| Index | Metric | Description |
|---|---|---|
| 1 | Domain & Scope | • General Purpose<br>• High Level Programming<br>• Low Level Programming<br>• Creating small and large systems and applications |
| 2 | Compatibility | • Generate Code in different programming languages like C, C++ , C#, Java & Python |
| 3 | Programming Paradigm | • Imperative<br>• Procedural<br>• Object-Oriented Programming<br>• Functional |
| 4 | License & Cost | • Free Open Source (GPL version 2) |
| | | • Exception allows developing commercial applications without releasing the source code |
| 5 | Development Time | • Increased using the keyboard, Intellisense/Auto-complete, Cut/Copy/Paste/Search/Replace<br>• Interaction using Keyboard/Mouse<br>• Colors are useful for interaction<br>• Syntax Directed Editor (prevent errors)<br>• VPL Compiler help in discovering errors when the Syntax directed editor is off. |
| 6 | Size of programs | • Small – using TreeView |
| 7 | Readability | • Better Using Text, Colors, Tree & Froms<br>• Change the font name / size<br>• Time Dimension |
| 8 | Performance | • Depend on the application |
| 9 | Memory Usage | • Depend on the application |
| 10 | Portability | • Windows, Linux and Mac OS X |
| 11 | Security and Safety | • Depend on the application |
| 12 | Scalability | • Depend on the application<br>• Using Standard GUI controls like TreeView is faster than owner drawn controls. |
| 13 | Extension | Comes with Designers and a scripting language called creating new components or VPLs |
| 14 | Documentation | Complete documentation (Tutorials, Movies & Reference) |
| 15 | Support | Forums |

Table 4. Requirements for General-Purpose VPL

## 4.2 Graphical Code Replacement (GCR)

Many researchers believe that Visual Programming Languages are inevitable tools to attract more people to the programming world. In the past half century, many Visual Programming Languages were developed. A large number of the developed languages are domain specific while only few are general purpose. In this chapter, PWCT (Programming Without Coding Technology, a novel visual programming language) is presented. The main objective behind developing this tool was to create not only a general purpose VPL but one that possesses textual language's capabilities. A language that can be used to develop similar programs to the ones developed in C++ or Java for example. As the name indicates, PWCT requires no coding at all. A person needs only to know basic programming concepts to be able to use the tool. PWCT has many attractive features such as: General Purpose, Open source, easy to use, Exports code to different languages, allows customizing visual components and creating new ones, uses keyboard shortcuts in addition to the mouse, and generates automatic documentation

Most of the successful VPLs are domain-specific languages. For example, LabVIEW (Laboratory Virtual Instrumentation Engineering Workbench), although it is claimed to be general purpose it is only popular in data acquisition and control applications [12]. Ladder logic is a VPL used to develop software for Programmable Logic Controllers (PLCs) in industrial control applications [13]. In the multimedia domain, OpenMusic is an object-oriented visual programming environment for musical composition based on Common Lisp [14]. In the educational domain, Alice [15] and Scratch [16, 17] are reputable VPLs that allow even little kids get familiar with the programming world. Besides being domain specific, developing software with VPLs becomes much harder as the software gets larger and more complex due to their inherent limitations and graphical representation.

One main limitation of VPLs is known as Deutsch Limit. It states that "The problem with visual programming is that you can't have more than 50 visual primitives on the screen at the same time"[41, 105]. And so, for any VPL to be comparable to textual programming languages it needs to overcome this limitation at least. In fact, the authors believe that a superior VPL should have the following features to be a rival to text-based languages:

- Pure visual language: the language should require no coding at all and no knowledge of syntax.
- General purpose: the language could be used to create any software just like text-based language.
- Suitable graphical representation: the graphical representation used in the language should be

suitable for small and large software development.

- Ease of use: the language should be easy to use and require short time to get mastered.
- Open source: an open source VPL encourages people to use it and allows users to tailor it to their needs.
- Capability of creating new visual components: a developer should be able to create new components easily.
- Capability of producing (exporting) code in different programming languages such as C# and java.
- Provides multi-lingual interface.
- Utilizes both mouse and keyboard interaction.

In this chapter, we propose a novel technique called Graphical Code Replacement (GCR) to overcome Deutsch's limit and to provide the needed scalability and flexibility. GCR was used in building a new VPL called PWCT. The proposed technique facilitates the application construction through a set of sequential interactions with visual components. GCR depends on a generated Tree of Steps (Steps Tree) that describes the program under development. The manipulation of TS is done through the interaction with visual components without the need to deal with text based code. The code generation process is managed in the background by the visual components to hide the complexity from the user. The visual representation of the program is constructed using traditional GUI controls (Tree, labels, textbox, list box ...etc).

GCR replaces the common drag-and-drop technique to create an application by the give input-get generated representation technique. The input is accepted by visual components through interaction pages (Forms to get users' inputs).The drag-and-drop method enable the user to select a component from a toolbox then drag it to be dropped in a design region. Here the visual representation of the component is determined from the beginning and attributes of the object can be changed from options available in the component representation as in scratch and Alice or from properties window as in limnor. But, in the GCR the visual representation is completely dynamic and is generated later after providing the options to the user through interaction pages (forms) and after that we get the visual representation.

For example the if—elseif-else-endif component in GCR is one component which provides the optional elseif and else operations. And based on the user selection of these operations in the interaction page (forms) through checkbox the visual representation will be generated to include elseif/else or not. In scratch we have separate components for the “else” operation.

The proposed work introduces a new general purpose visual programming language by combining the strengths of the current successful VPLs. PWCT is completely visual. It uses a standard graphical notation for building a program. The source editor is replaced by a TreeView control with an adaptive and semantic representation. PWCT provides benefits for both novice and expert programmers. It pays special attention to how expert programmers can easily adapt to this new visual programming methodology. In addition, PWCT is built based on an open architecture where users can easily add/modify his own components.

## 4.3 Graphical Code Replacement Method

NOTE: that the words "step", "control", and "component" are used interchangeably and they all mean a graphical control also the words "goal" and "module" are used interchangeably to mean a programming module.

The basic idea behind the Graphical Code Replacement (GCR) was to find a way to make the visual programming language fully equivalent to a traditional textual programming language and to overcome all the limitations and weaknesses of VPLs. Looking at the structure of code written in textual programming languages, it can be easily seen that the code contains a set of instructions and commands written as statements in a specific syntax. The statements are arranged in nested structures similar to a computer tree with one or more nodes/leaves. A complex statement such as “For Loop” can be represented with a node that can be expanded, while a simple statement such as a “printf()” can be represented with a leaf.  For example, the code shown in Figure 10 is written in C. This code can be seen as a tree of statements or steps (a step here is defined as a visual component) arranged in a way to print numbers from 1 to 3, print a message, and then continue Printing 4 to 10. Keep in mind that each instruction in any programming language consists of two parts: the operation, and the required data for the specification of that operation.

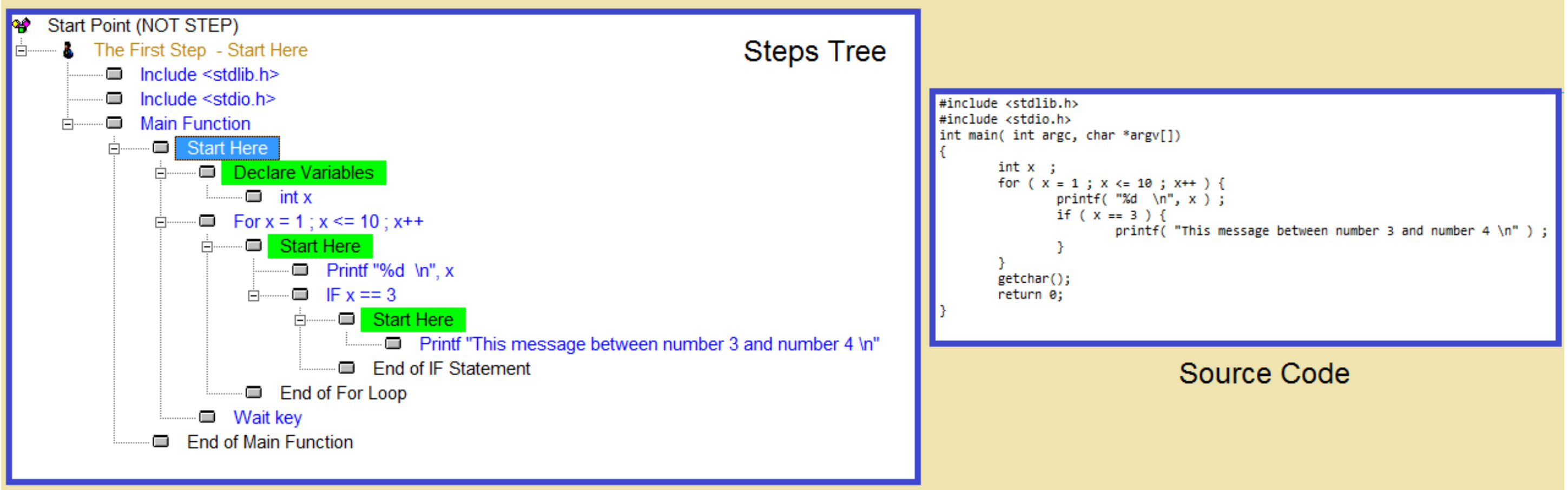

Figure 10: Source Code (in the right) and the equivalent Steps Tree (left).

To better describe the concept, the following equivalences can be used:

- Text code structure IS EQUIVALENT TO Nested structures of Instructions (Steps Tree)
- Instruction IS EQUIVALENT TO a function F(X), where F is the operation or the function and X is the required data.
  - Text code structure IS EQUIVALENT TO Nested structures of F(X). Notice that F can be any function and X can be any input data.

When the above equivalences are applied to the C code shown in Figure 10, we can have a clear view of the essential concept of the GCR technique.

| | |
|---|---|
| Line of code (Source Code) | printf( " This message between number 3 and number 4 \n") ; |
| The operation/function (F) | Print text in new line |
| Data = X | " This message between number 3 and number 4 " |
| Step's Description/Label | Print Text "This message between number 3 and number 4" |

Table 5. Shows the replacement process for one line of code.

At this point many questions need to be answered. How the steps will be generated automatically? How will they determine what the program will do? How can the developer/programmer control and manage the program as she does with traditional languages? GCR uses the Nested Tree (Steps Tree) to represent the structure of the program. Instead of using F(X) which is related to the syntax of the programming language, GCR uses natural language (English, Arabic, etc) to describe the steps (i.e. the name and description of the step or control which could be in any language). Therefore, it’s easy to understand and remember. The last row in Table 5. Implies that a step (visual component) was added to the tree and it was labeled “Print Text “This message between number 3 and number 4””.

## 4.4 PWCT architecture

PWCT is composed of three main layers: VPL Abstract Layer; Middleware Layer; and System Layer. Figure 11. Shows each layer and its constituent components within PWCT’s architecture. In the following sections, each layer and its components are described in more detailes.

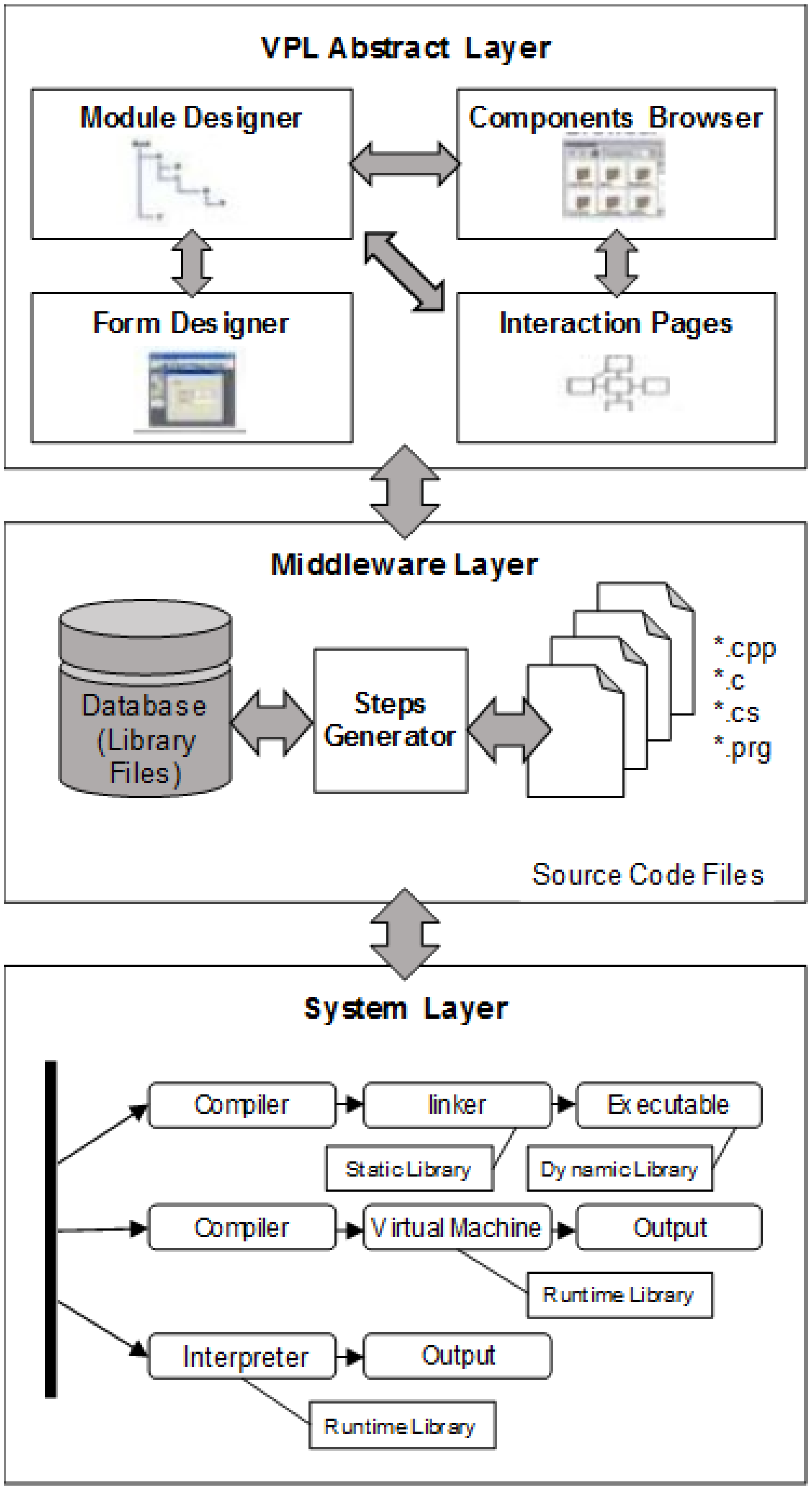

Figure 11: PWCT Architecture.

<u>4.4.1. VPL Layer</u>

The VPL layer is the presentation layer. It provides all the functionality needed to allow a user to perform a specific task. It is composed of four important components: the Goal/Module Designer; the Components Browser; the Interaction Pages; and the Form

Designer. A user would interact with at least three components to produce a meaningful program. For example, a user starts with the Goal/Module Designer to create a new goal (in other words module). Next, the user might choose a visual component from the Components Browser. Then, the user might work with the Interaction Pages to provide the needed data or to specify some attributes. If the user wants to develop a GUI, they can use the Form Designer to do so.

4.4.1.1. Goal/Module Designer

The Goal Designer is the main component in the VPL layer. It can be thought of as an equivalent one to the code editor in a textual programming language. Through the Goal Designer, a developer can manipulate her program using the Steps Tree. The tree structure was chosen to simplify the replacement process. Furthermore, the Treeview control has some special advantages which distinguish it from other controls. These advantages are listed below [99]:

- Hierarchical data structure and relationships between elements can be represented and digested more clearly.
- All controls and their children (data, control structure, etc.) can be entirely seen together in a single window.
- Navigation and paging to focus on a particular part of data are easily manageable with expansion and collapsing functionalities when compared to alternative forms of viewing (i.e. tabular form view, text editor view, or list view).
- The developer can arrange the controls in the run-time. She has the ability to move, update, delete and add any kind of program's control according to her needs.

Developers could have more than one goal and each goal could include more than one step as shown in Figure 12. These steps provide a visual representation of the code which is generated in the background. The developer doesn't write the steps, instead, she selects them from the Components Browser and manipulates them using the Interaction Pages.

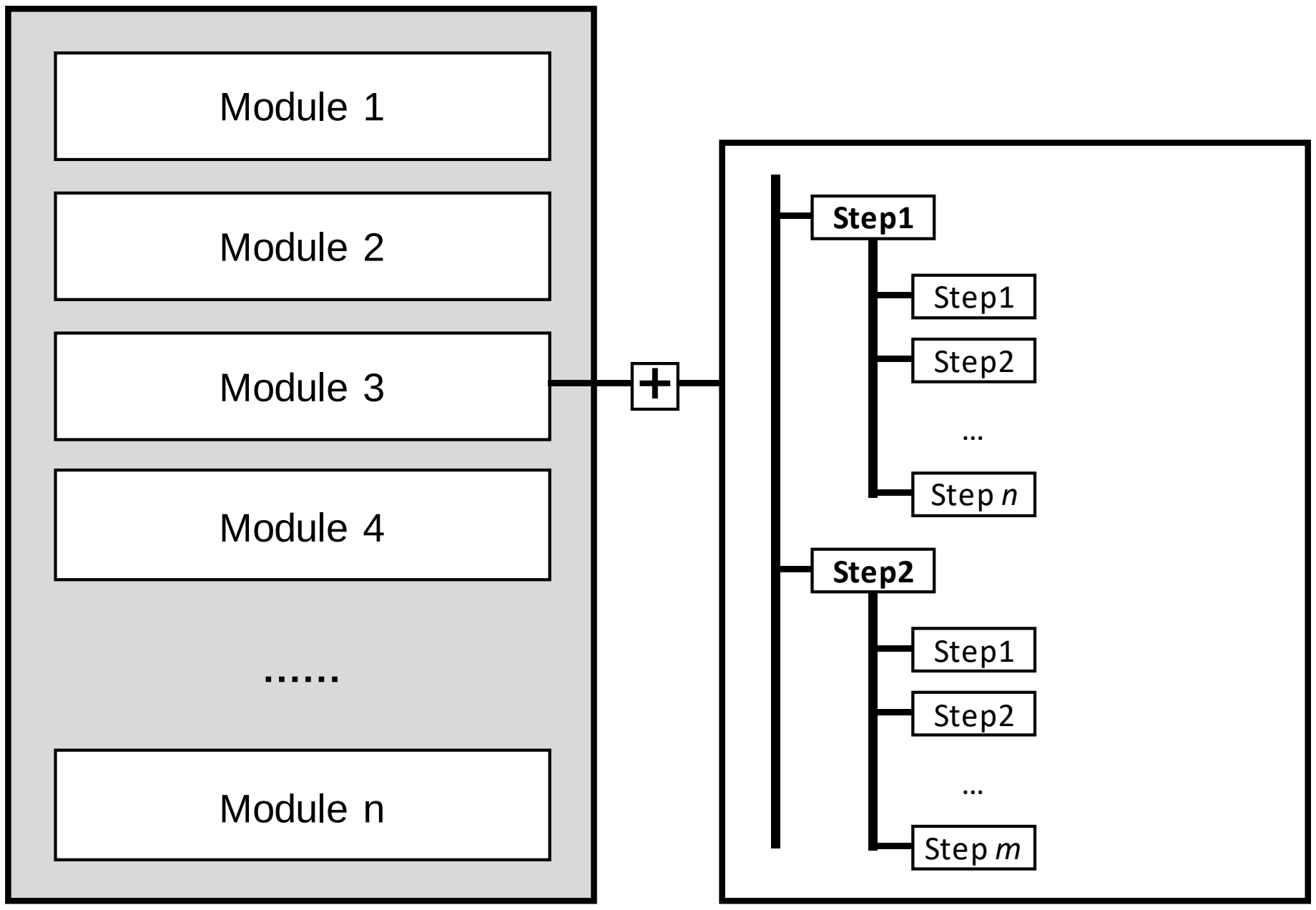

Figure 12: A program in PWCT can have many goals/modules

To make it easier to understand, table 6 shows a comparison between few text-based language's terms and their PWCT counterparts.

| *TPL's Terms* | *PWCT's Terms* |
|---|---|
| Problem | Goal/Module |
| Statement/ Line of code | Step |
| Coding | Selecting & interaction with components |
| Code Editor | Goal/Module Designer |

Table 6: Textual Programming Language's vs Programming Without Coding Technology's terms

In the following, the use case diagram in figure 13 describes the behavior of the Goal Designer when the developer interacts with it. The use case diagram is composed of four main use cases addressing the interaction process. Goal Designer allows developers to efficiently add, delete, or modify both goals and steps. Along with data processing, the developer can arrange the program steps according to his needs, even in the run-time (move-up, move-down, cut, paste, copy, etc.). The developer would need to open the form designer, when he/she defines a new window, to deal with the components and tools and modify their properties according to the requirements.

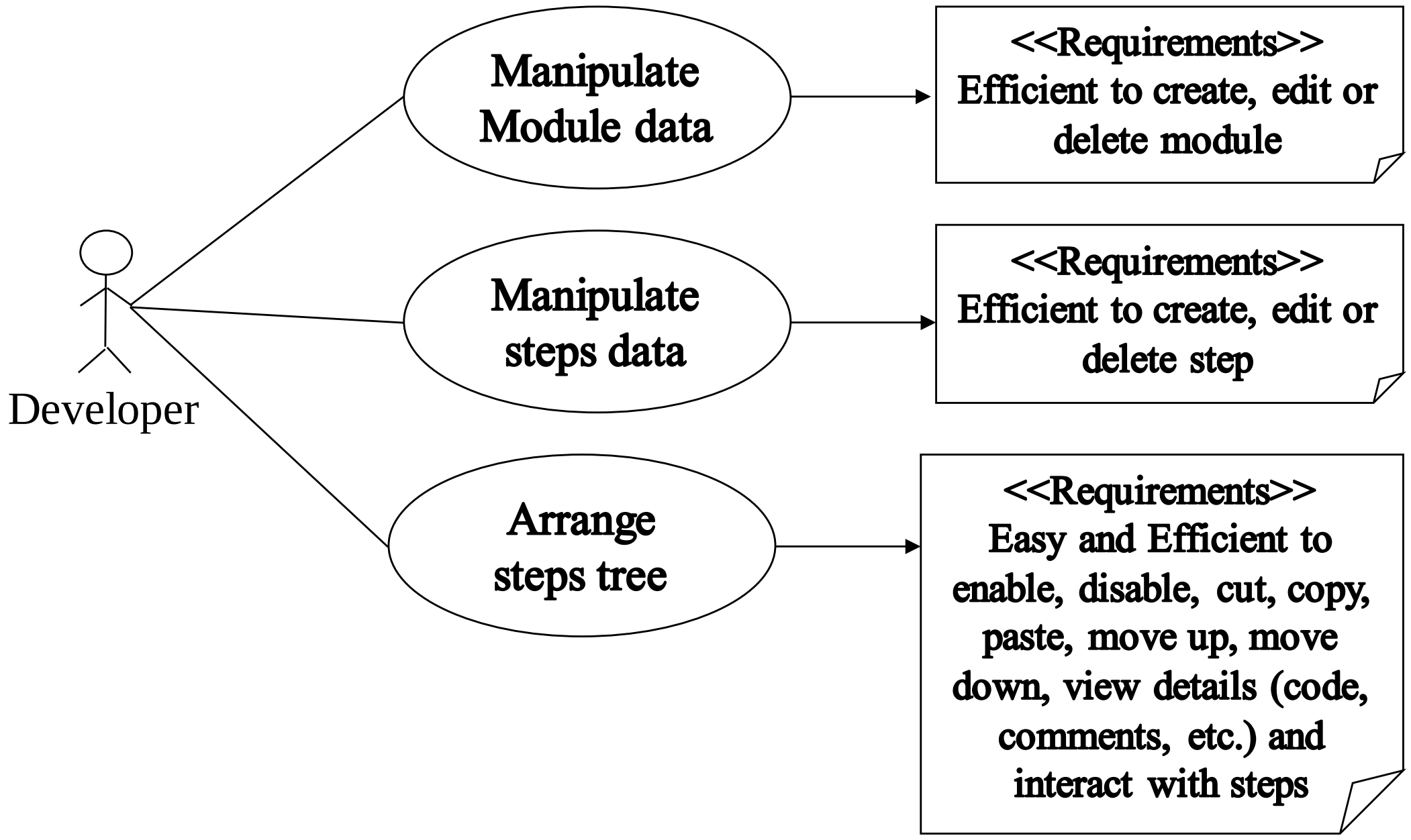

Figure 13: Essential Use Case Model for Goal/Module Designer

4.4.1.2. Components Browser

There are several definitions of the word component in the literature. In [100], components are defined as units of independent production, acquisition, and deployment that interact to form a functioning system. The authors of [101] define components as executable units of code that provide a set of services through a specific interface. Lastly, components were defined as self-contained replaceable parts of a system that encapsulate their data and processes [102]. In PWCT, the last definition is adopted.

The second component of the VPL layer is the Component Browser. Developer uses the Components Browser to search, select, and navigate amongst components in the components' library. Figure 14 shows how a programmer might interact with Components Browser.

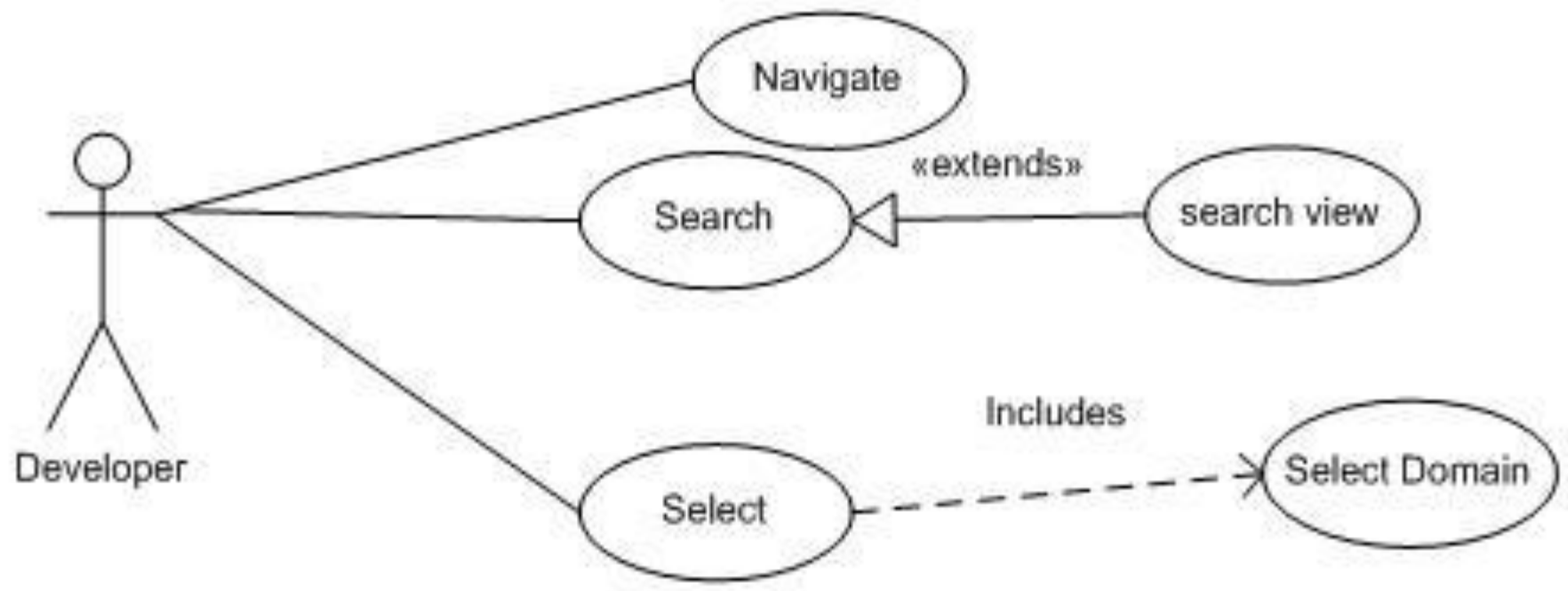

Figure 14: Components Browser Use Case Diagram

In PWCT, a large number of components (496) were built and stored in a library. The developer can search the library for a specific component, select the component to be included in her program, or just navigate through the library to see the available components. The developer can use both the mouse and the keyboard to do such operations. Furthermore, the developer can limit her search to a specific category of components by selecting the required category. In addition to that, the developer can create a new component or modify an existing one if needed.

In general, the components in PWCT are either controls (such as GUI controls like Windows, Labels, Buttons, Textbox) or software modules (such as Database, Threads and Sockets components) that perform a specific service needed by the programmer. A component could be designed to generate code in any language such as Harbor, C#, Python, etc. To better organize the components, several categories were created. Figure 15. Shows the most important categories.

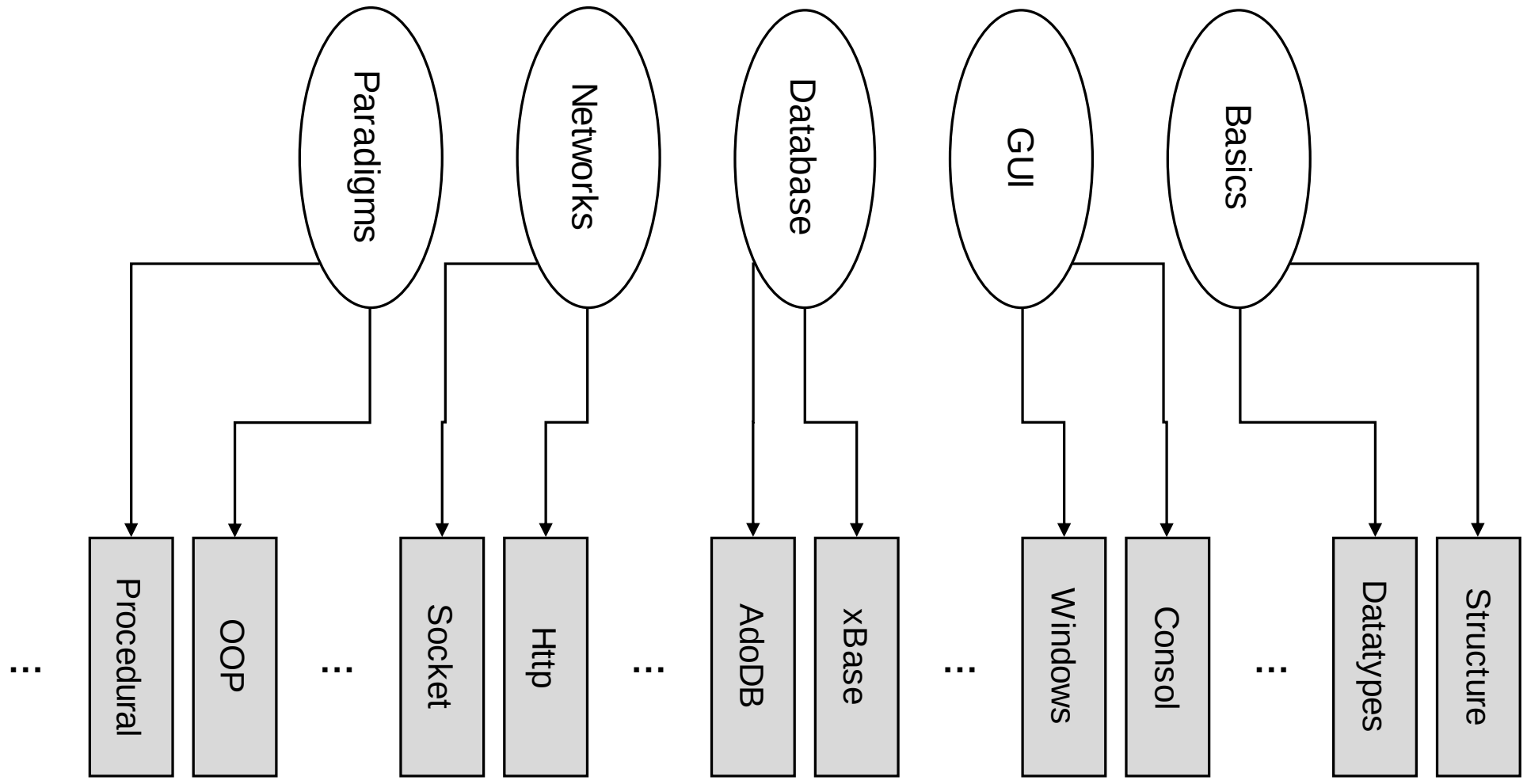

Figure 15: Components' Categories in PWCT Library

When selected, a component will be inserted as a step/s into the Steps Tree. Simultaneously, a link will be created between the generated step in the goal designer and a file of the selected component indicative of the step. The link information is stored in the data repository independently. Users may change/delete some parameters without affecting the component file.

#### 4.4.1.3. Interaction Pages

The interactive pages are acting like a properties panel in traditional languages, but they differ in their simplicity and ease of use. In addition, through the interactive page a user can change and adjust not only a component properties but also its functionality.

For example we have the console application component that can be used to do different tasks related to console applications like (Print text, clear screen, draw rectangles, get input, and menus).

Also we have the classes' components where each class is represent by one component contains three pages to (Set Attributes, Get Attributes and Invoke methods).

#### 5.3.1.4. Form Designer

When a Graphical User Interface (GUI) is needed, a developer can use the Form Designer to create forms for her application. Similar to those that exist in Visual Studio; Eclipse; and other programing environments, the Form Designer in PWCT allows the user to build a form and add to it any typical component such as: a button, textbox, label, etc.

There are components for controls events that determine the procedure/method that will be invoked when an event is fired. Using the events components the user can select and event (click event for example) then determine the procedure that will be called when the user click on the control. The procedure name can be written or selected from the list of names that appears during typing (intellisense). Later the user define the procedure using the define procedure component and add steps to the procedure that will be executed when the procedure is called.

### 4.4.2. PWCT's Middleware Layer

A Middleware is software that connects applications with the underlying operating system or database. The software consists of a set of services that allows multiple processes running on one or more machines to interact [103]. PWCT's middleware connects the user's view (i.e. the four components in the VPL Layer) with the system's view (i.e. processes in the System Layer).

This layer (Middleware Layer) has four main functionalities: interpretation of user's interactions with components in the VPL Layer; generating the Steps Tree according to user's interactions; replacing variables in the script that generate the code with the values entered during the user's interactions with components; and generating the source code.

As demonstrated in figure 11, the Middleware layer consists of three units; the Step Generator, a database, and the source code files. The most important part of the middleware layer is the Step Generator. Generating the code in this layer if fairly simple and straight forward, Algorithm 1 illustrates the mechanism to generate the source code and to store it in a data repository.

*Lib_file* ¬ open library file (*.TRF)

*InteractionPageTable* ¬ load interaction pages data from (*Lib_file*)

*CodeMaskTable* ¬ load code mask data from (*Lib_file*)

*CodeMatchTable* ¬ load code match data from (*Lib_file*)

**if** *InteractionPageTable is not empty* **then**

  *tmpDataTable* ¬ empty

  **for** *each page in InteractionPageTable,* **do**

    *page_file* ¬ open library file(*.IDF)

    *tmpDataTable* ¬ load controls name and values from (*page_file*)

    close the interaction page file

  **endfor**

  **for** *each item in CodeMatchTable*, **do**

    *Control_Name* ¬ load controls name from *CodeMatchTable*

    *Control_Variable* ¬ load controls variable from *CodeMatchTable*

    *Control_Value* ¬ load controls value from *tmpDataTable(Control_Name)*

    *Codemask_Variable* ¬ *Control_Value*

  **endfor**

**endif**

**for** *each item in CodeMaskTable,* **do**

**if** *the item* is *an instruction* **then**

**Execute** *the instruction* with respect to GCR Rules

**else**

**Add** step code

**endif**

**endfor**

close the library file

Algorithm 1. Code Generation Algorithm

To illustrate how this layer works, a simple scenario is presented below:

When a developer selects a component using the components browser, an interaction page will open which asks her to provide more details. For example, if the user selects a print component, the interaction page will ask the user to insert the value she wants printed. This interaction will be interpreted and fed to the Step Generator the results of the interaction process will be an input for the process of generating new steps or updating an existing one in the steps tree as describe in algorithm 1. The component may have more than one interaction page. It depends on the type of the component, for examples if the selected component is a class, it needs three interaction pages; set properties, get properties and the data. Ultimately, when the data table is ready with the required variables, the process of generating code starts by matching the variables with values then masking this operation to get the new/updated code line. For example, printing “Hello World” will proceed as shown in the following tables:

| *Variable* | *Value* |
|---|---|
| Page1_Text1 | "Hello World” |

Table 7: Sample of the variables and values

The Code Mask Simple Script for printing text looks like the following:

<**RPWI:NEWSTEP**> *Print Text – New Line – ( <T_V1> )*

*cout << <T_V1> << "\n" ;*

The component file include a matching table to match between interaction pages variables and Code Mask Variables.

| *Interaction Page Variable* | *Code Mask Variable* |
|---|---|
| Page1_Text1 | T_V1 |

Table 8: Sample of the variables of Interaction Page Variable and Code Mask

The Code Mask is written using simple rules which enables controlling and determining the generated code. Table 9 depicts these rules.

| *Rule* | *Description* |
|---|---|
| Put code mask variables between '<' and '>' | Syntax: <variable name> Example: <Name> |
| <RPWI:POSITIVE> | This command used to ensure Generating the code between <RPWI:TEST> and <RPWI:ENDTEST> if the result of the test is true. |
| <RPWI:NEGATIVE> | This command used to ensure Generating the code between <RPWI:TEST> AND <RPWI:ENDTEST> if the result of the test is false. |
| <RPWI:VALUE> | Determine that value that will be compared with the value of the variable after <RPWI:TEST> , if the two values are typical , the test result is true, if not, the test result is false. |
| Nested <RPWI:TEST> and <RPWI:ENDTEST> | |
| <RPWI:INFORMATION> | Used to generate output in Information tab of the step details, in goal designer |
| <RPWI:NOTE> or <*> | This is not more than comment |
| <RPWI:NEWSTEP> | To create step in Goal Designer |
| <RPWI:TEST> <Variable><br>more instruction<br><RPWI:ENDTEST> | |
| <RPWI:PUTMARK> NUMBER_FROM_1_TO_30 | Save the ID of the current step (active step) to special area (Range 1-30) |
| <RPWI:SETMARK> NUMBER_FROM_1_TO_30 | Set a step as the active/parent using a saved ID (Range 1-30) |
| <RPWI:IGNORELAST> <ANY CHARACTER> | Used for deleting the last occurrence character from the generated code |
| <RPWI:NEWVAR> VARIABLE_NAME | Used for creating new variable |
| <RPWI:SETVARVALUE> VARIABLE_VALUE | Used for setting the value of the variable (Active Variable) |
| <RPWI:SELECTVAR> VARIABLE_NAME | Set the active variable |
| <RPWI:REPLACEVARSWITHVALUES> | Replace variables with corresponding values |

Table 9: GCR Rules to write a Code Mask

### 4.4.3 PWCT's System Layer

This layer is a low-level layer that deals with the source code generated by the upper layer. PWCT generates and executes code in four text-based programming languages namely C#, Java, Harbour, and Python and hence, the System Layer had to support the different mechanisms in all four languages to produce executable code.

Programs that were written by text-based languages are implemented either by an interpreter, a compiler or a combination of both [104]. While Python uses only an interpreter and C# uses a compiler, Scala uses both a compiler and an interpreter. . Consequently, the System Layer support compilers and interpreters at the same time. After generating the source code in any of the supported text-based language (C#, Java, Harbour and Python), executing the code follows the procedures of that specific language. PWCT doesn't impose any restrictions on the process of building the source code.

# Chapter 5: Implementation and Validation

## 5.1 Usability and Capability Evaluation

To assess the usability and capability of PWCT, a free online remote learning course was offered. The goals of the course were to get feedback about the tool and to monitor the learning and progress of the participants. A total of 298 participants were accepted. For each registered participant, the team provided a folder where the participant's work is stored. The course outline was very simple. At the beginning, participants were asked to download the tool and install it based on a short tutorial. Next, Participants were asked to watch videos that show how to develop different programs using PWCT. At first, the programs are very simple and later they become more advanced. After watching the video, the participants were asked to develop the program they just saw in the video. If the participant was successful, he/she was asked to send the files which were saved to the participant's folder. Those who faced problems during the development of the project sent their questions to the project's team and got the needed help. The lessons were classified into three levels of difficulty. The first level was an introduction to programming which contained 32 lessons. The second level was an introduction on how to develop database applications which contains 18 lessons. The third level was more advanced and introduced developing database applications using templates which contained 20 lessons.

Table 14. In the Appendix B. shows some information about the all the participants such as the number of developed applications as well as the level of expertise of the participant. All of the developed applications can be found at URL: http://sourceforge.net/projects/doublesvsoop/files/PWCT%20For%20MS-Windows/PWCT%20Students/

The main results of the course can be summarized as follows:

- All students completed their first application without problems.
- 41 level-3 (advanced) students were able to develop real world database applications using PWCT.
- 25 level-2 (intermediate) students were able to develop simple database applications using PWCT.
- 232 level-1 (beginner) students were able to develop simple applications using PWCT.

46 level-1 students created only one application.

From the progress of the students. The main results are presented in table 10 below.

| Ease-of-use Results | |
|---|---|
| Students that don't like PWCT | 15% (Students created only one application) |
| Students like to learn PWCT | 85% (Students created more than one application) |
| Gained Knowledge Results | |
| Students learned programming basics only (Level 1) | 78% |
| Students learned basics of database programming (Level 2) | 8% |
| Students learned how to create professional database applications (Level 3) | 14% |
| Average Learning Curve (Time required for watching the Course) | |
| Time required to learn programming basics according to the course (Level 1) | 4 hours, 41 minutes and 42 seconds |
| Time required to learn basics of database programming according to the course (Level 2) | 2 hours, 39 minutes and 39 seconds |
| Time required to learn how to create professional database applications according to the course (Level 3) | 4 hours, 37 minutes and 22 seconds |

Table 10: results from students

It is very clear from the data above that the majority of the participants liked the tool. 78% of the participants were able to learn the basics of programming by developing few applications only. And finally the average learning curve was extremely short even for the advanced lessons.

## 5.2 Efficiency Evaluation

To make sure that PWCT's performance was comparable to that of other VPL's and even textual languages, a programming experiment was carried out. In this experiment PWCT competed against two well-known general-purpose VPLs (Tersus and Limnor), and two reputable VPE (Visual Studio and Net Beans). In the evaluation One professional programmer for each language was hired to develop programs in that specific language. All programmers were asked to develop 20 simple to intermediate applications. The authors monitored the time and memory needed to complete each task. Table 7 shows the averages of all tasks. It is extremely obvious that PWCT's outperformed the other four languages by a considerable margin. Table 11. Shows that PWCT's time was 8.5% less than Visual Studio (the second best) and 55.6% better than Net Beans. This criterion is very important as it hints at the productivity of the language. Similarly, while PWCT used only 4.5 MB of RAM on average, the second best (Tersus) used 12.1 MB. Figure 16 shows that PWCT outperforms all other languages as far as time and memory are concerned.

| *Criteria\ Alt.* | *General Purpose Visual Programming Languages* | | | *Visual Programming Environments* | |
|---|---|---|---|---|---|
| | PWCT | Limnor | Tersus | Visual Studio | Netbeans |
| Average time required to create and test the application | 106.66 seconds | 144 seconds | 140 seconds | 115.66 Seconds | 165 seconds |
| The steps to create and test the application | 41 steps | 54 steps | 46 steps | 26 Steps | 33 steps |
| Average time required for each step | 2.6 seconds | 2.66 seconds | 3.04 seconds | 4.45 seconds | 5 seconds |
| Average Time spent using keyboard in user steps | 106.66 seconds | 13 Seconds | 18 seconds | 31.1 seconds | 30 seconds |
| Memory used before application design starts | 13356 Kbyte | 13460 Kbyte | 50268 Kbyte | 51276 Kbyte | 156052 Kbyte |
| Memory used after the application design | 17880 Kbyte | 29292 Kbyte | 62392 Kbyte | 109420 Kbyte | 173352 Kbyte |
| Memory consumed in application representation during design time | 4524 Kbyte | 15832 Kbyte | 12124 Kbyte | 58144 Kbyte | 17300 Kbyte |

Table 11: The results of simple applications development

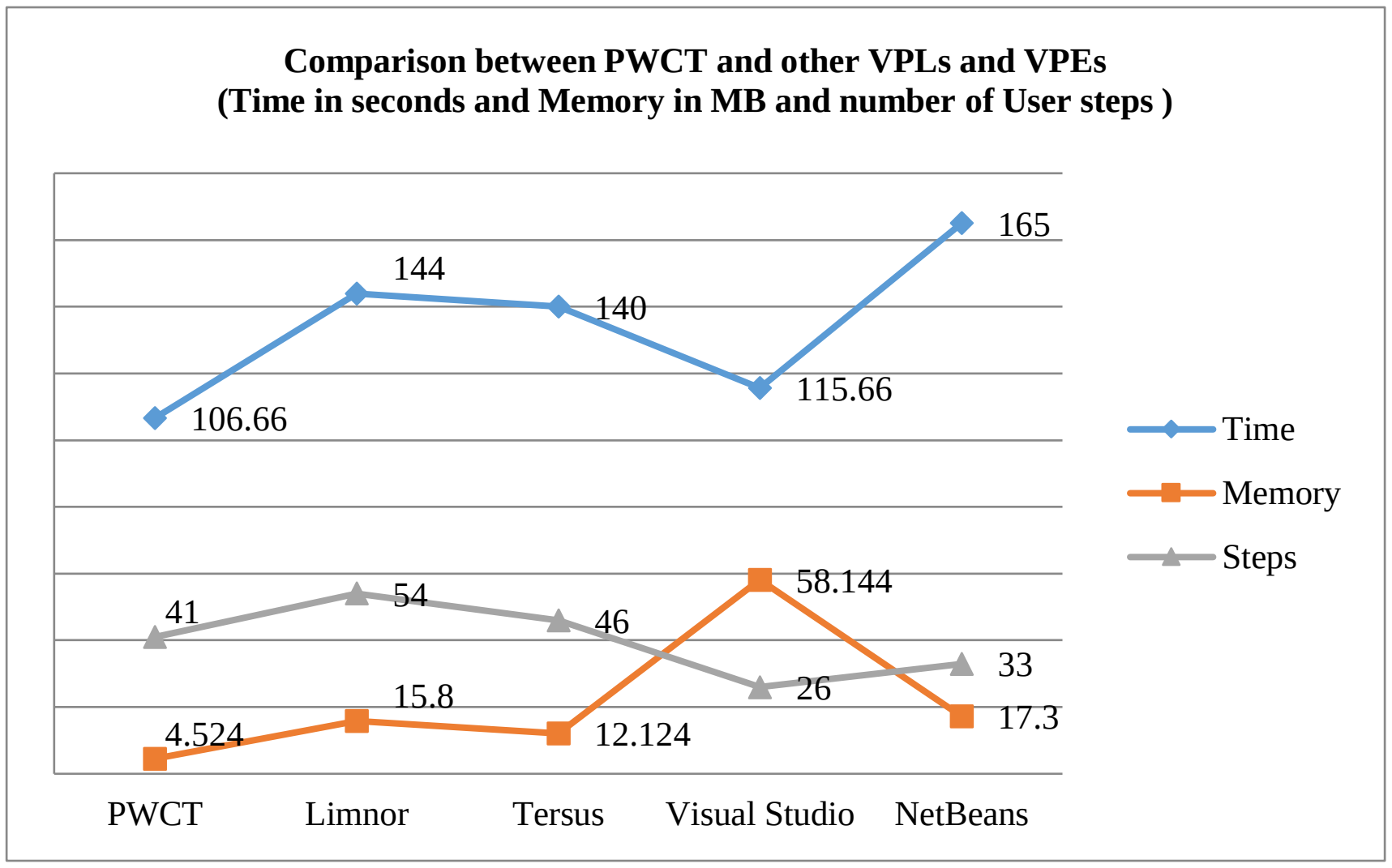

Figure 16: Time in seconds and Memory in MB and Number of user steps

To make sure that PWCT outperforms Visual Studio even on large applications, the authors carried out a second experiment to develop one large application in both PWCT and VS (The application manage projects flow in the Riyadh Valley Company Occupation).

Table 12 below illustrates the results of this experiment.

| | *PWCT* | *Visual Studio (CS + ASP.NET)* |
|---|---|---|
| Lines of source code | 20477 lines | 20606 lines |
| No of source files | 96 files | 102 files |
| No of forms | 26 forms | 32 forms |
| No of Generated steps (Visual Source) | 7616 | NA |
| No of Interaction Processes | 4723 | NA |
| Time required for writing source code line in visual studio or doing Interaction process in PWCT | 2 * 1.12(Average time for user step) = 2.24 seconds | 2.07 seconds |
| Pure Coding/Development Time - Effort Time only | 2.24 * 4723 = 10579.52 seconds = 2.94 hours | 2.07 * 20606 = 42654.42 seconds = 11.85 hours |

Table 12: The results of large application development

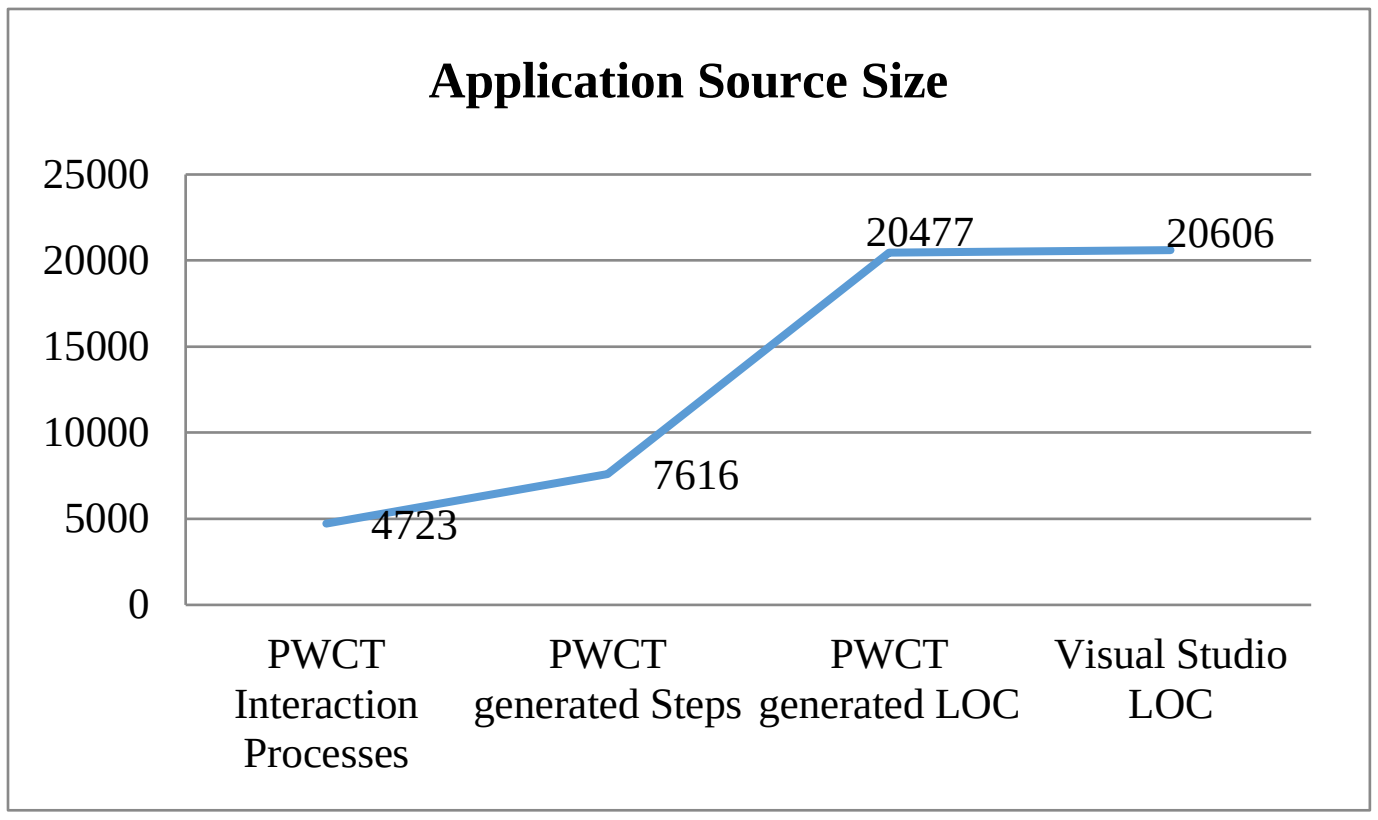

Figure 17: The differences of Application source size between PWCT and Visual Studio

In Figure 17 visual studio LOC are 20606 lines, which have been written during the development time. They also represent the source code managed by the programmers during maintenance and any future development. On the other hand, PWCT LOC got almost the same number 20477, and we expect this result as we developed the same application. Nevertheless,

LOC in PWCT generated, managed, and maintained in the background. The generated steps that actually represent the application are 7616 which approximately near to one third of the LOC. Thus, source in PWCT is more efficient to read and maintain.

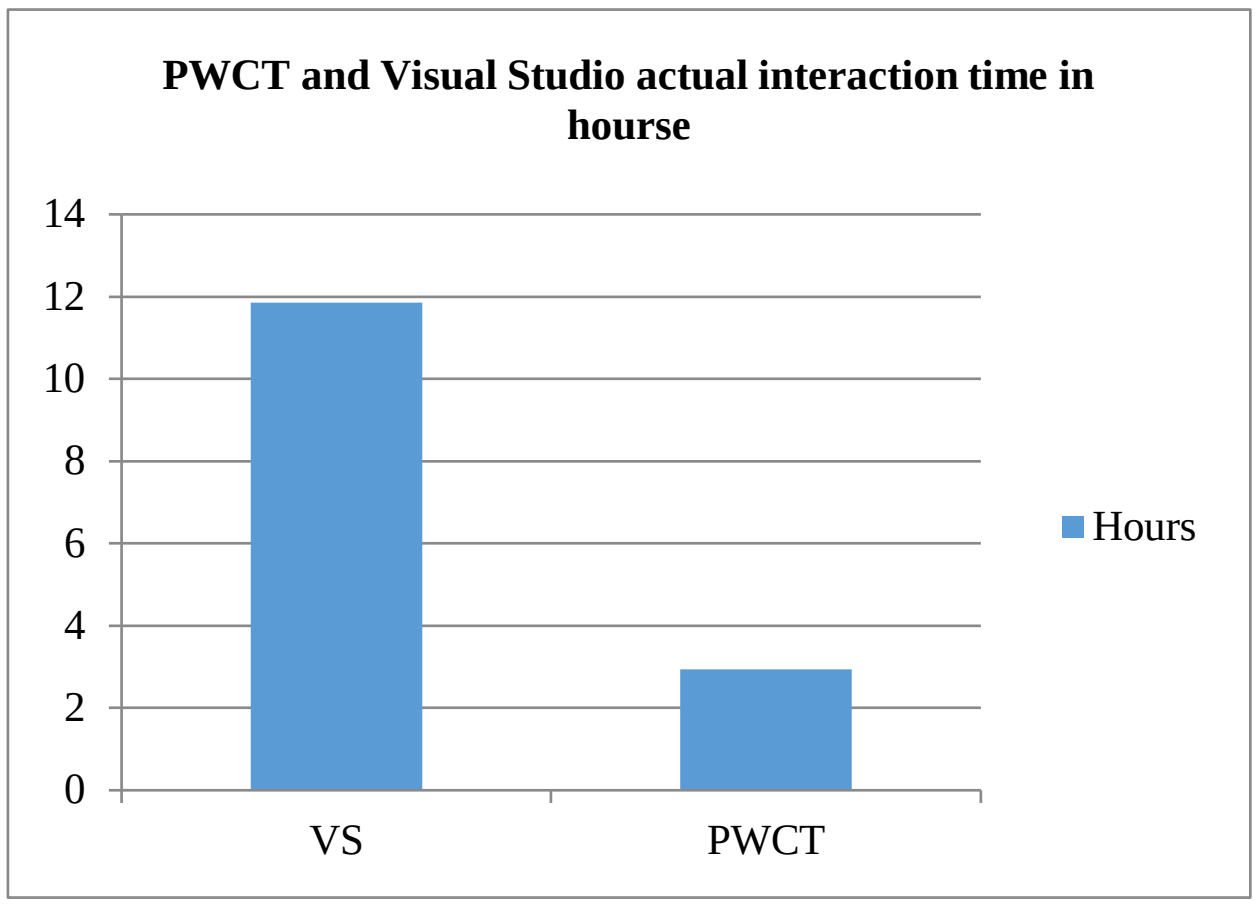

Figure 18: PWCT and Visual Studio actual interaction time in hours

Figure 18 depicts the actual average time consumed in the interaction with both VS and PWCT. It is obvious that the average time to develop the application with PWCT was much less than that of VS. in fact, only 25.38% of the time was needed to develop in PWCT.

## 5.3 Reputation Evaluation

PWCT is very popular on Sourceforge. Out 300,000 projects on Sourceforge, PWCT rank is 3 in the science and engineering category as of April 2016. The number of weekly downloads (environment, samples and tutorials) according to Sourceforge is more than 70,000 downloads.

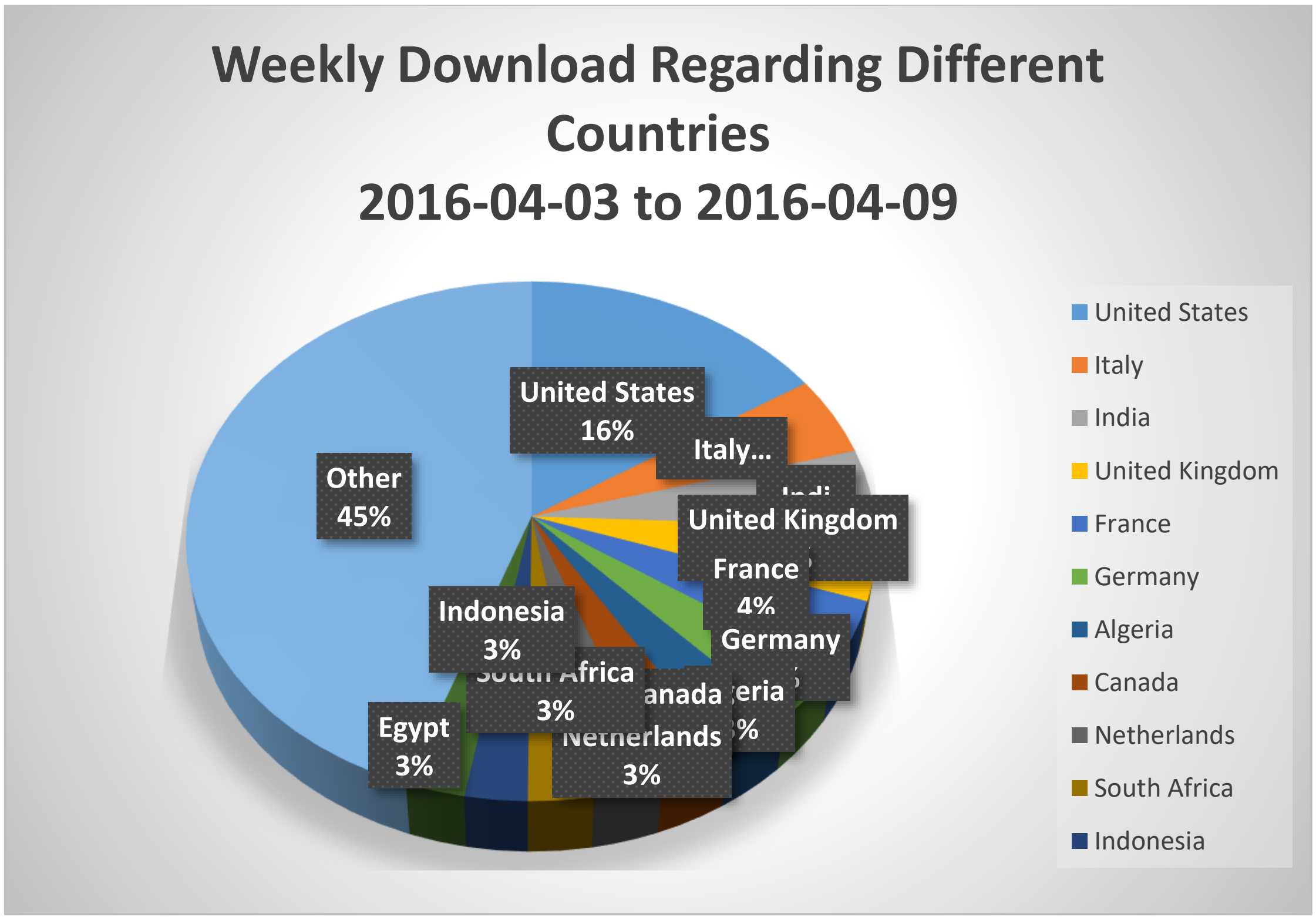

Figure 19: Weekly downloads according to www.sourceforge.net

According to Sourceforge's information which is depicted in figure 19, USA has the highest download rate of all countries followed by the Italy, India, United Kingdom, France, Germany, Algeria and Canada in the 8th place. Moreover, according to Sourceforge's user's satisfaction information, 93% of all users are satisfied with PWCT while only 7% are not as illustrated in Figure 20.

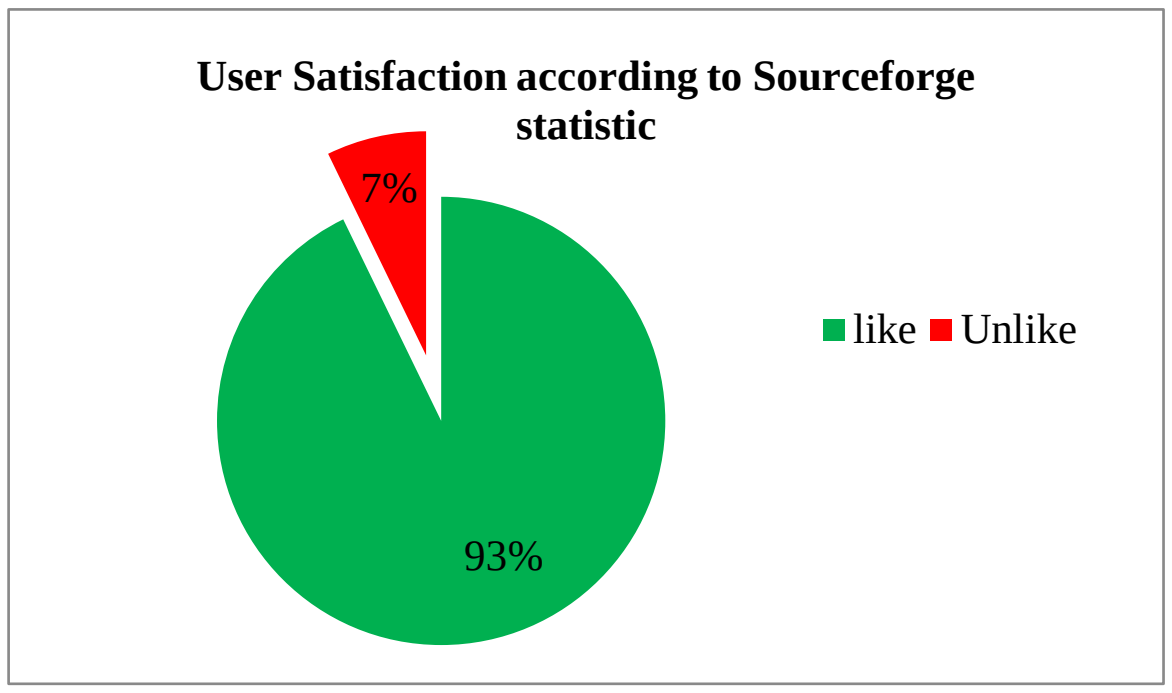

Figure 20: User satisfaction according to [www.sourceforge.net](http://www.sourceforge.net)

Many applications were developed using PWCT in several domains, such as: multi-media, database, network, business management systems, and Mathematics (for more information please see http://doublesvsoop.sourceforge.net/pwcthelp/main.htm). Figure 21 shows two applications made by PWCT's users. Both applications are related to multimedia. The first one is a movie player and the second is an audio player.

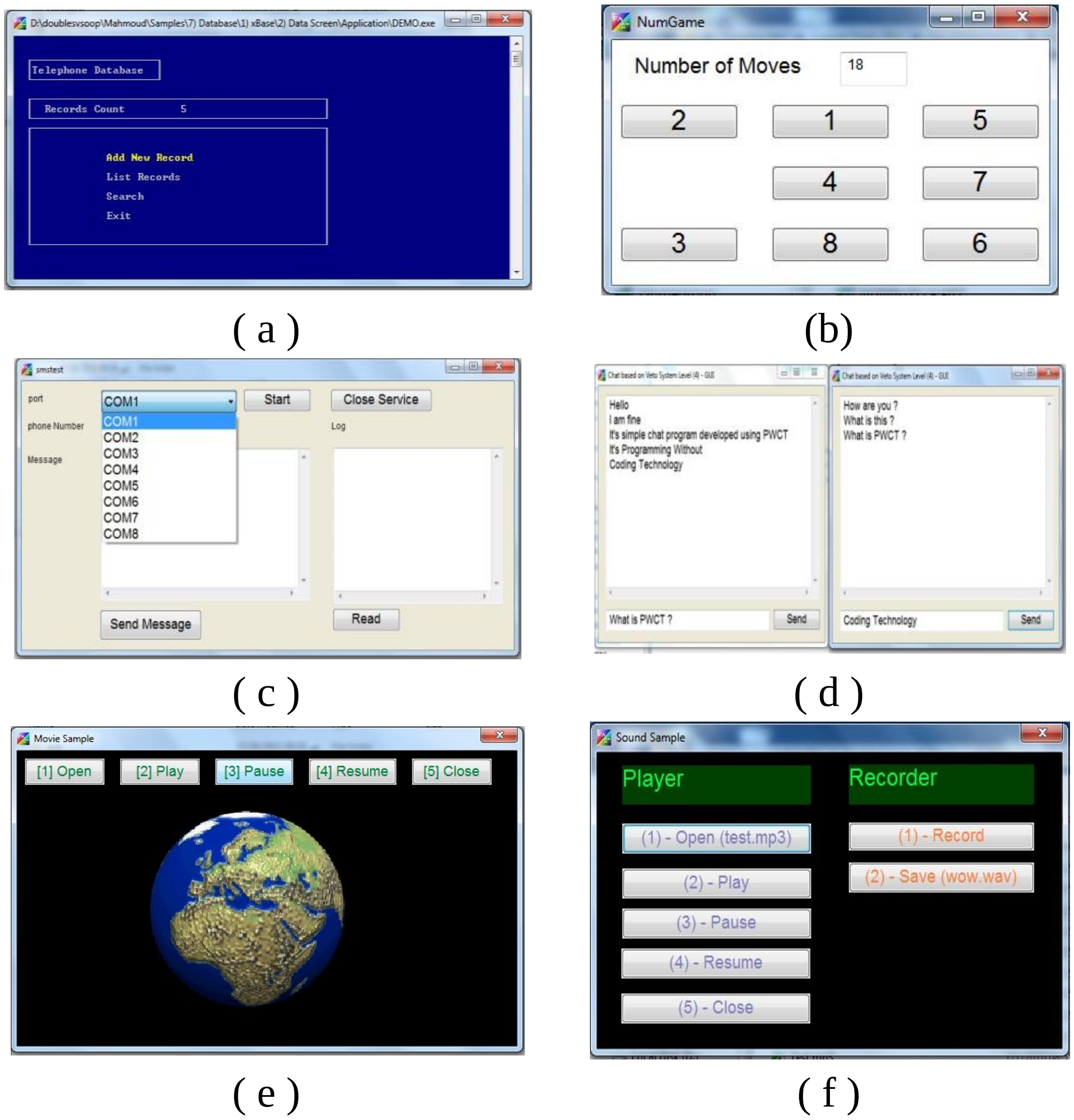

Figure 21: Multimedia applications screen snapshots developed using PWCT

Using PWCT, one can create a console application and this feature is not available in the other VPLs even the general-purpose ones as shown in the first screen snapshot in Figure 21. The second snapshot depicts a famous puzzle game application, which was built by one of the users of PWCT. The third application shows a computer-interfacing application to send SMS message using an SMS Gateway connected to the computer through serial communication. The fourth application is chatting software that uses TCP/IP protocol for establishing a connection between two nodes, one works as client and the other works as server.

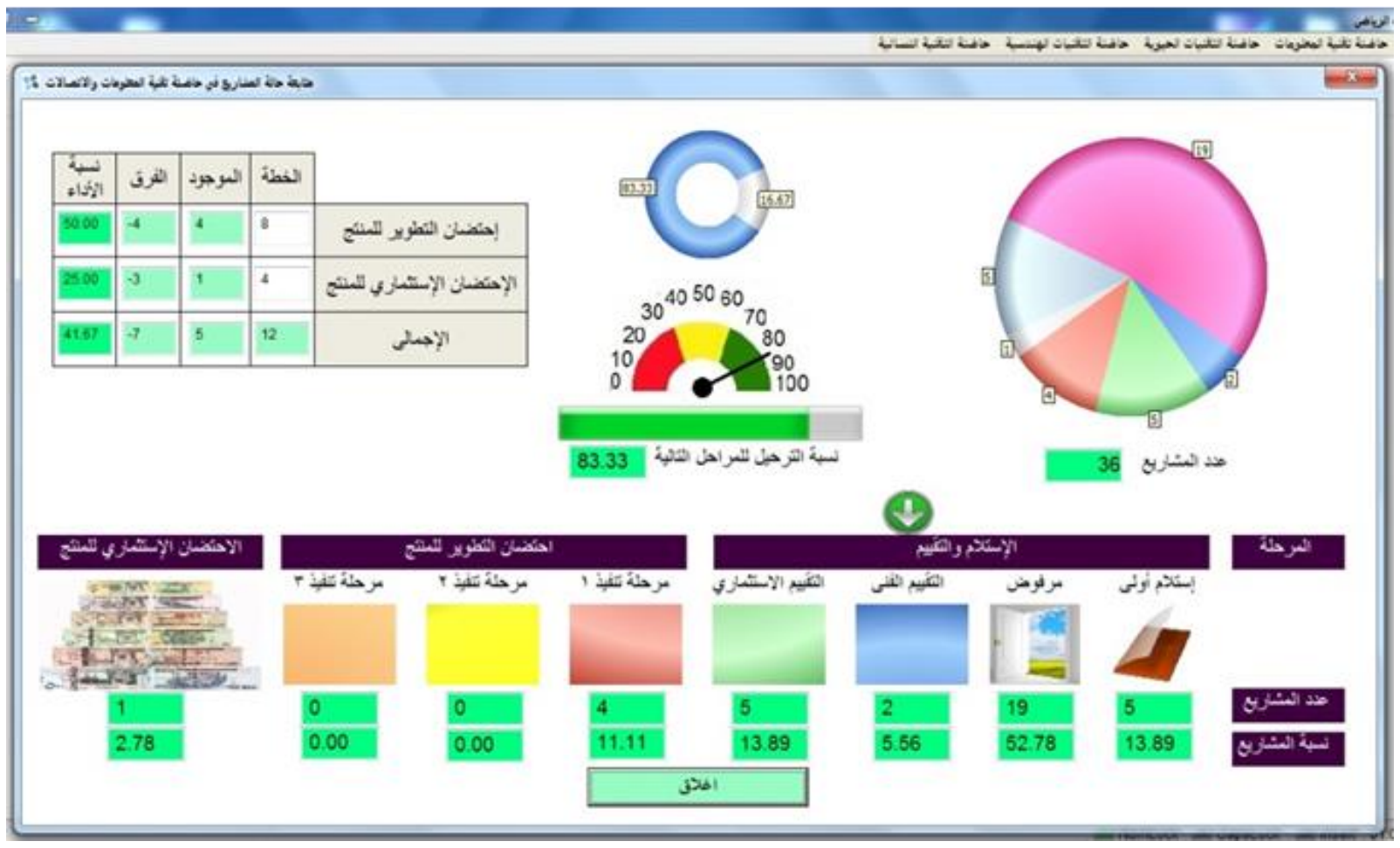

Figure 22: RVC Business applications screen snapshots developed using PWCT

PWCT has not ignored the business management systems, where a group of programmers in Riyadh Valley Company, built a business application (Projects Management Application) for the company to follow-up its projects and monitor the performance of its divisions as showing in the dashboard in Figure 22. The screen snapshot in Figure 23. Depicts a multi-user client-server database application for managing car-hiring companies. This software achieved by one of the active users of PWCT in Netherlands.

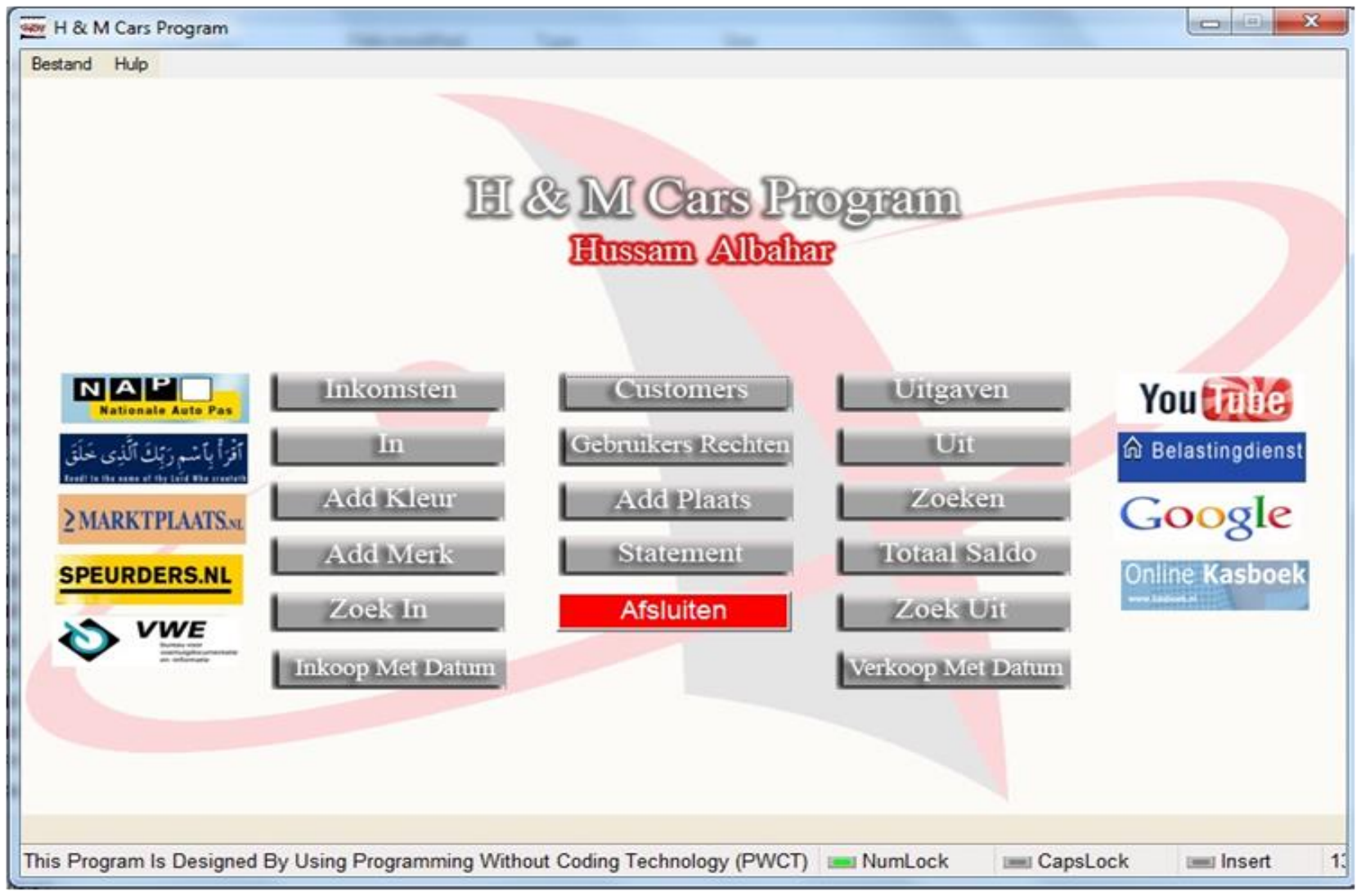

Figure 23: Multi-user client-server database application developed using GCR method

The screen snapshot in Figure 24. Depicts a multi-user client-server database application for a factory to manage orders data and their progress (Production Follow-up Application). The application include items data and their amounts in the store. The application is designed based on the requirements of the Al-Jabreen Factory in (Riyadh).

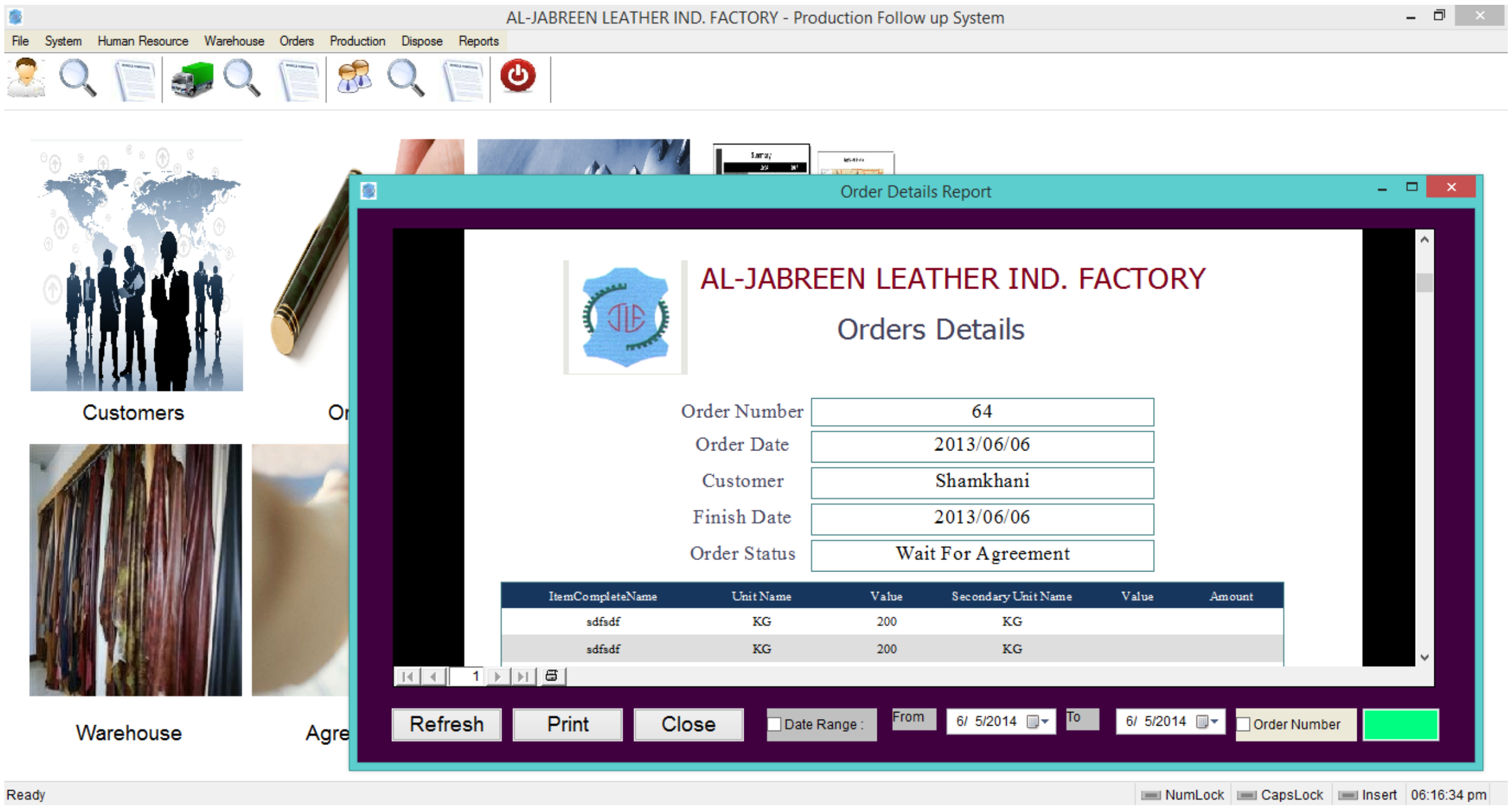

Figure 24: Production Follow-up Application developed using the GCR method.

The screen snapshot in Figure 25. Depicts a multi-user client-server database application for a training center (Courses Management Application) includes data store & analysis of (Courses, Trainers, Students, Courses review by students and Certifications). The application is designed based on requirements of Izdehar Training Center in (Jeddah).

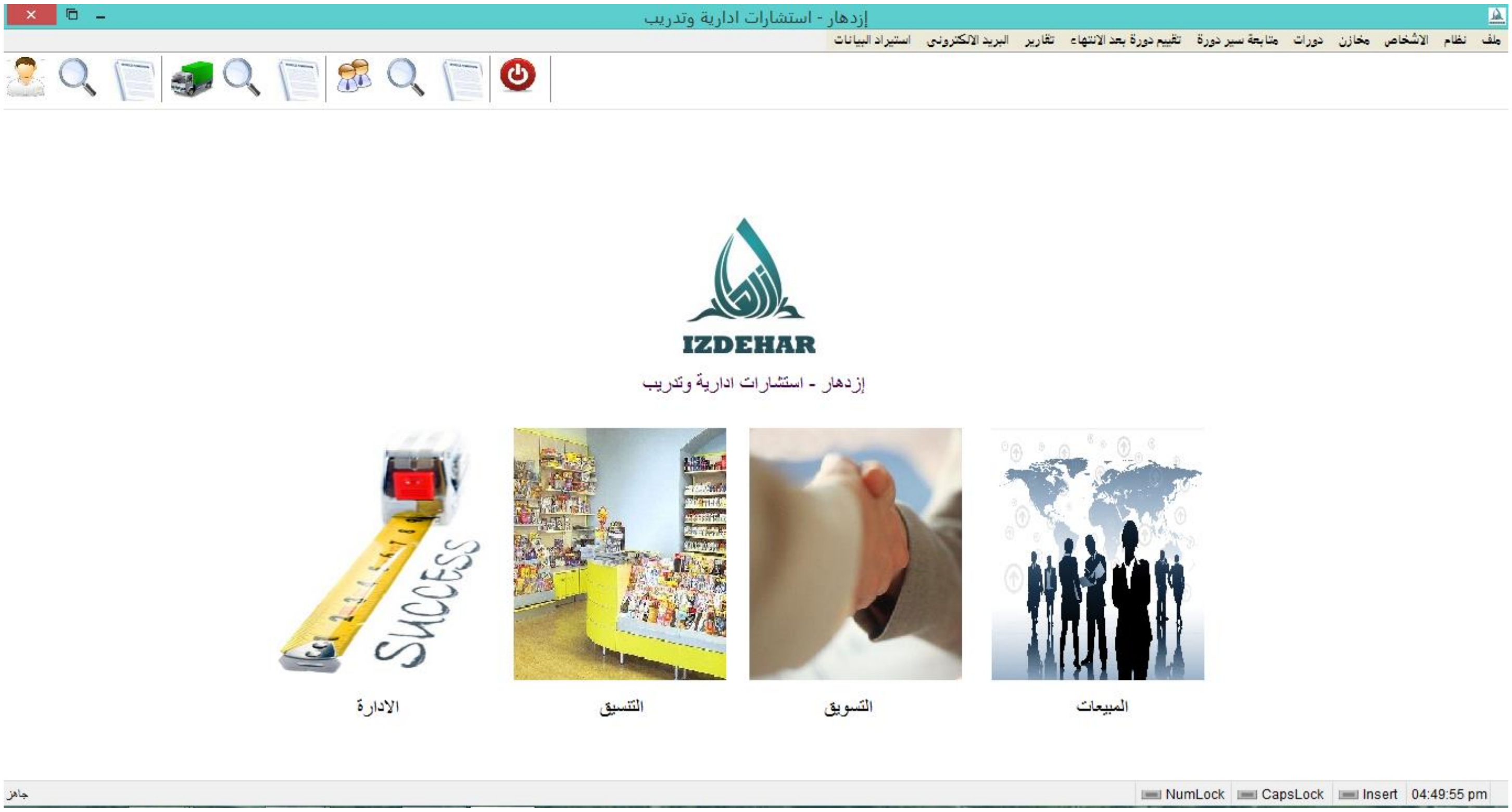

Figure 25: Courses Management Application developed using the GCR method.

Each application (Projects Management Application, Production Follow-up Application and Courses Management Application) can work on one machine (for testing) and can be deployed in a Client-Server environment for production. Applications comes with a control panel to define users and set permissions. Also applications support changing the database management system (DBMS) and tests are done using MS-Access, Microsoft SQL Server and MySQL. The applications are real world application based on real requirements of the company that will uses the application and during the design and the development we used the Object-Oriented Programming (OOP) paradigm where each application contains a group classes and components.

The screen snapshot in Figure 26. Depicts an Implementation of a new Localized Algorithm for detection of Critical Nodes using PWCT.

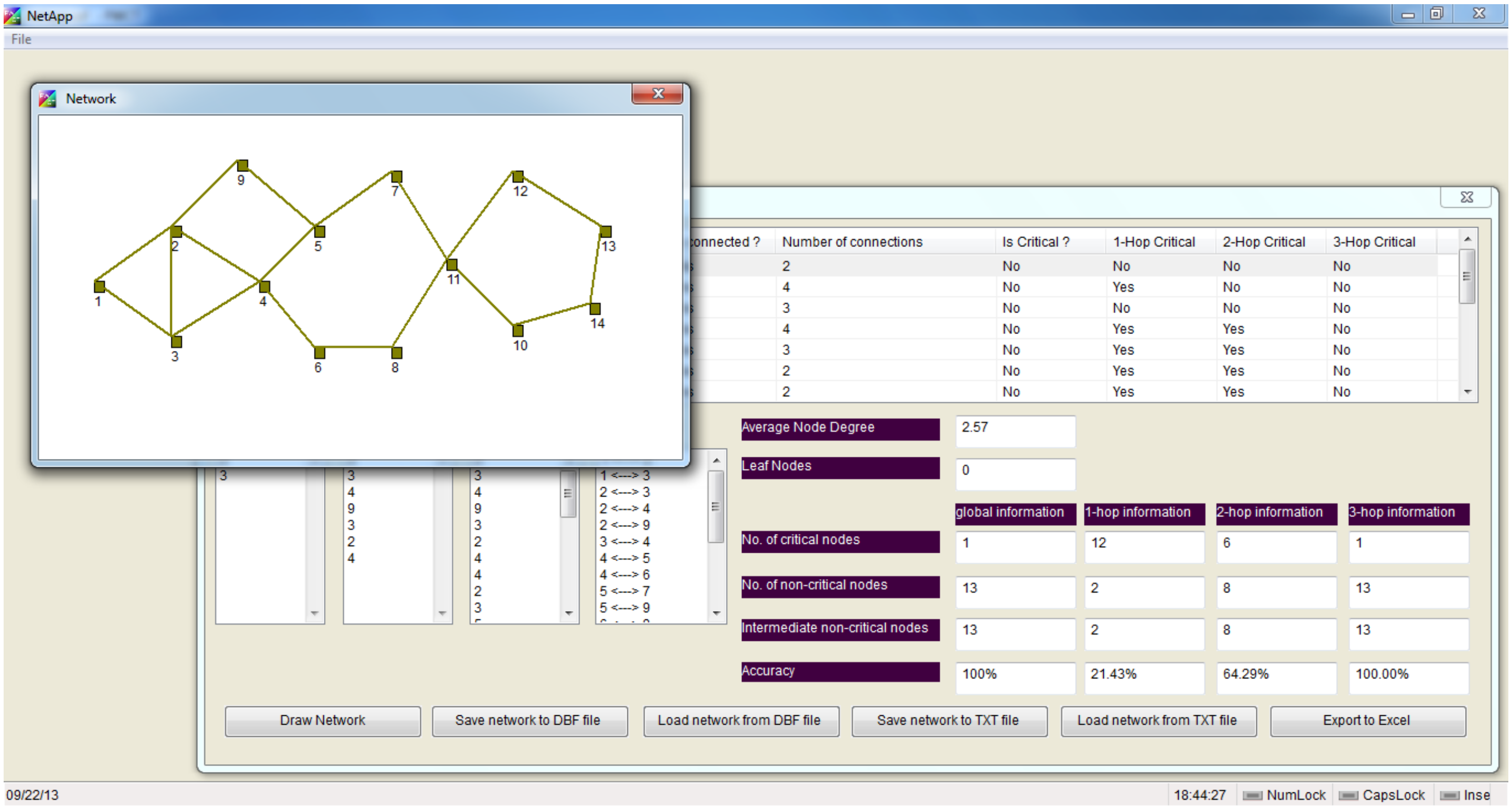

Figure 26: Critical Nodes Detection Application using the GCR method.

The screen snapshot in Figure 27. Depicts an Implementation of Natural Programming Language called Supernova using PWCT based on the Harbour Language through the HarbourPWCT visual programming language.

The screen snapshot in Figures 28, 29 and 30. Depicts an Implementation of Scripting Language called Ring using PWCT based on the C Language through the CPWCT visual programming language.

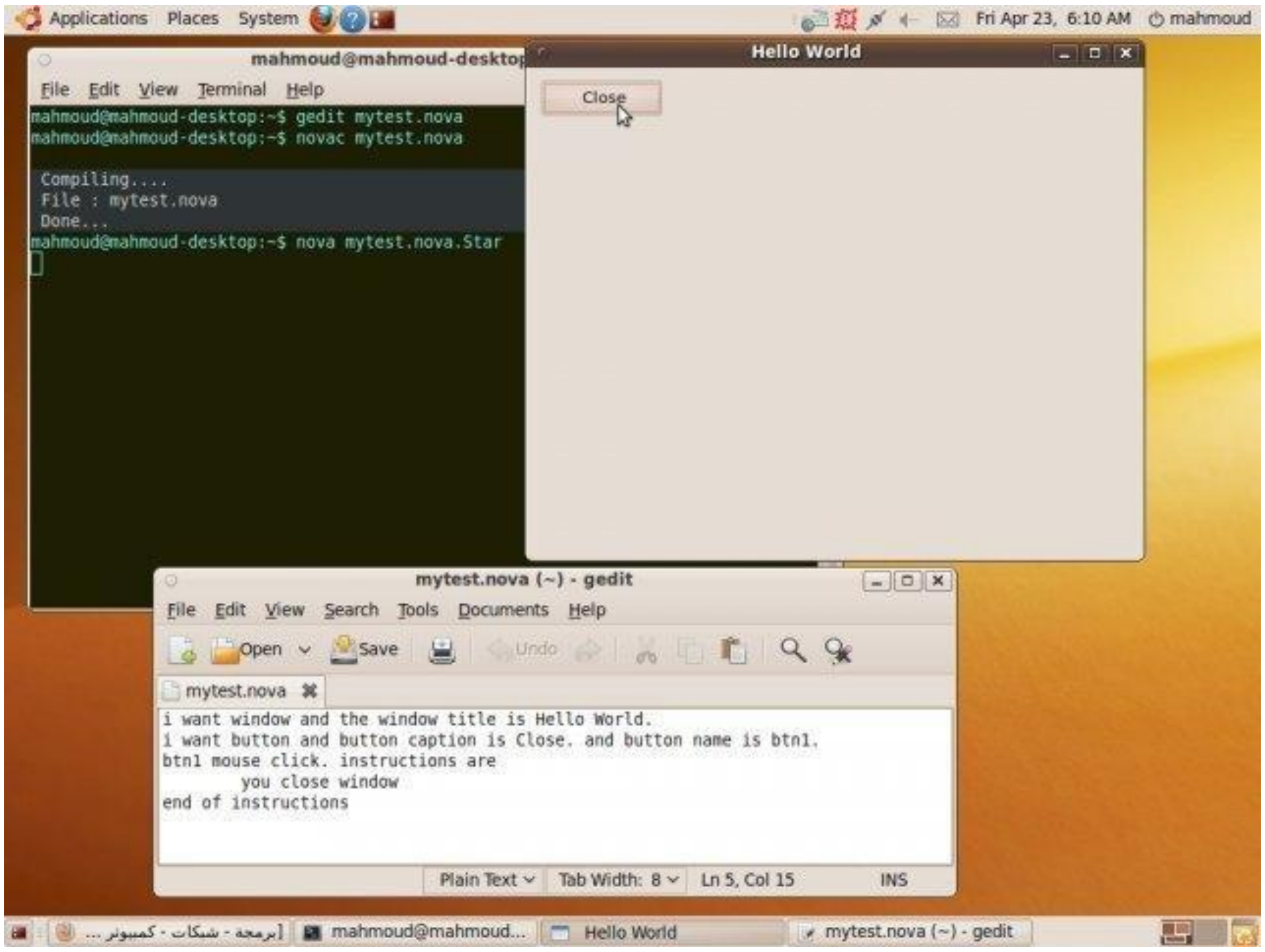

Figure 27: The Supernova programming language developed using the GCR method.

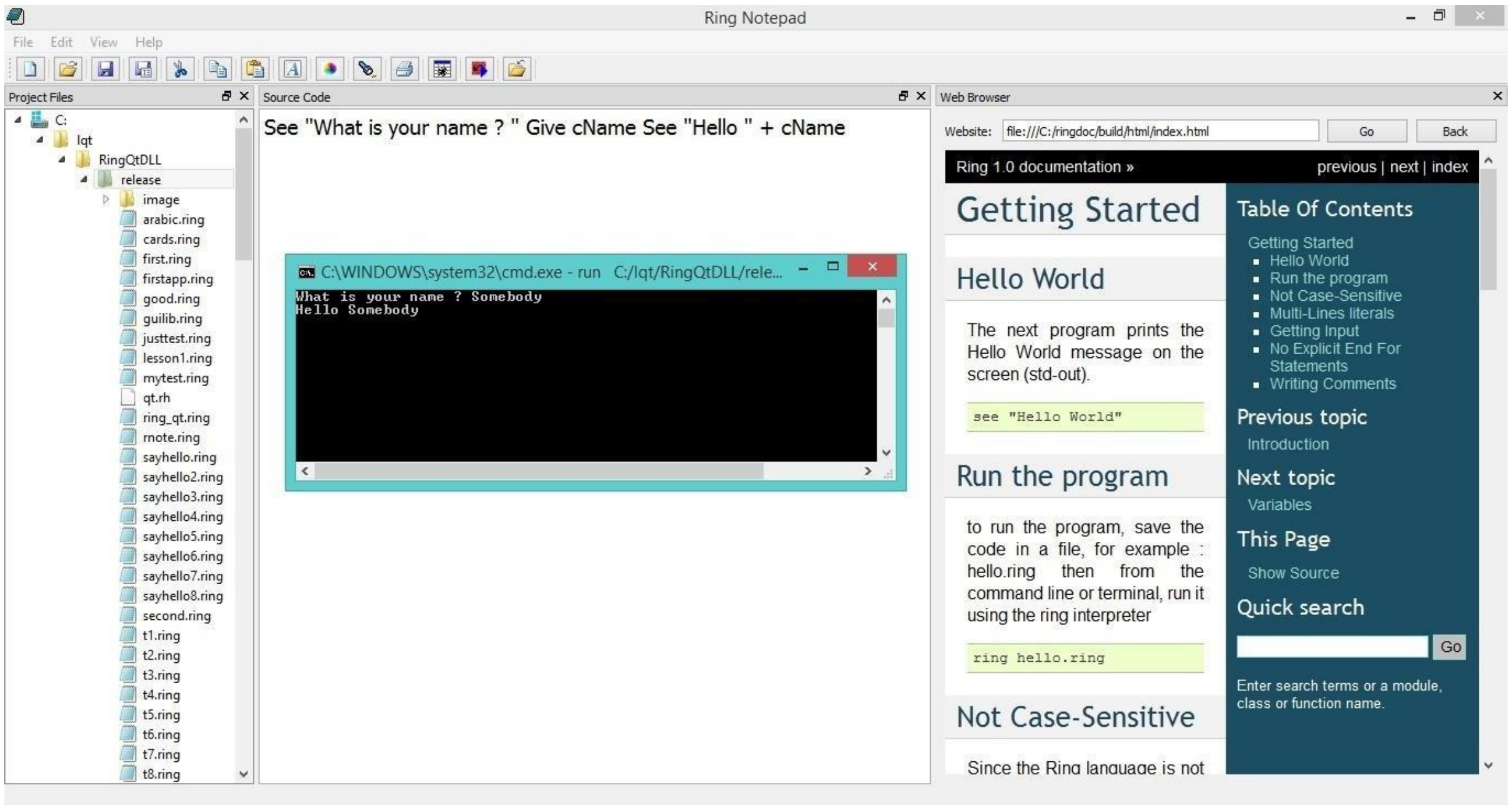

Figure 28: The Ring programming language developed using the GCR method.

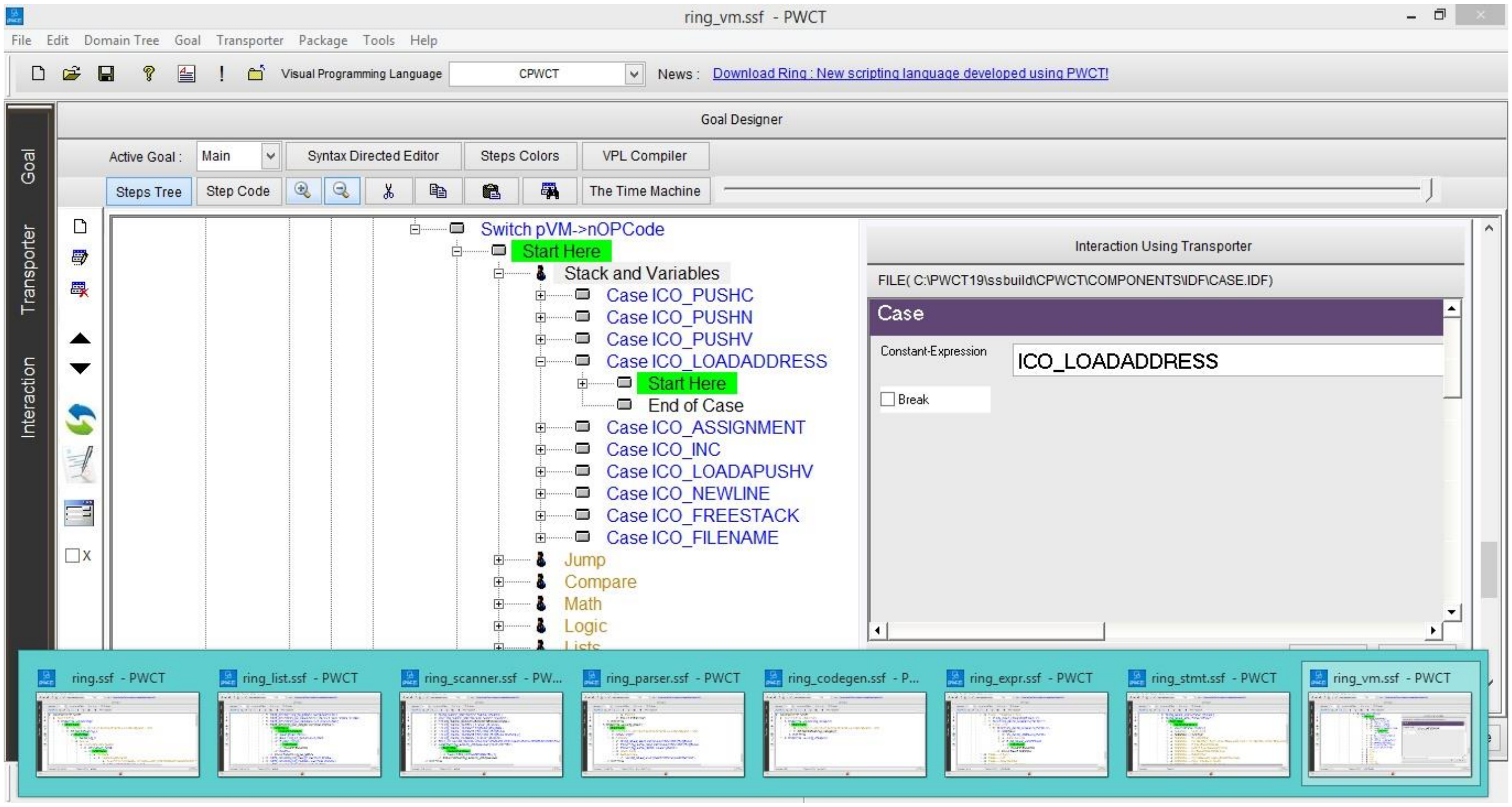

Figure 29: The Ring Virtual Machine developed using the GCR method.

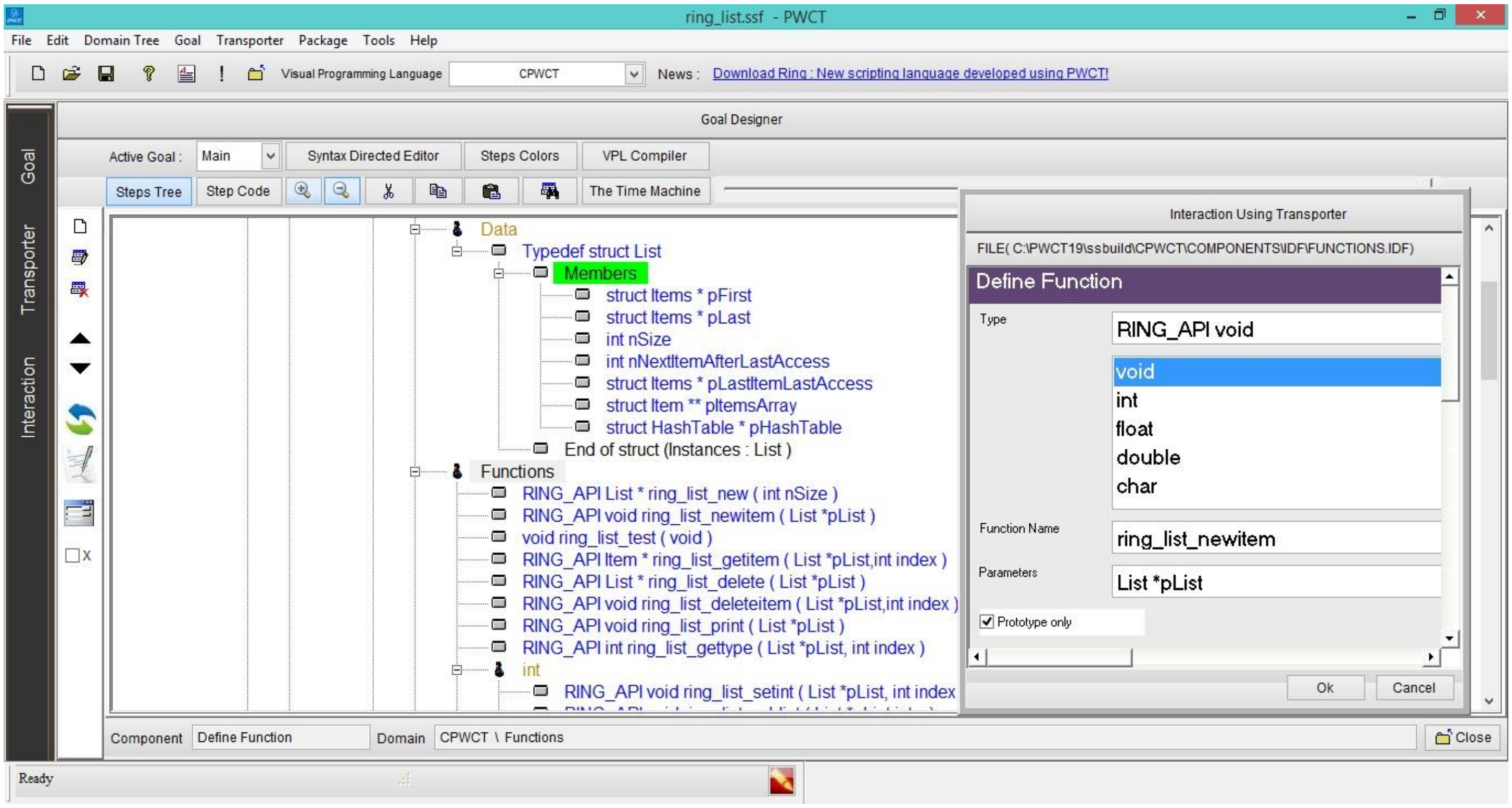

Figure 30: The Ring Lists Implementation developed using the GCR method.

# Chapter 6: Summary and Conclusion

In this thesis we talked about the progress in the visual programming languages, the success in different programming domain, and the features of the most popular visual programming languages. According to our study we see a gap in attracting mainstream programmers to using visual programming languages in each new programming task. The current VPLs are not equivalent to the current textual programming languages at the practical side. Yes, some limitations are a result of missing features that can be added but this doesn't count only for "what we can do?" but for "how we can do things with respect to the productivity of the programmer during software development and the quality of the final software".

We provided the requirements of a general purpose visual programming tool that can be used as replacement for the usage of popular programming languages like ( C,C++, C# and Java) + (visual programming environments like Visual Studio and Netbeans). We expect that after the implementation and marketing of this new tool, many new software projects can be developed using Visual Programming Language in less time with high quality. Also we expect attracting more users to real programming industry after the existence of this tool.

Also in this thesis, PWCT was presented. PWCT is a novel general purpose VPL that was designed to compete with textual programming languages such as C++ and Java. The novelty of this work comes from the invention of a technique called GCR. GCR uses graphical components to replace textual code in an easy and seamless process. PWCT's architecture, design, and implementation were covered in this thesis.

Three types of evaluations were used to make sure that PWCT is a competitive VPL. Usability and capability evaluation shows that PWCT is extremely easy to use and requires a very short time for learning. Moreover, Efficiency evaluations show that PWCT requires less time and memory to develop applications than the other used languages (VS, NB, Tersus and Limnor). Finally, data from Sourceforge indicates that PWCT is a popular, well accepted, and highly downloadable VPL.

The development of many business applications and some programming languages like Supernova and Ring using PWCT demonstrates that the new VPL can be used in serious development for creating real world projects.

PWCT include many features such as generating code in multiple languages and providing record/replay functionality. Also PWCT allow the users to create their own VPLs based on GCR. In the future, we will expand PWCT to support more platforms, web and mobile development.

## 6.2 Future Direction

A more advanced version of PWCT is under development. It will support more platforms and will provide code generation in more textual languages. The components will be redesigned to provide more productivity. An interesting feature will be the support of our new textual programming language that we designed for the development of PWCT 2.0. This language called Ring. Using the Ring language we will have behind the Visual Programming Language, a scripting language that uses Natural Language Programming and Declarative Programming Paradigms. This will provide high-level of abstraction when the developer decide to switch from visual programming and look at the generated source code. Also the other textual programming languages will still be supported when the developer select them.

# Appendix A: Guided tour

## A.1 Hello World Example

In this section, the concept of GCR is illustrated through a Hello World program using PWCT. The application development process starts from the Goal/Module Designer where the application contains one or more of visual source files and each visual source file contains one or more of goals and each goal contains one or more of steps; the step maybe a visual representation that present a computer instruction that do something or the step maybe a comment written by the programmer to describe the process.

From the next figure we see that the active visual programming language is HarbourPWCT and the active goal name is “main” and this goal contains a tree of steps that includes Start Point (NOT STEP) which is not more than a visible root node for the steps tree. Inside the steps tree by default there are a step called (The First Step) which is a step of type (comment) written by the programmer to indicates that this is the first step in our program where we can add more steps the do something (computer instructions).

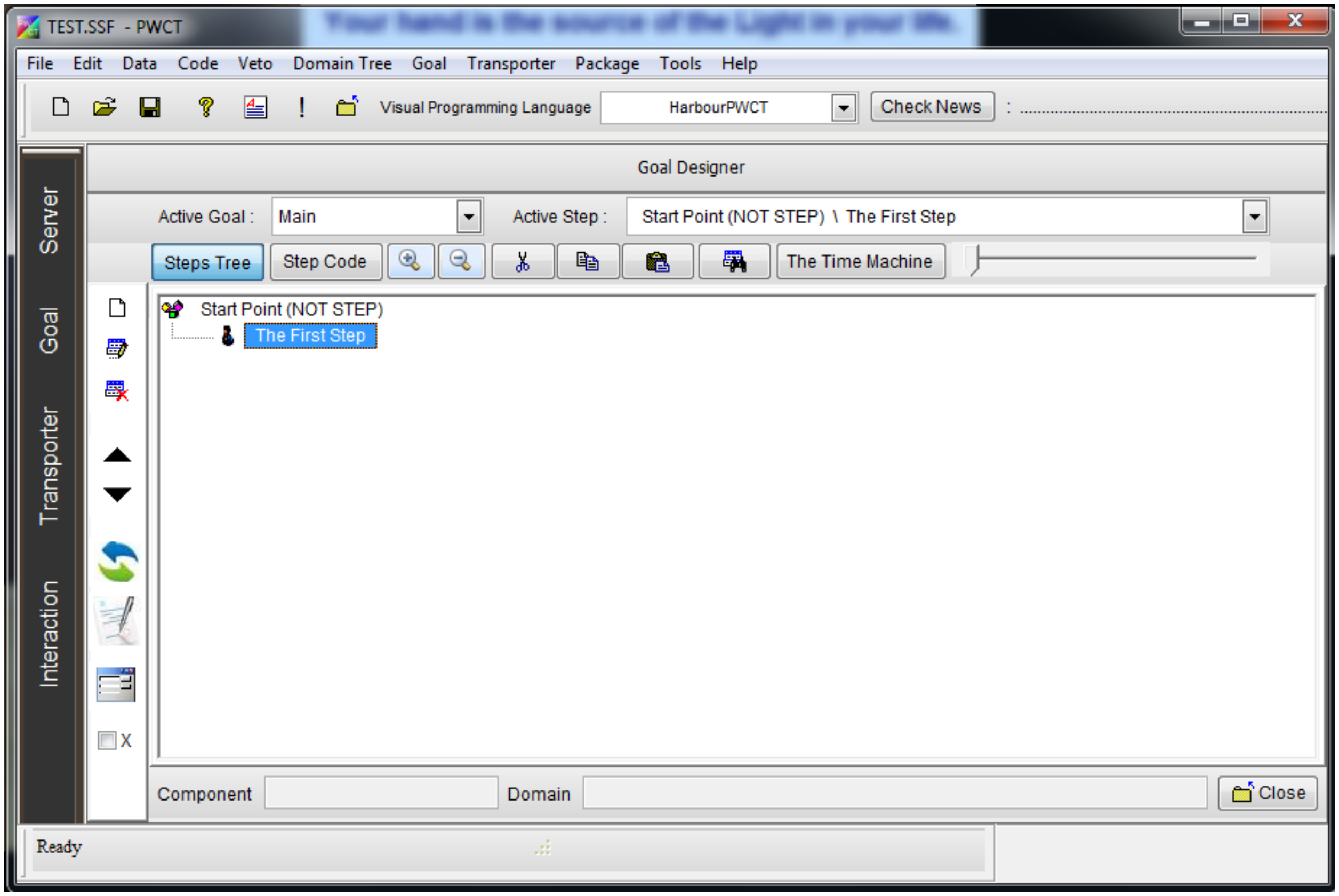

Figure 31: PWCT – Goal/Module Designer

The next table describe the function of the most important buttons in the Goal/Module designer window

Table 13: Goal/Module Designer buttons and their description

| Index | Button Icon | Description |
|---|---|---|
| 1 | 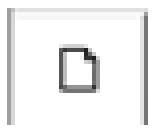 | Create new step of type comment – the step name is written by the programmer to describe something in the program |
| 2 | 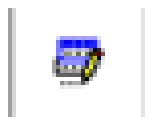 | Edit step name – the step name is modified by the programmer – if the step type is a comment then this feature is used to modify the comment but if the step type is a computer instruction then this feature is used to provide better description for the instruction. |
| 3 | 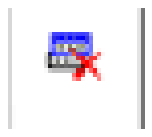 | Delete step from the steps tree to remove a comment or a computer instruction. |
| 4 | 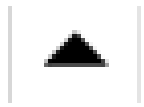 | Move step up in the steps tree to organise the visual source and/or to control the flow of instructions execution during the runtime |
| 5 | 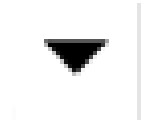 | Move step down in the steps tree to organise the visual source and/or to control the flow of instructions execution during the runtime |
| 6 | 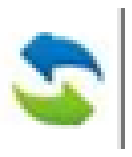 | Start new interaction process to generate new steps inside the steps tree. These steps will do something during the runtime because it's from the type (computer instruction) |
| 7 | 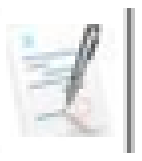 | Modify a step. This features is available for steps of type computer instructions to modify something in the interaction process by changing attribute values |
| 8 | 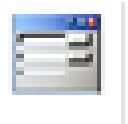 | Open the form designer. This feature is available for steps that are related to GUI programming (Windows, controls,...etc) |
| 9 | 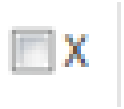 | Ignore/Enable a step. This feature is available for steps of type computer instructions to include/exclude the step from the designed program during runtime |
| 10 | 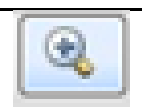 | Increase the font size of the steps names in the steps tree |
| 11 | 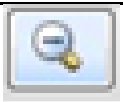 | Decrease the font size of the steps names in the steps tree |
| 12 | 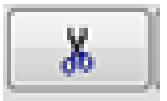 | Cut a step, remove the step from the steps tree and keeping it in the memeory |
| 13 | 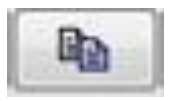 | Copy a step to the memory without removing it from the steps tree |
| 14 | 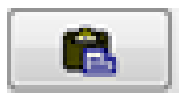 | Paste a step from the memory to be a child step to the selected step in the steps tree |
| 15 | 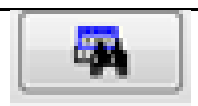 | Search in the visual source by step name |

Toward creating hello world program, after selecting the step (The First Step) which is selected by default

Click the interact button which carry the icon 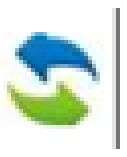 to start new interaction process.

After clicking the interact button the components browser window will be opened
The components are classified into different domains were each domain contains group of components and using the component browser you can select a component to use.

From the domain tree select the domain (Print Text) then from the components list select the component (Print Text to Console) Then click the Ok button.

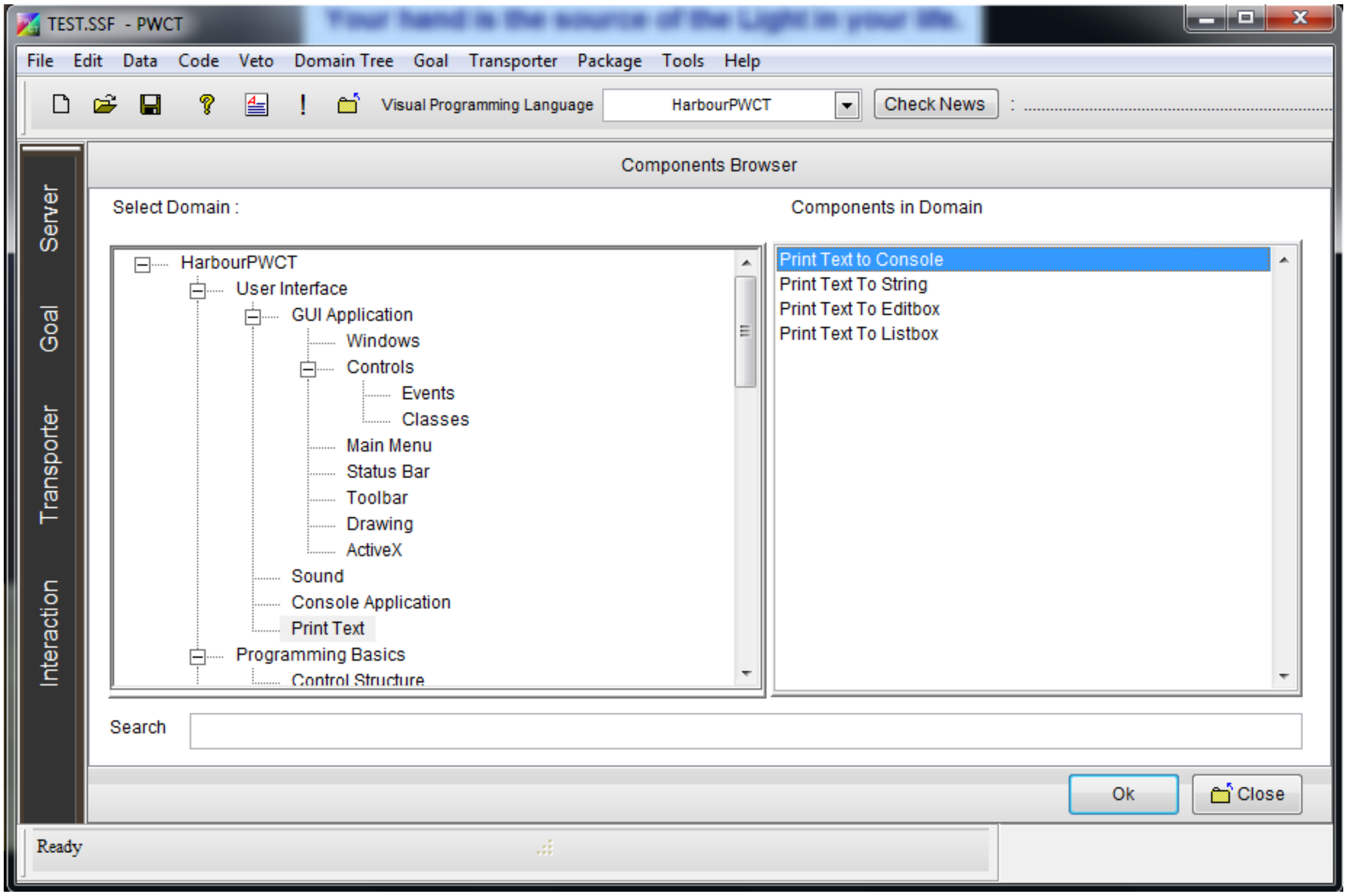

Figure 32: PWCT - Components Browser

After selecting the component using the components browser and clicking Ok, The interaction using transporter window will be opened to enter the data required for the selected component. The transporter is another name of the component inside PWCT.

The component (Print Text To Console) presents simple interaction page (data entry form) that ask the used to enter the text that will be printed on the screen. In our example we will write the text “Hello World”. Then we will click Ok button.

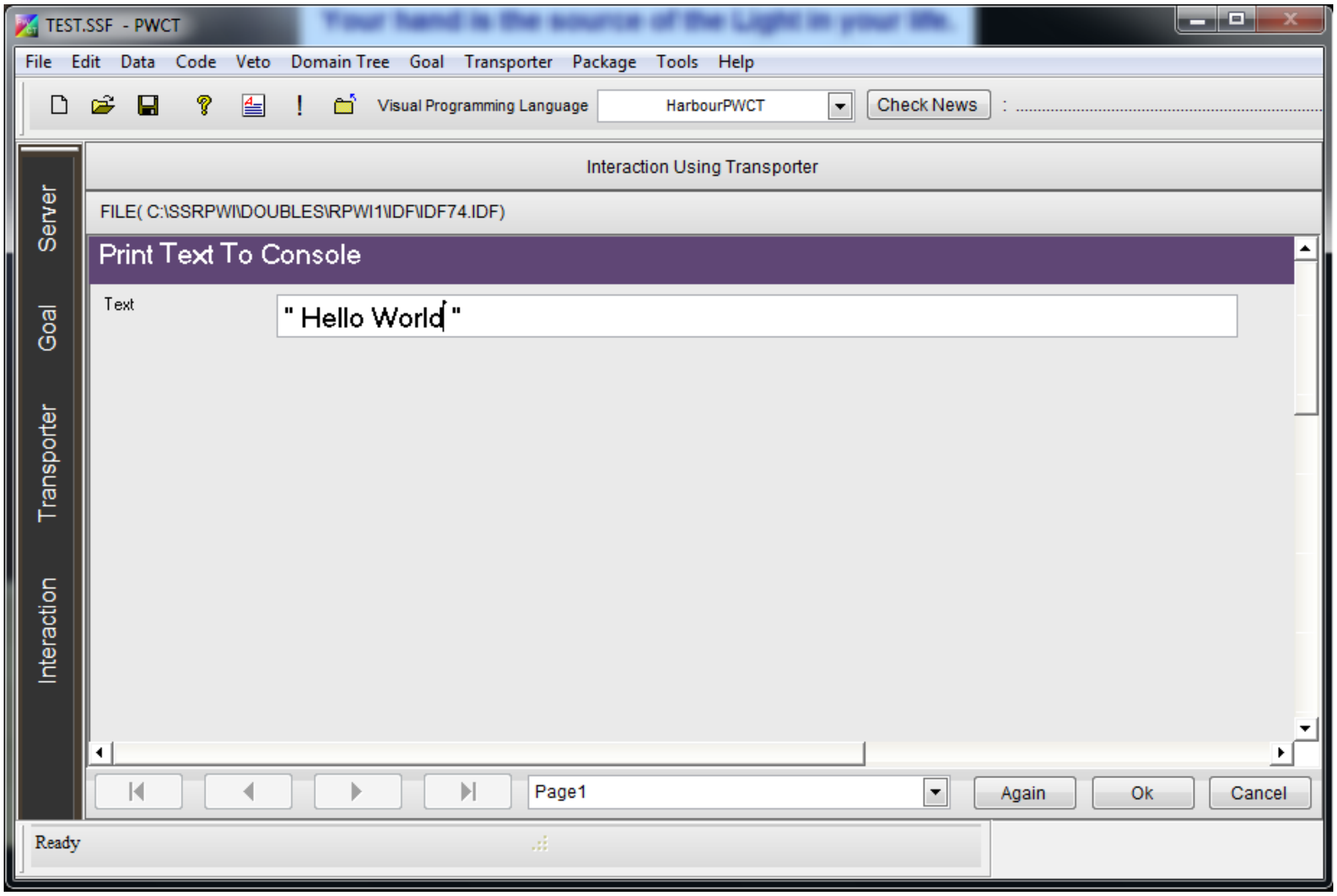

Figure 33: PWCT – Interaction Using Transporter window

After entering the text and pressing Ok button we will return again to the goal designer window but this time will notice that a new step is added to the steps tree. And the name of the new step is Print text ("Hello World") to console. The new step name can be changed by the programmer using the edit step button

At this stage we can run our program. But running the program at this stage will not give us the chance to see the message because the program will be executed and the execution will done and the console screen will disappear before seeing the message.

To see the message we can run the program from the windows command line window or we can add another step to the program to wait for a number of seconds to see the results. To modify the program by adding a step for waiting for number of seconds we will do another interaction process with another component but this time we will control the interaction process using the keyboard instead of using the mouse to clear that the Coding Simulation Method is designed to give us the ability to use the keyboard to get more productivity.

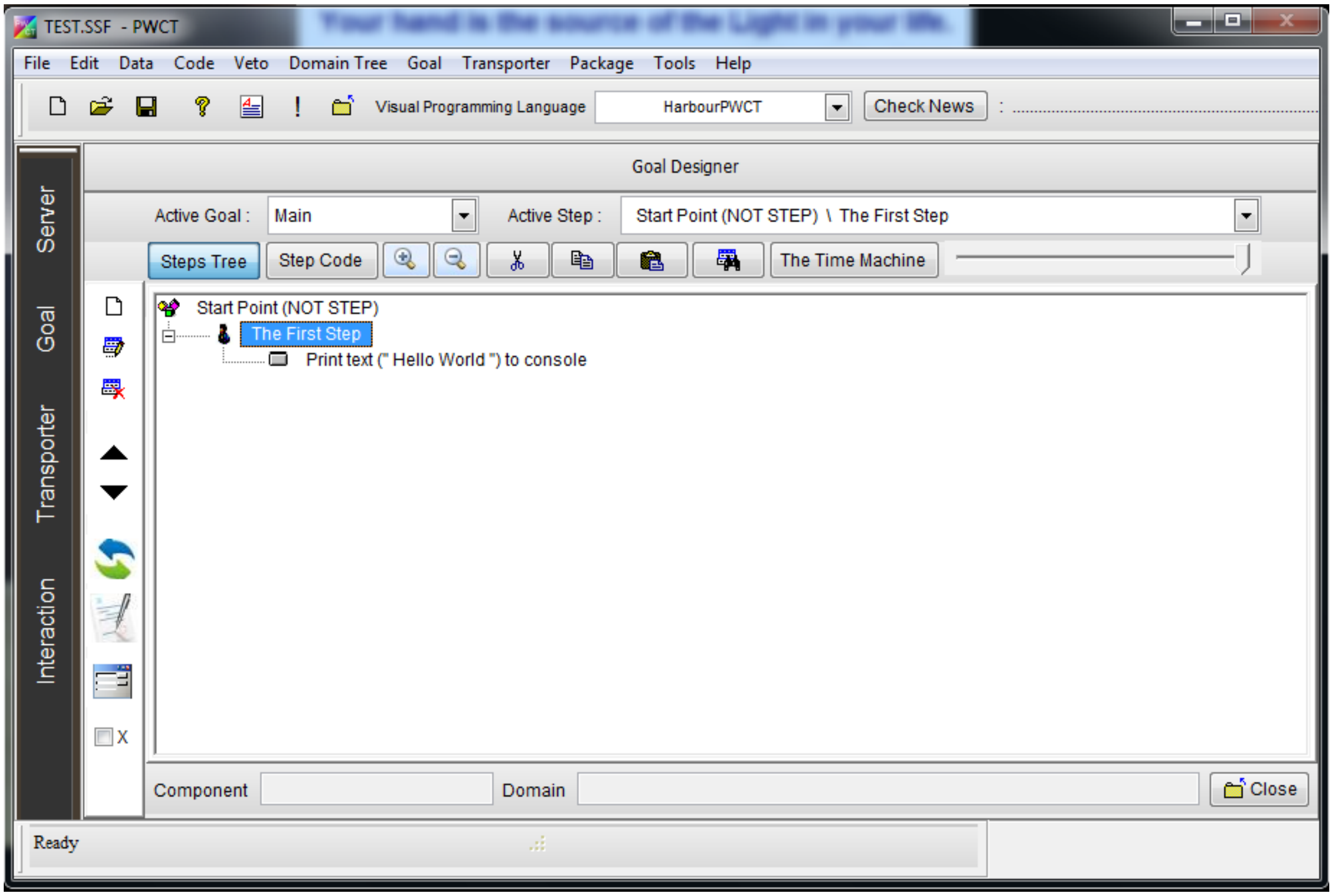

Figure 34: PWCT – Goal/Module Designer – Generated step as a result of interaction process

To use the keyboard to start new interaction process we can press CTRL+T or we can write the first letter of the component name while we still in the Goal Designer and the first step is selected.

In our case we will press the letter “W” because the required component name is (Wait Key/Seconds)

After pressing the letter “W” the components browser window will be opened and this letter will be available in the search line where we can continue the search process by typing more letters.

After typing the second letter in the component name, in our case it’s the letter “A” we will see that the required component (Wait Key/Seconds) is selected.

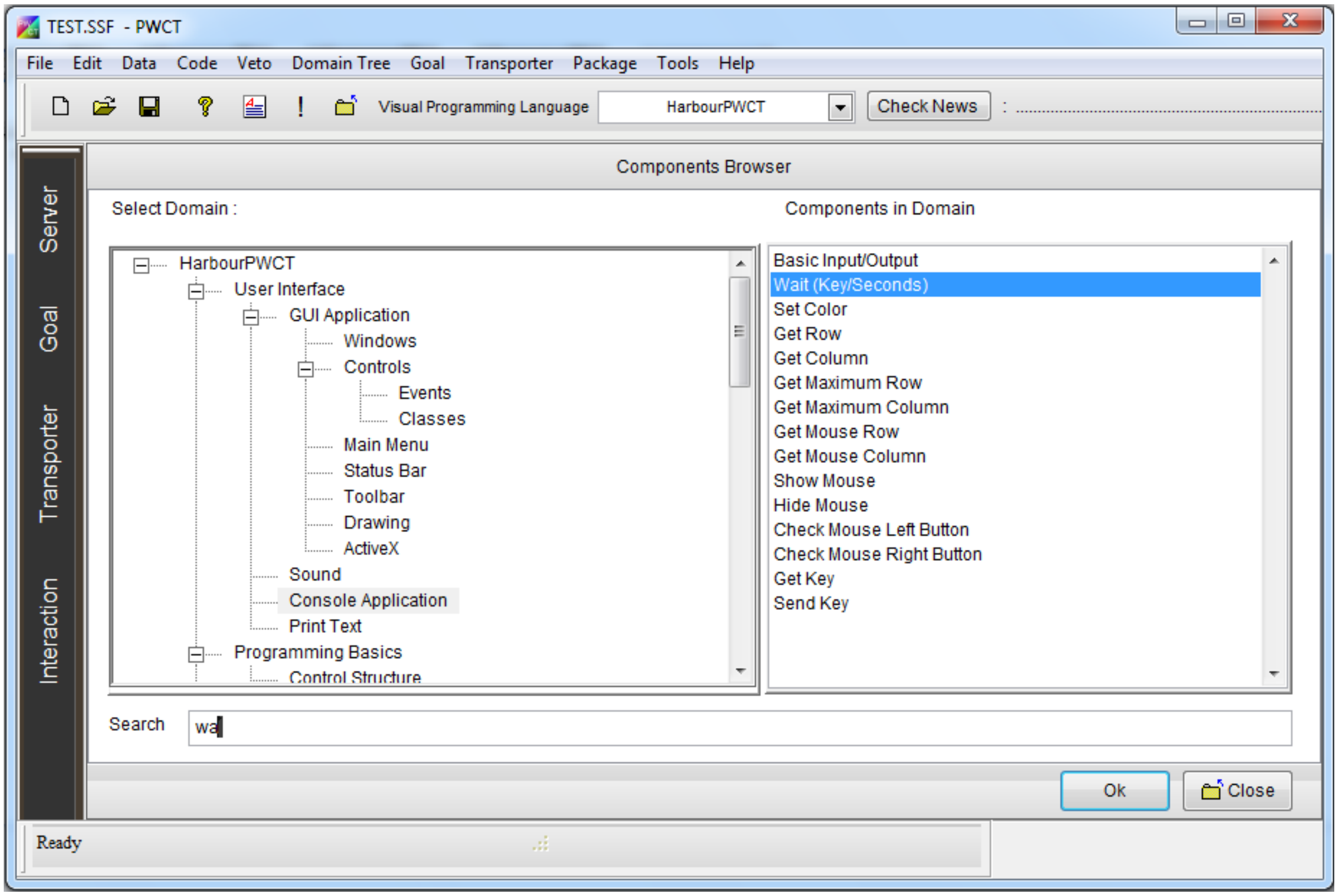

Figure 35: PWCT - Components Browser – Using Keyboard to quickly find a component

Instead of clicking the ok button we will press Enter to use the selected component.

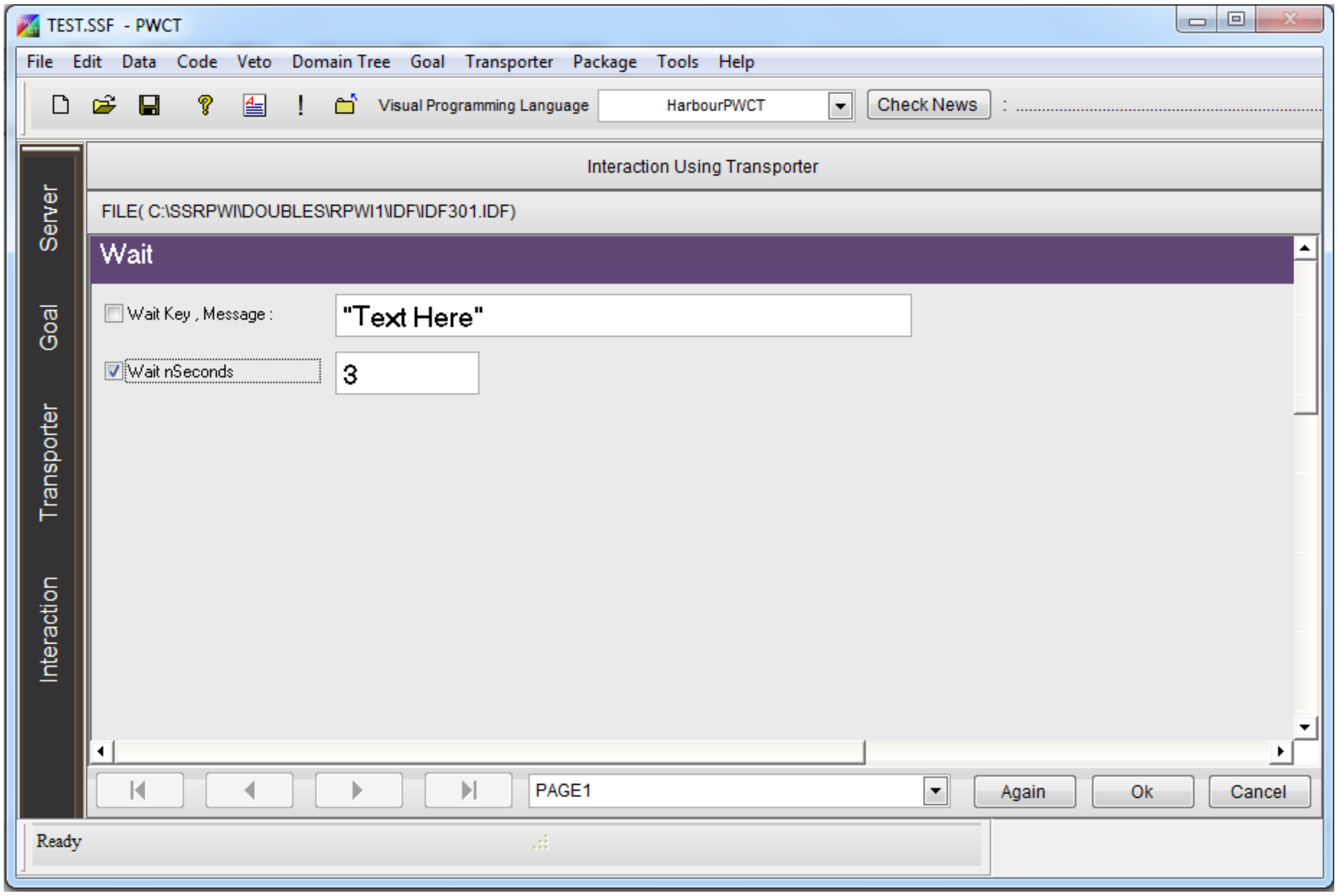

Figure 36: PWCT – Interaction using transporter window

Inside the interaction using transporter window and using arrows we can move to the checkbox (Wait nSeconds) and this will be done after pressing the down arrow for two times. Then we will press enter to set the checkbox (Wait nSeconds) to On.
Then press enter to move the textbox to determine the number of seconds
In our example we will type 3. Then press CTRL+W to end the interaction process to return to the Goal Designer.

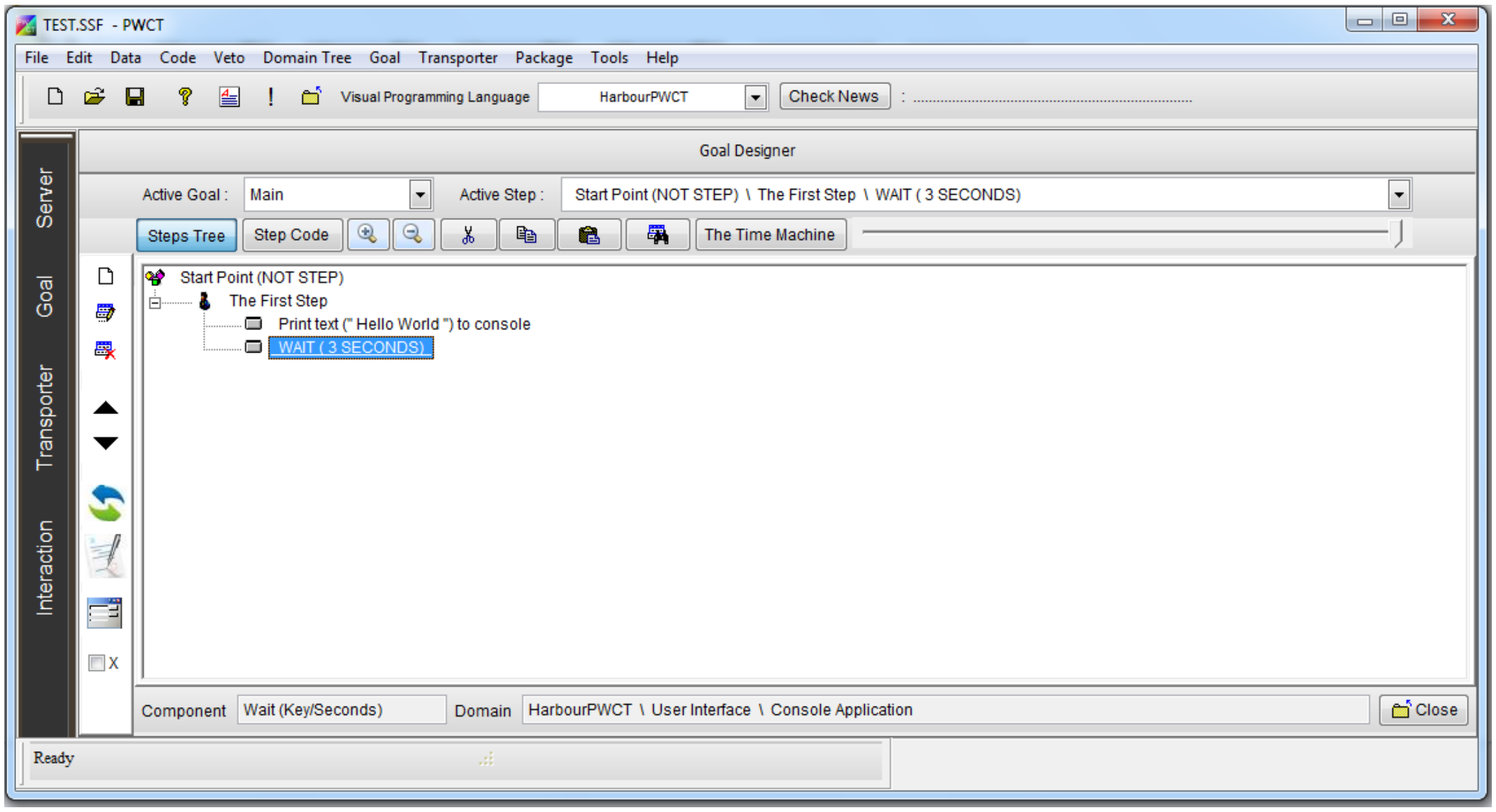

Figure 37: PWCT – Goal/Module Designer – Generated steps as a result of interaction process.

After returning to the Goal designer we notice a new step added to the steps tree and this step carry the name Wait (3 Seconds).

Now we can run the program by pressing CTRL+R or by click the (i) button from the toolbar.

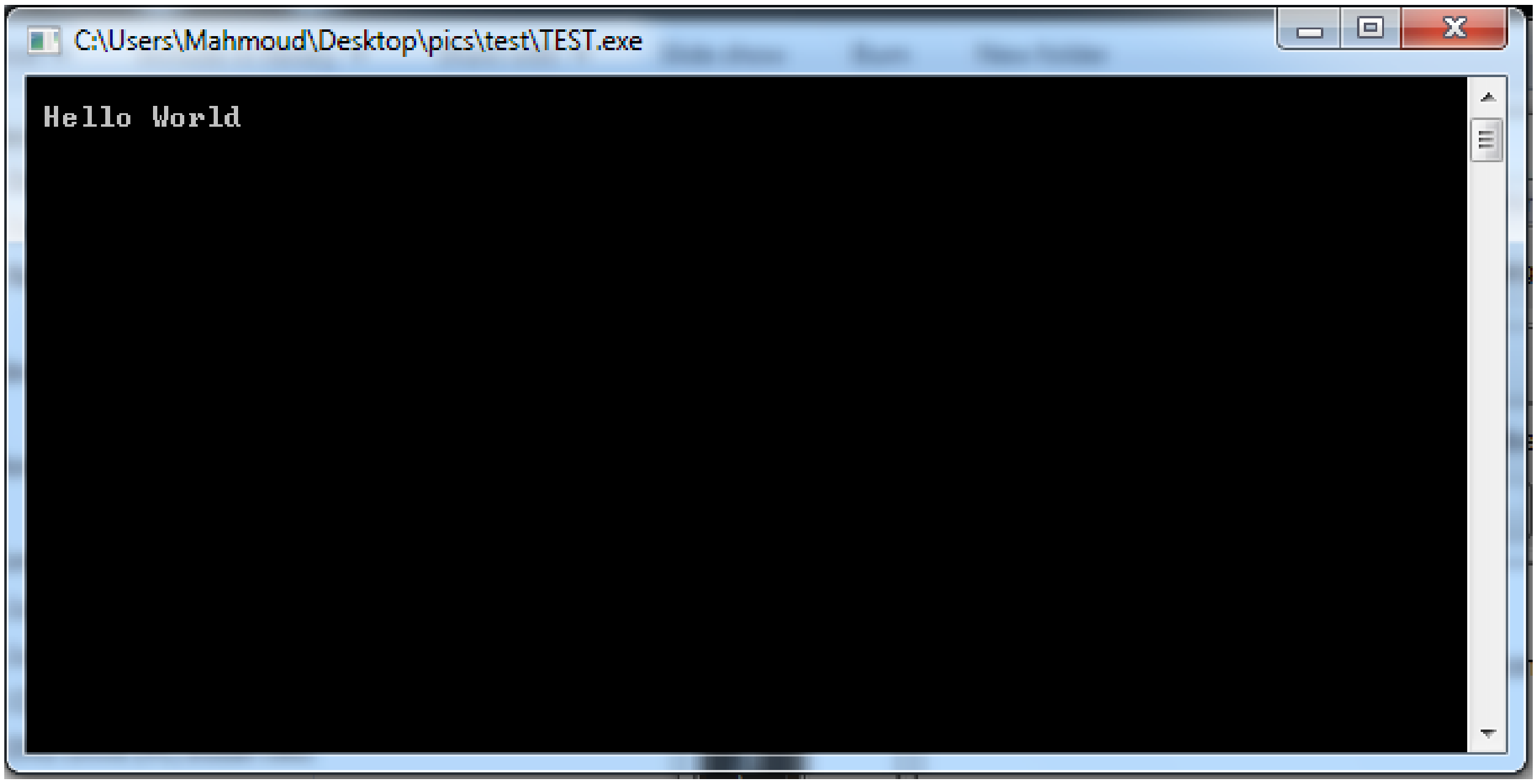

Figure 38: Application during the runtime

A.2 How to maintain programs

Using the Goal/Module designer buttons we can add, edit and delete steps

And using the Modify button 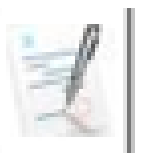 after selecting a step the interaction using transporter window will be opened where we can change the values (entered data) and this change will be reflected in the step name inside the goal designer.

## A.3 GUI Applications

The GCR is flexible enough to represent different types of applications and GUI applications can be developed and represented simply using the steps tree and interaction pages.

Also we can use a form designer to set controls positions and sizes

The next figure presents the steps tree of a simple GUI application

This applications contains a window carry the title (Welcome to PWCT) and this window contains a label carry a caption the same as the window title. Also the window contains a button to close the window.

When the button is clicked by the user a procedure called (myclose) will be invoked and this procedure will close the window.

All of these steps are a result of interaction processes with visual components like (Define new window, Label, Button, Button Events, Define Procedure and Window Class)

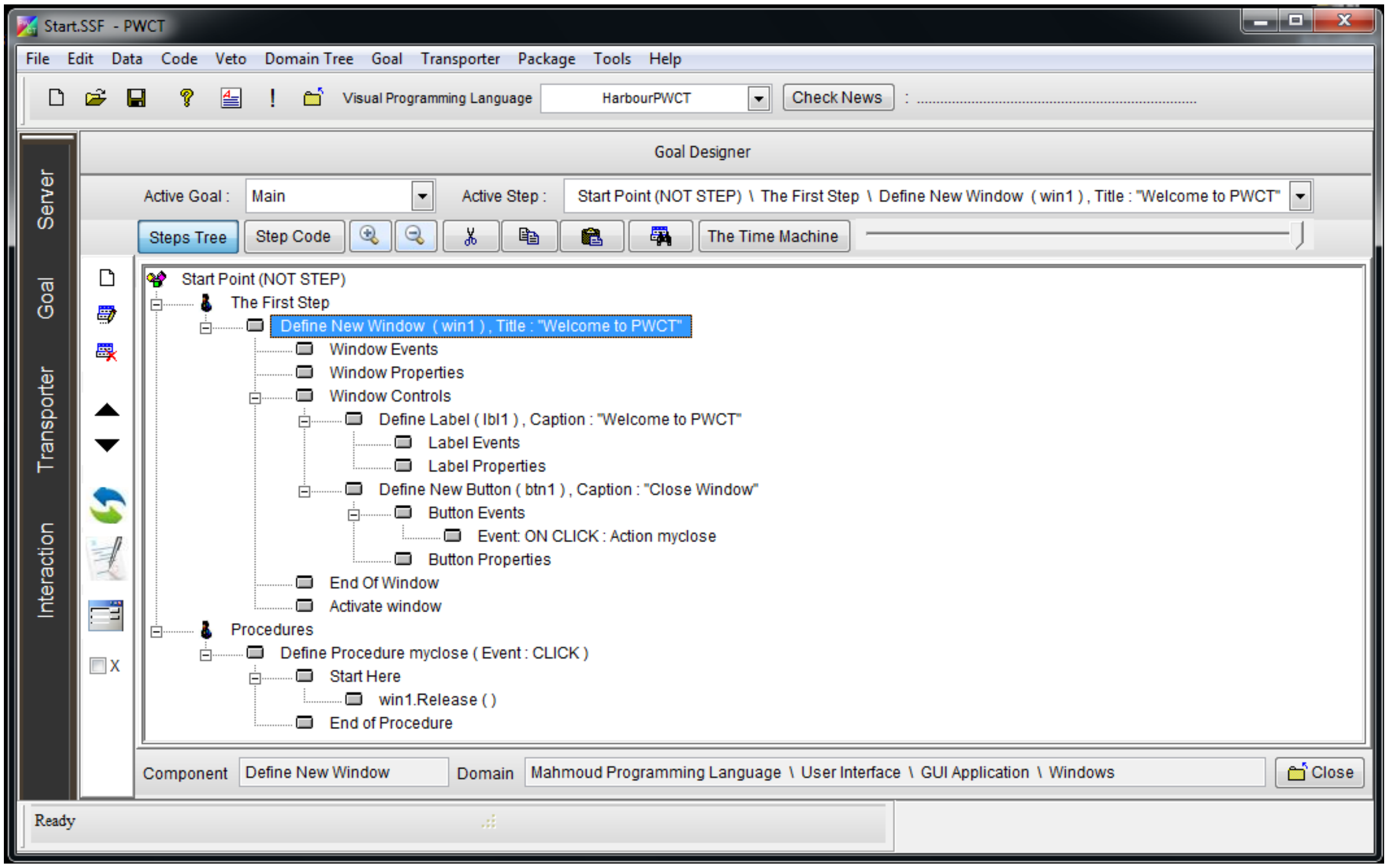

Figure 39: PWCT – Goal/Module Designer – Steps Tree for simple GUI application

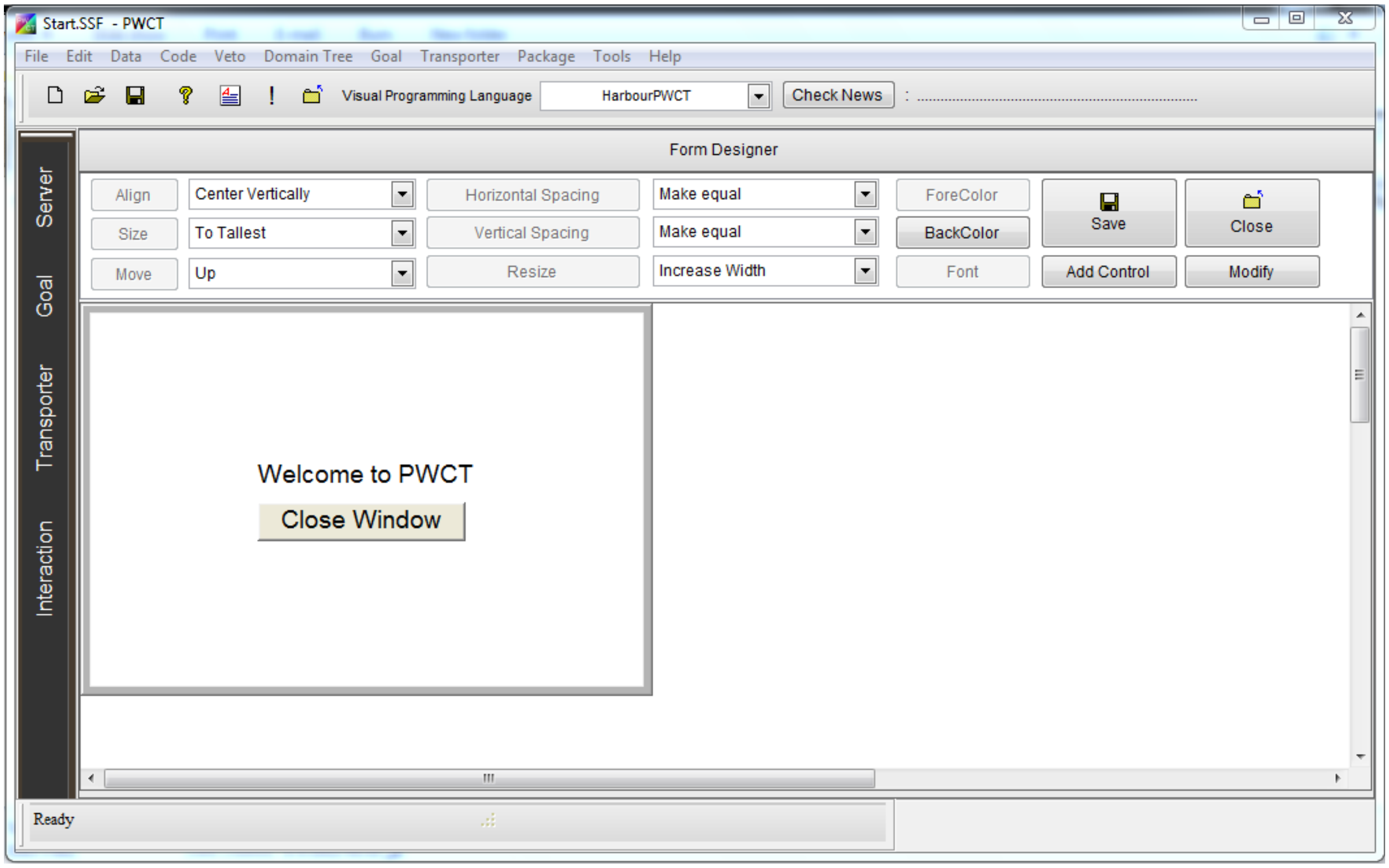

Figure 40: PWCT – Form Designer

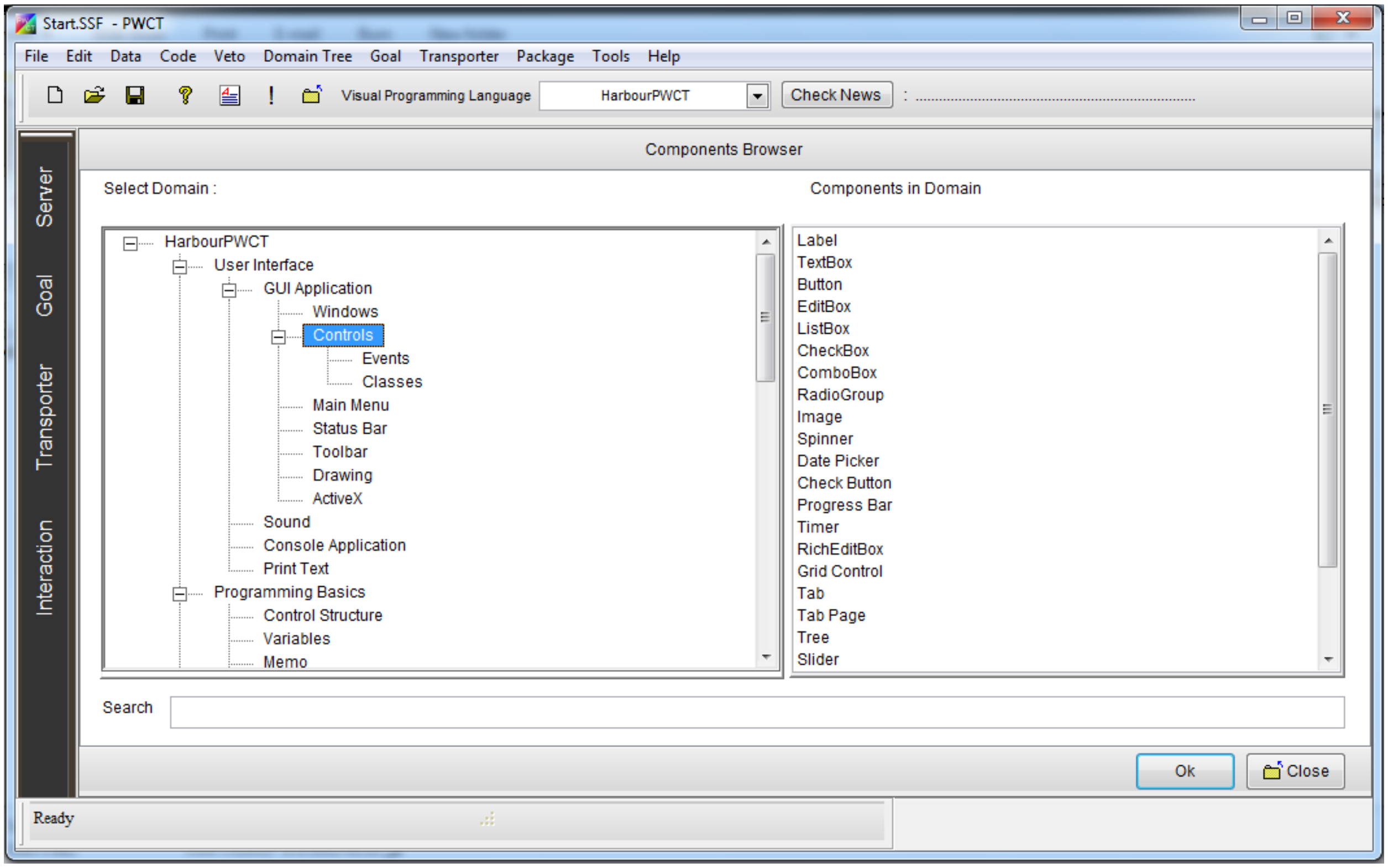

Figure 41: PWCT – Components Browser – Components inside the domain “Controls”

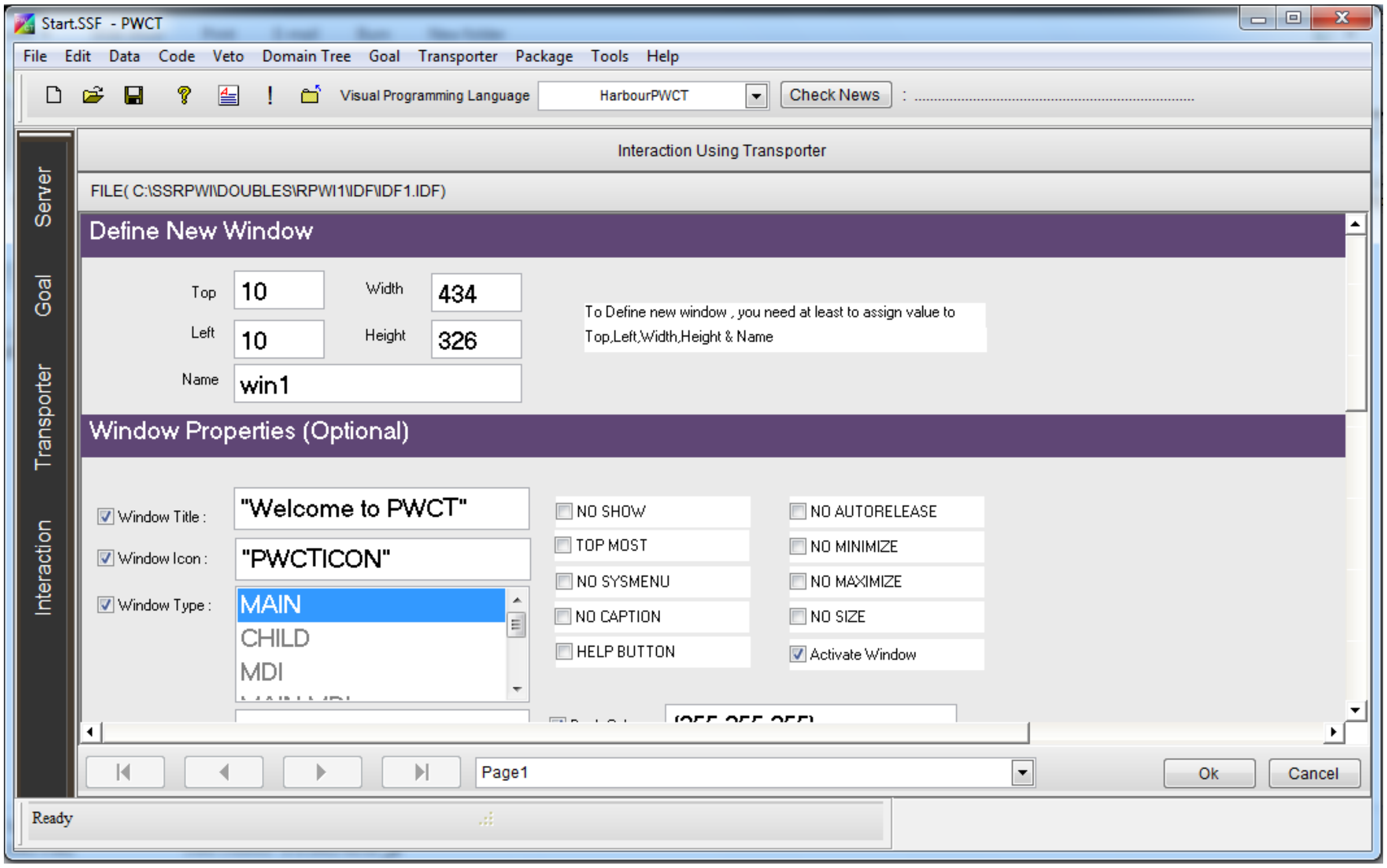

Figure 42: PWCT – Interaction using transporter window for the component "Define new window"

The interaction page of the component (Define new window) is an example about designing a large component with many features.

Note: the same component may contain a group of interaction pages like the (Window Class) component which contain three interaction pages (Set Property page, Get Property Page & Invoke method page).

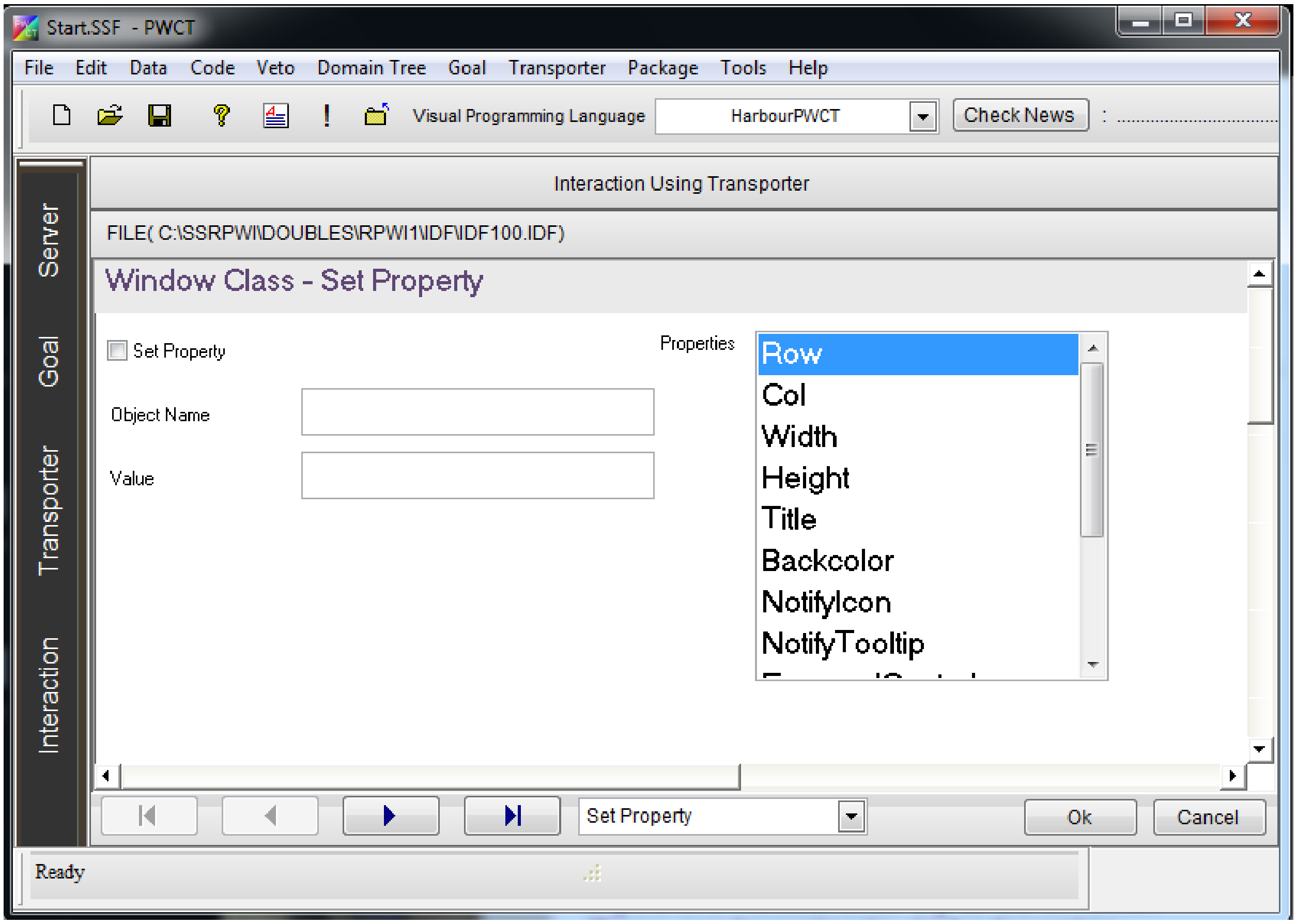

Figure 43: PWCT – Interaction Page for Multi pages component

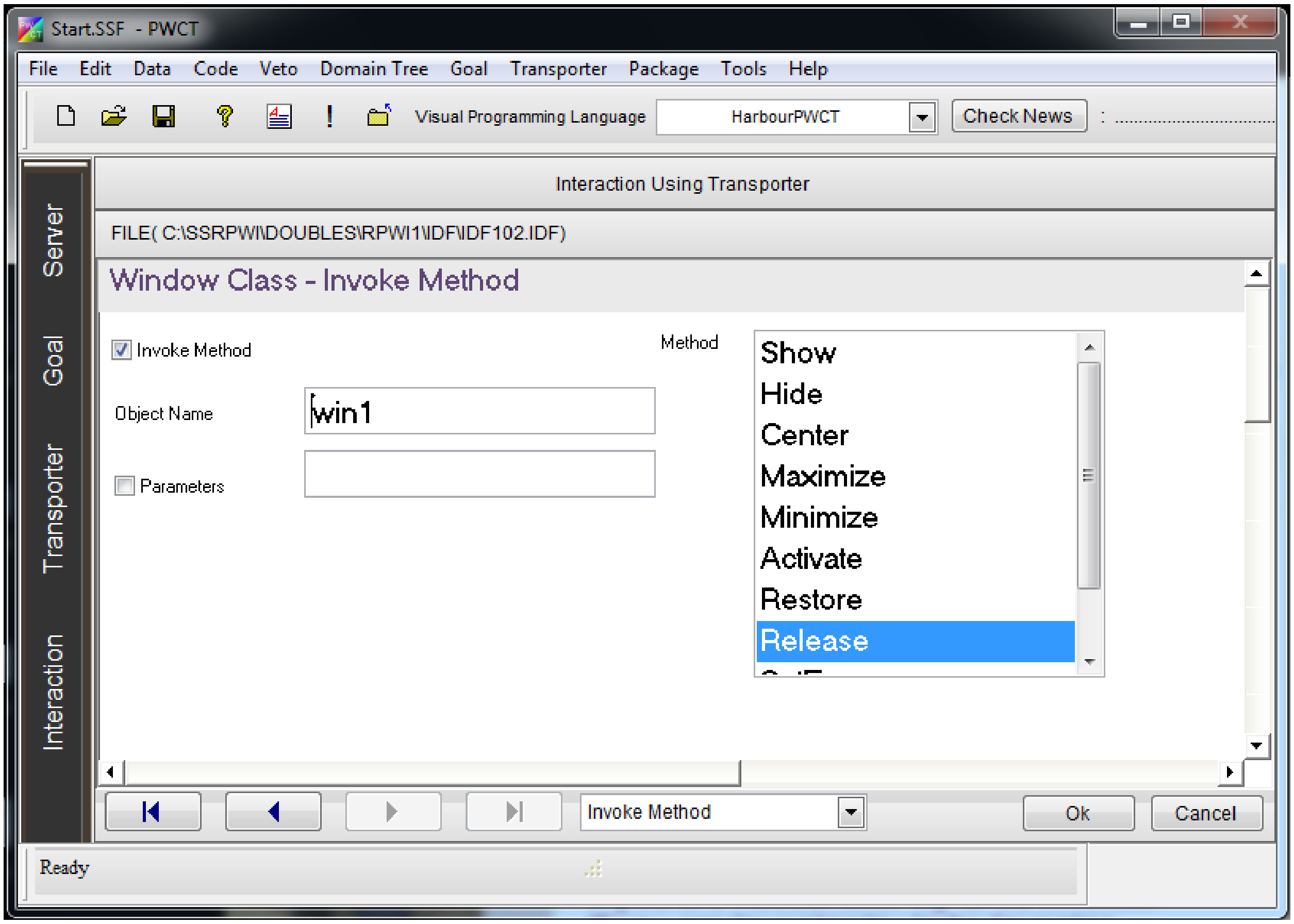

Figure 44: PWCT – Interaction Page for Multi pages component

## A.4 Search in Visual Source

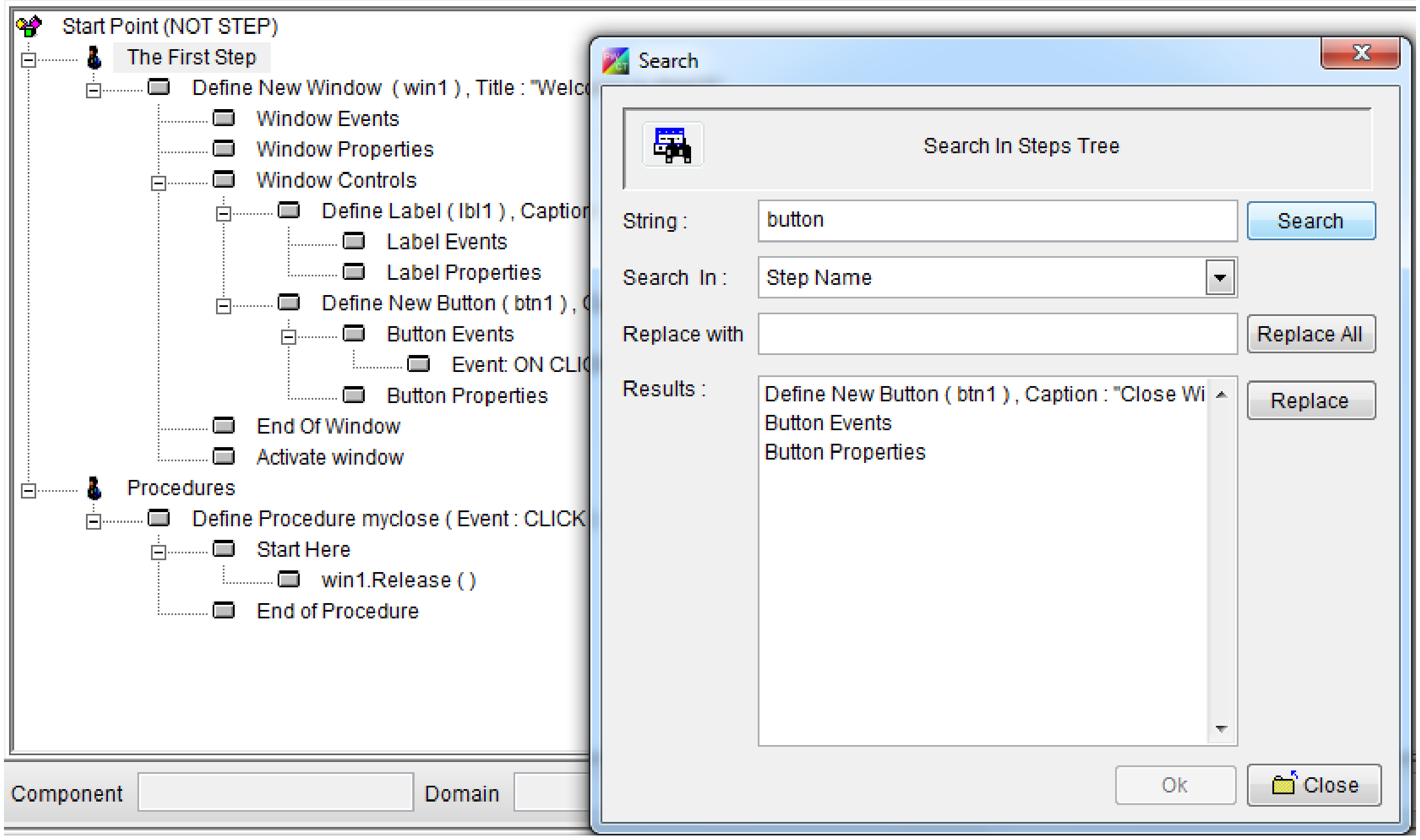

Figure 45: PWCT – Search for a step inside the goal designer

The programmer can do search operation in the visual source to quickly find and select a step. The search process can be done in the step name (goal designer) or in the step data (data entered through the interaction using transporter window).

## A.5 The Time Dimension

One of the characteristics of visual programming is programming in more the one dimension. One of the available dimensions for PWCT users is the time dimension where the programmer can move forward/backward during the application construction process and the programmer can run the application at any point on the time dimension to test the programmer before or after adding a step to the steps tree. This can be done easily using a slider control available in the Goal/Module Designer.

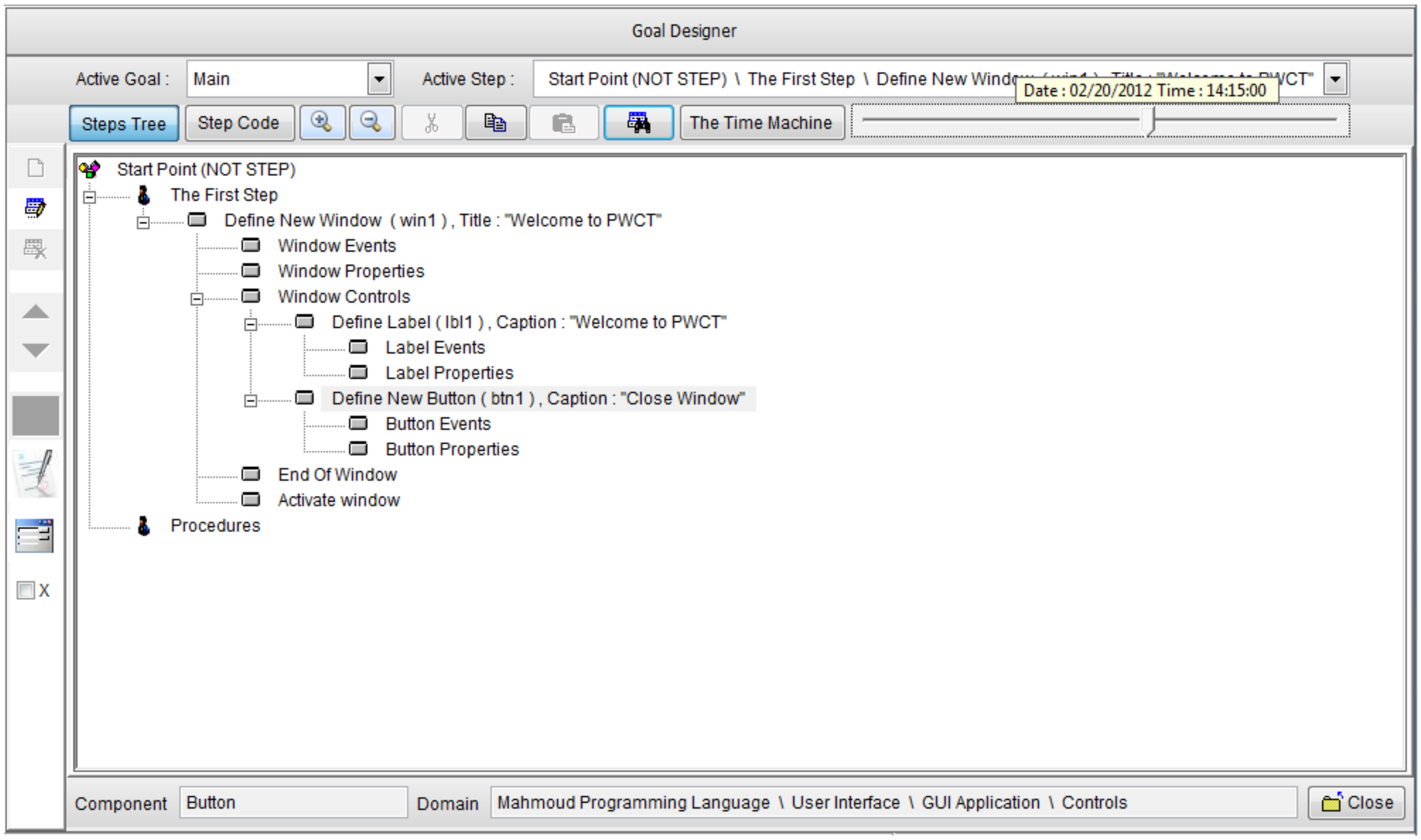

Figure 46: PWCT – Using the Time Machine to go backward at a time in the past during the design process

## A.6 See the code behind every step

PWCT is a hybrid system where we can design programs visually without writing source code by hand but we still can see the generated textual source code which is generated and managed in the background.

We can see the textual source code behind each step in the goal designer and we can get the source code behind all the steps when we run the application or using a utility comes with PWCT and called the Code Extractor application.

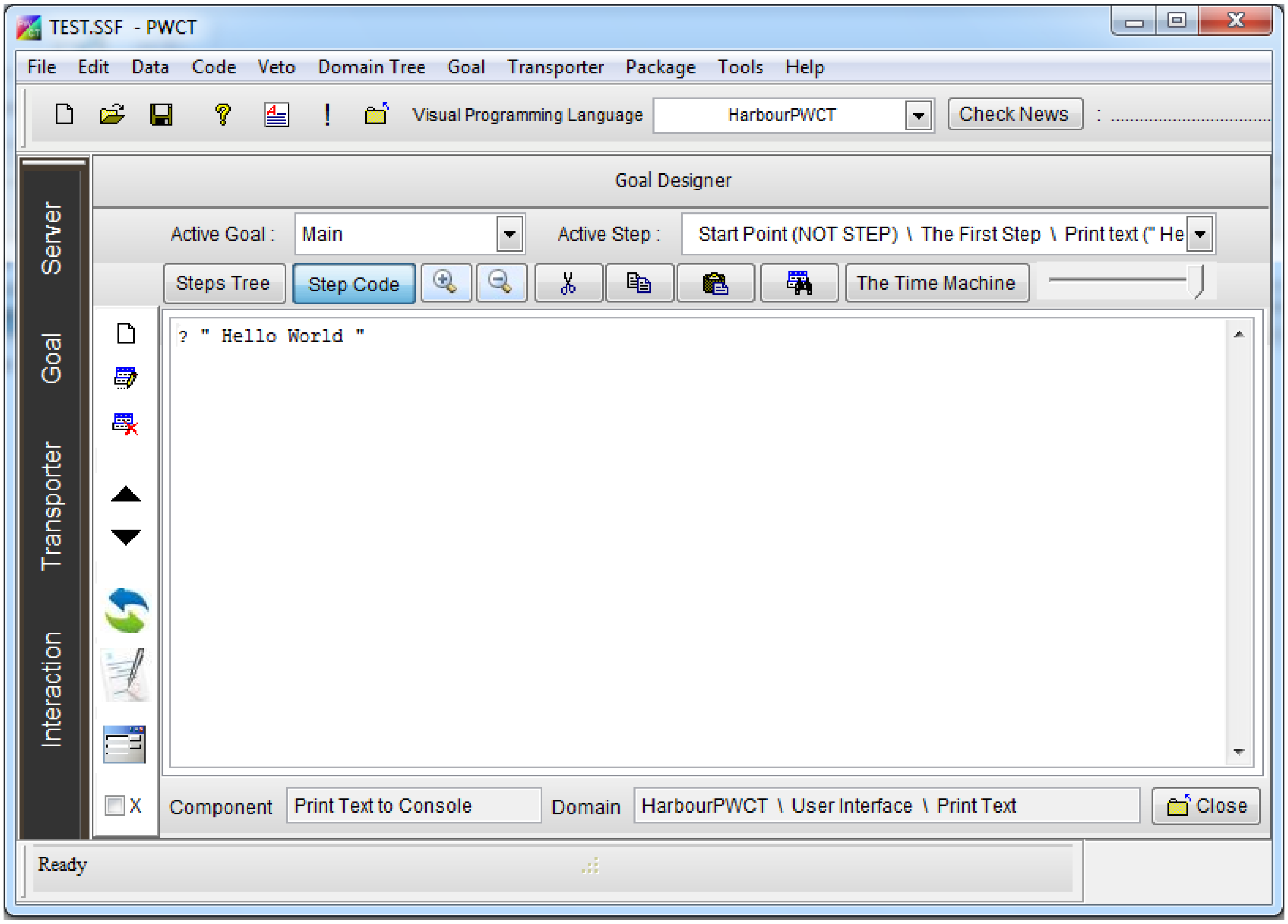

Figure 47: PWCT – Looking at the code behind a step

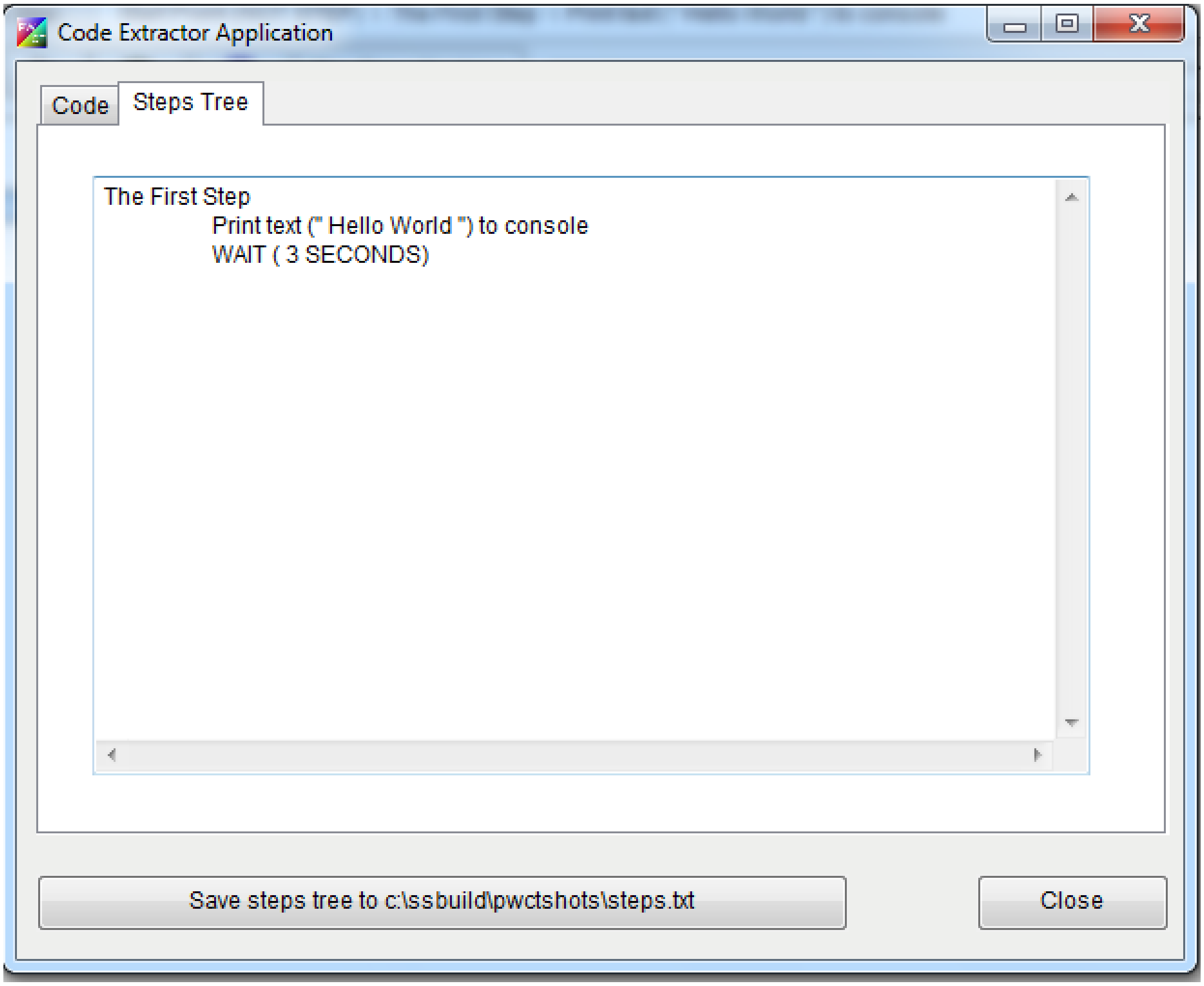

Figure 48: PWCT – Code Extractor Application

## A.7 Dealing with multiple steps at the same time

PWCT Contains a tool called the Goal/Module Viewer which contains some of the features available through the goal designer but enable the programmer to work with more than one step at the same time where each node in the steps tree contains a checkbox for selecting the step.

Using checkboxes more than one step can be selected then we can do different operations of the selected steps like (Cut, Copy, Insert, Delete , Move Up, Move Down and Enable/Disable).

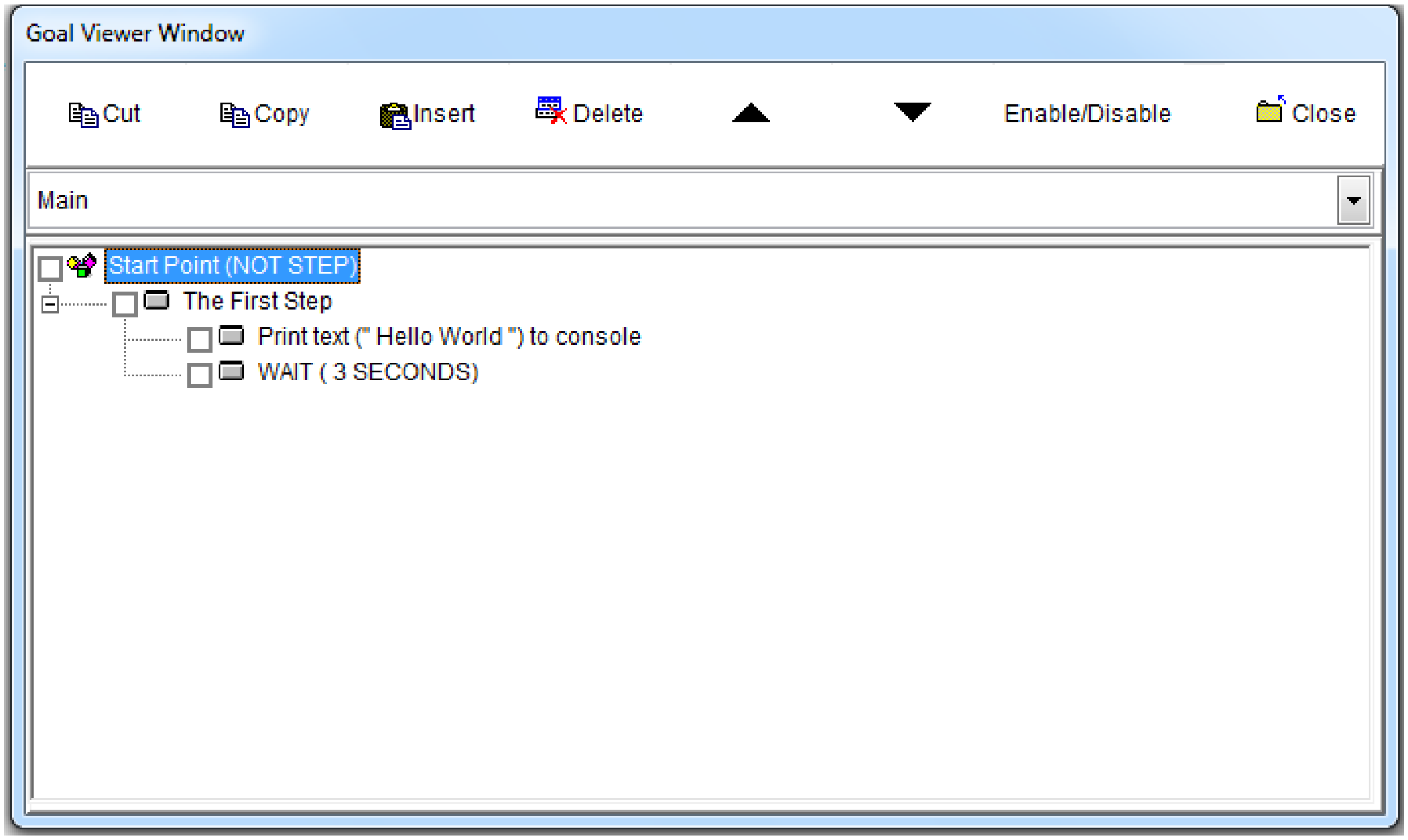

Figure 49: PWCT – Goal/Module Viewer

## A.8 The Time Machine and playing programs as movies

When you open any project/application developed using PWCT and instead of looking at the design to see how to create like this program, simply you can see a movie that teach you how to design the program using PWCT step by step

To do that from the Goal/Module designer press The Time Machine button then select the item (Play as Movie from the first time frame).

Then the steps will disappear from the Goal/Module designer and you will see how every step will be added by selecting the required component then entering the data to the interaction pages.

This feature reduces the need for sample documentation because every sample developed using PWCT is a movie that can be played.

Also this feature help in program understanding where the program can use this feature and keep watching instead of manually moving through the visual representation

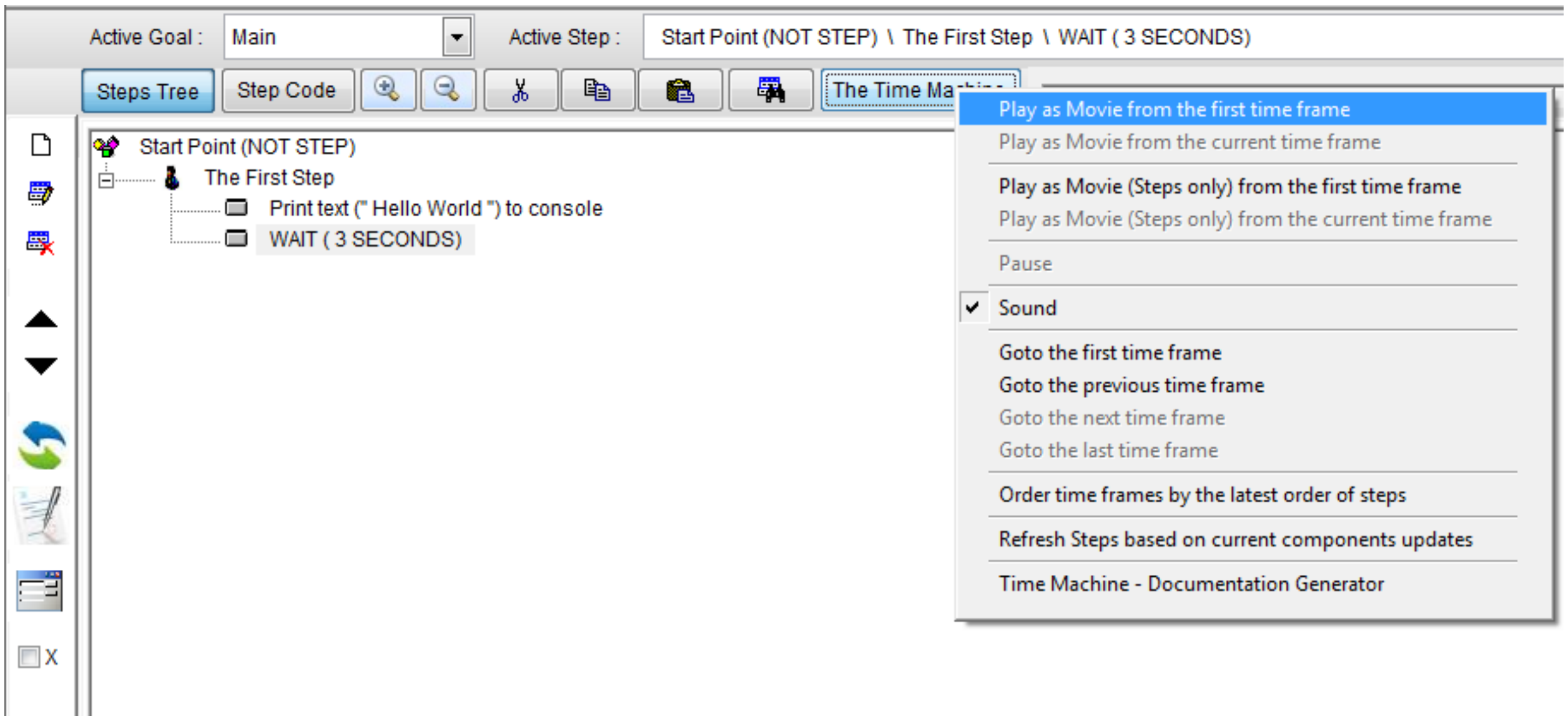

Figure 50: PWCT – Goal/Module Designer - Using the Time Machine

## Appendix B: Students Progress In Learning PWCT

In Table 14 we see the students' progress in learning PWCT through the free remote learning course. The first column in the table represent the student ID (instead of the student name) and the second column contains the number of applications made by the student using PWCT. The third column contains the student level.

In figure 50 we see the graph that represent the number of students and the student level and in figures 51, 52 and 53 we see the statistics of the YouTube Playlists for each level in the course.

Table 14: Students progress in learning PWCT

| Student ID | No. of Applications | Level |
|---|---|---|
| 1 | 1 | 1 |
| 2 | 7 | 1 |
| 3 | 47 | 2 |
| 4 | 29 | 1 |
| 5 | 2 | 1 |
| 6 | 4 | 1 |
| 7 | 64 | 3 |
| 8 | 24 | 1 |
| 9 | 19 | 1 |
| 10 | 4 | 1 |
| 11 | 6 | 1 |
| 12 | 18 | 1 |
| 13 | 2 | 1 |
| 14 | 1 | 1 |
| 15 | 7 | 1 |
| 16 | 9 | 1 |
| 17 | 4 | 1 |
| 18 | 7 | 1 |
| 19 | 1 | 1 |
| 20 | 42 | 2 |
| 21 | 25 | 1 |
| 22 | 10 | 1 |
| 23 | 46 | 2 |
| 24 | 3 | 1 |
| 25 | 19 | 1 |
| 26 | 4 | 1 |
| 27 | 1 | 1 |
| 28 | 1 | 1 |
| 29 | 11 | 1 |
| 30 | 51 | 3 |
| 31 | 66 | 3 |
| 32 | 3 | 1 |
| 33 | 5 | 1 |
| 34 | 6 | 1 |
| 35 | 62 | 3 |
| 36 | 64 | 3 |
| 37 | 13 | 1 |
| 38 | 18 | 1 |
| 39 | 35 | 2 |
| 40 | 64 | 3 |
| 41 | 6 | 1 |
| 42 | 3 | 1 |
| 43 | 1 | 1 |
| 44 | 16 | 1 |
| 45 | 7 | 1 |
| 46 | 4 | 1 |
| 47 | 5 | 1 |
| 48 | 27 | 1 |
| 49 | 1 | 1 |
| 50 | 11 | 1 |
| 51 | 3 | 1 |
| 52 | 3 | 1 |
| 53 | 7 | 1 |
| 54 | 34 | 2 |
| 55 | 2 | 1 |
| 56 | 59 | 3 |
| 57 | 6 | 1 |
| 58 | 4 | 1 |
| 59 | 64 | 3 |
| 60 | 5 | 1 |
| 61 | 9 | 1 |
| 62 | 8 | 1 |
| 63 | 1 | 1 |
| 64 | 1 | 1 |
| 65 | 5 | 1 |
| 66 | 11 | 1 |
| 67 | 60 | 3 |
| 68 | 1 | 1 |
| 69 | 43 | 2 |
| 70 | 61 | 3 |
| 71 | 46 | 2 |
| 72 | 1 | 1 |
| 73 | 24 | 1 |
| 74 | 4 | 1 |
| 75 | 6 | 1 |
| 76 | 7 | 1 |
| 77 | 38 | 2 |
| 78 | 20 | 1 |
| 79 | 5 | 1 |
| 80 | 5 | 1 |
| 81 | 1 | 1 |
| 82 | 60 | 3 |
| 83 | 4 | 1 |
| 84 | 5 | 1 |
| 85 | 45 | 2 |
| 86 | 4 | 1 |
| 87 | 5 | 1 |
| 88 | 1 | 1 |
| 89 | 53 | 3 |
| 90 | 8 | 1 |
| 91 | 1 | 1 |
| 92 | 1 | 1 |
| 93 | 61 | 3 |
| 94 | 7 | 1 |
| 95 | 10 | 1 |
| 96 | 2 | 1 |
| 97 | 1 | 1 |
| 98 | 61 | 3 |
| 99 | 1 | 1 |
| 100 | 65 | 3 |
| 101 | 4 | 1 |
| 102 | 9 | 1 |
| 103 | 64 | 3 |
| 104 | 4 | 1 |
| 105 | 1 | 1 |
| 106 | 3 | 1 |
| 107 | 5 | 1 |
| 108 | 65 | 3 |
| 109 | 23 | 1 |
| 110 | 2 | 1 |
| 111 | 3 | 1 |
| 112 | 62 | 3 |
| 113 | 17 | 1 |
| 114 | 64 | 3 |
| 115 | 62 | 3 |
| 116 | 65 | 3 |
| 117 | 49 | 2 |
| 118 | 70 | 3 |
| 119 | 9 | 1 |
| 120 | 4 | 1 |
| 121 | 5 | 1 |
| 122 | 4 | 1 |
| 123 | 27 | 1 |
| 124 | 6 | 1 |
| 125 | 49 | 2 |
| 126 | 22 | 1 |
| 127 | 23 | 1 |
| 128 | 26 | 1 |
| 129 | 48 | 2 |
| 130 | 22 | 1 |
| 131 | 2 | 1 |
| 132 | 1 | 1 |
| 133 | 68 | 3 |
| 134 | 2 | 1 |
| 135 | 5 | 1 |
| 136 | 70 | 3 |
| 137 | 8 | 1 |
| 138 | 10 | 1 |
| 139 | 1 | 1 |
| 140 | 4 | 1 |
| 141 | 4 | 1 |
| 142 | 63 | 3 |
| 143 | 2 | 1 |
| 144 | 4 | 1 |
| 145 | 3 | 1 |
| 146 | 4 | 1 |
| 147 | 4 | 1 |
| 148 | 12 | 1 |
| 149 | 15 | 1 |
| 150 | 47 | 2 |
| 151 | 2 | 1 |
| 152 | 3 | 1 |
| 153 | 64 | 3 |
| 154 | 41 | 2 |
| 155 | 1 | 1 |
| 156 | 4 | 1 |
| 157 | 8 | 1 |
| 158 | 4 | 1 |
| 159 | 70 | 3 |
| 160 | 7 | 1 |
| 161 | 2 | 1 |
| 162 | 30 | 1 |
| 163 | 65 | 3 |
| 164 | 65 | 3 |
| 165 | 66 | 3 |
| 166 | 20 | 1 |
| 167 | 8 | 1 |
| 168 | 7 | 1 |
| 169 | 4 | 1 |
| 170 | 7 | 1 |
| 171 | 5 | 1 |
| 172 | 6 | 1 |
| 173 | 1 | 1 |
| 174 | 26 | 1 |
| 175 | 7 | 1 |
| 176 | 12 | 1 |
| 177 | 1 | 1 |
| 178 | 24 | 1 |
| 179 | 64 | 3 |
| 180 | 6 | 1 |
| 181 | 2 | 1 |
| 182 | 2 | 1 |
| 183 | 1 | 1 |
| 184 | 6 | 1 |
| 185 | 70 | 3 |
| 186 | 65 | 3 |
| 187 | 1 | 1 |
| 188 | 6 | 1 |
| 189 | 1 | 1 |
| 190 | 1 | 1 |
| 191 | 70 | 3 |
| 192 | 1 | 1 |
| 193 | 6 | 1 |
| 194 | 1 | 1 |
| 195 | 6 | 1 |
| 196 | 4 | 1 |
| 197 | 65 | 3 |
| 198 | 3 | 1 |
| 199 | 56 | 3 |
| 200 | 1 | 1 |
| 201 | 53 | 3 |
| 202 | 1 | 1 |
| 203 | 3 | 1 |
| 204 | 46 | 2 |
| 205 | 7 | 1 |
| 206 | 67 | 3 |
| 207 | 5 | 1 |
| 208 | 2 | 1 |
| 209 | 4 | 1 |
| 210 | 2 | 1 |
| 211 | 70 | 3 |
| 212 | 4 | 1 |
| 213 | 3 | 1 |
| 214 | 26 | 1 |
| 215 | 37 | 2 |
| 216 | 1 | 1 |
| 217 | 18 | 1 |
| 218 | 19 | 1 |
| 219 | 32 | 1 |
| 220 | 35 | 2 |
| 221 | 20 | 1 |
| 222 | 14 | 1 |
| 223 | 36 | 2 |
| 224 | 44 | 2 |
| 225 | 36 | 2 |
| 226 | 18 | 1 |
| 227 | 6 | 1 |
| 228 | 3 | 1 |
| 229 | 17 | 1 |
| 230 | 14 | 1 |
| 231 | 1 | 1 |
| 232 | 2 | 1 |
| 233 | 36 | 2 |
| 234 | 36 | 2 |
| 235 | 6 | 1 |
| 236 | 28 | 1 |
| 237 | 18 | 1 |
| 238 | 3 | 1 |
| 239 | 2 | 1 |
| 240 | 18 | 1 |
| 241 | 11 | 1 |
| 242 | 18 | 1 |
| 243 | 1 | 1 |
| 244 | 12 | 1 |
| 245 | 6 | 1 |
| 246 | 31 | 1 |
| 247 | 25 | 1 |
| 248 | 3 | 1 |
| 249 | 2 | 1 |
| 250 | 49 | 2 |
| 251 | 1 | 1 |
| 252 | 9 | 1 |
| 253 | 7 | 1 |
| 254 | 19 | 1 |
| 255 | 11 | 1 |
| 256 | 2 | 1 |
| 257 | 4 | 1 |
| 258 | 1 | 1 |
| 259 | 2 | 1 |
| 260 | 17 | 1 |
| 261 | 53 | 3 |
| 262 | 51 | 3 |
| 263 | 1 | 1 |
| 264 | 2 | 1 |
| 265 | 5 | 1 |
| 266 | 8 | 1 |
| 267 | 2 | 1 |
| 268 | 1 | 1 |
| 269 | 3 | 1 |
| 270 | 1 | 1 |
| 271 | 46 | 2 |
| 272 | 2 | 1 |
| 273 | 1 | 1 |
| 274 | 1 | 1 |
| 275 | 8 | 1 |
| 276 | 30 | 1 |
| 277 | 1 | 1 |
| 278 | 2 | 1 |
| 279 | 12 | 1 |
| 280 | 25 | 1 |
| 281 | 15 | 1 |
| 282 | 9 | 1 |
| 283 | 1 | 1 |
| 284 | 1 | 1 |
| 285 | 2 | 1 |
| 286 | 19 | 1 |
| 287 | 1 | 1 |
| 288 | 6 | 1 |
| 289 | 11 | 1 |
| 290 | 6 | 1 |
| 291 | 1 | 1 |
| 292 | 22 | 1 |
| 293 | 35 | 2 |
| 294 | 8 | 1 |
| 295 | 24 | 1 |
| 296 | 6 | 1 |
| 297 | 2 | 1 |
| 298 | 2 | 1 |

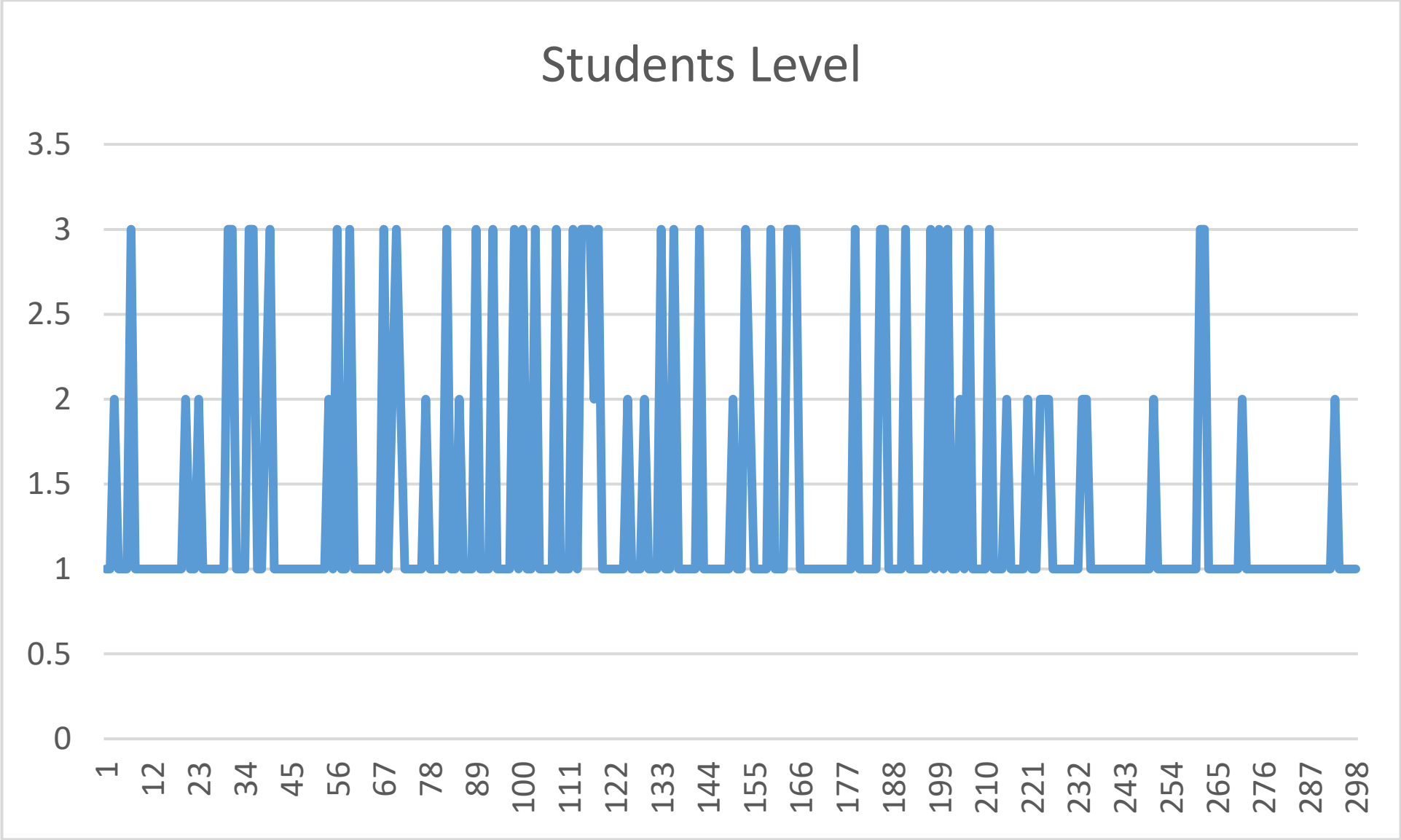

Figure 51: Students Level in PWCT course.

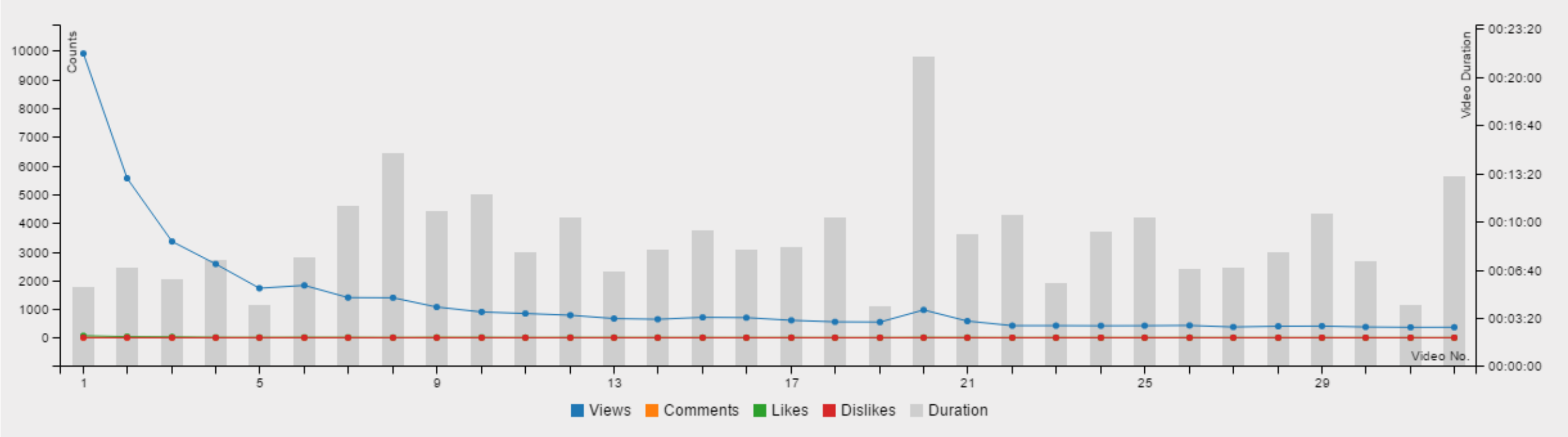

Figure 52: YouTube Playlist Statistics for Level 1 of PWCT Course.

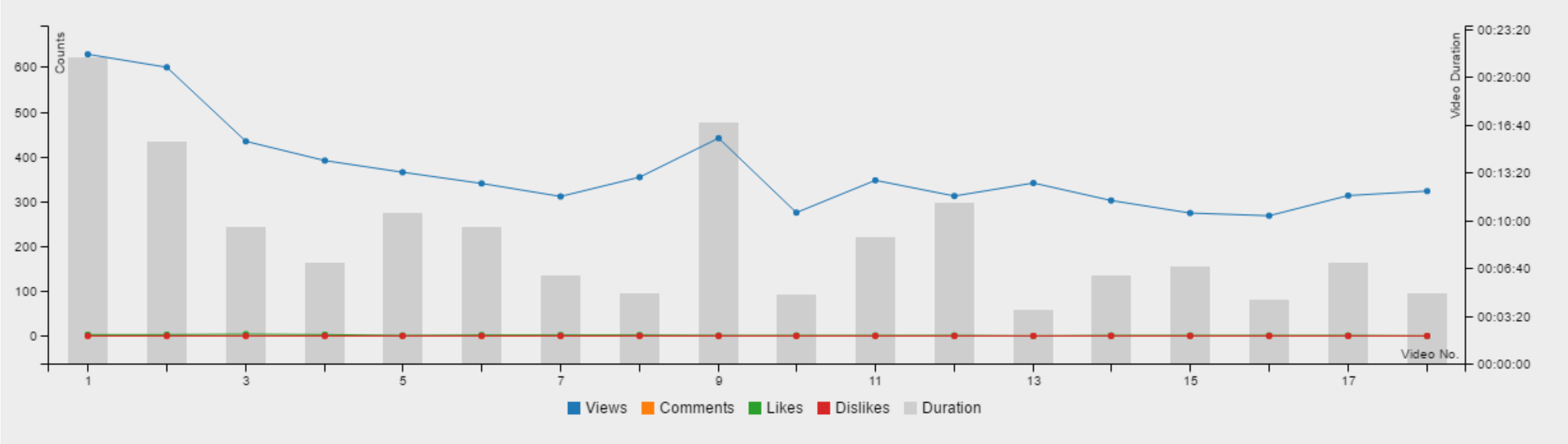

Figure 53: YouTube Playlist Statistics for Level 2 of PWCT Course.

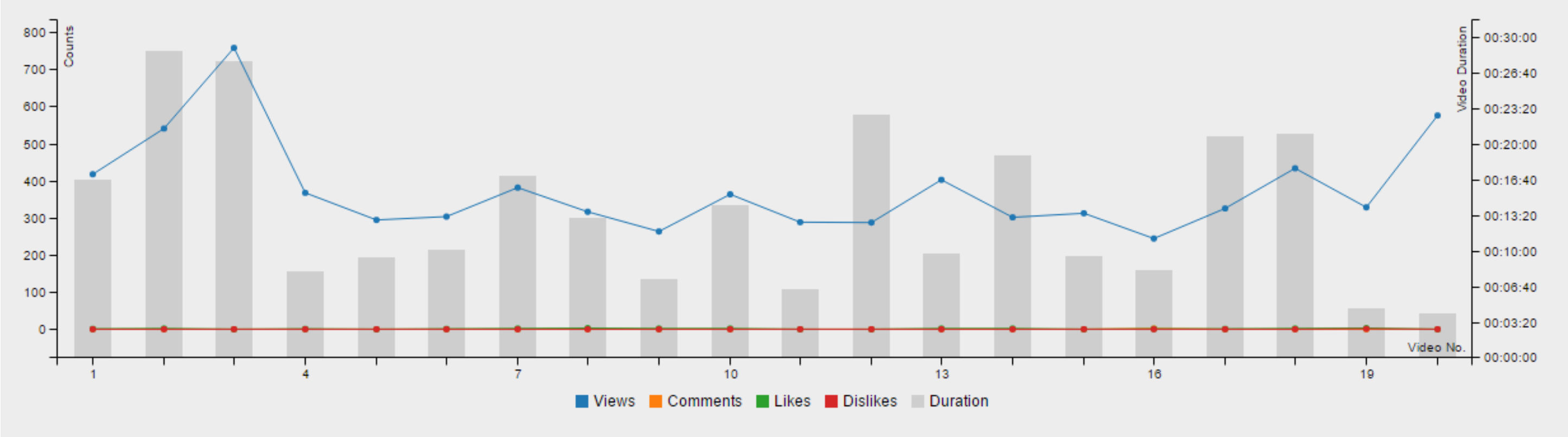

Figure 54: YouTube Playlist Statistics for Level 3 of PWCT Course.

## Appendix C: Designers in our Visual Programming Languages' framework

(1) Interaction Designer (Component User Interface Designer)

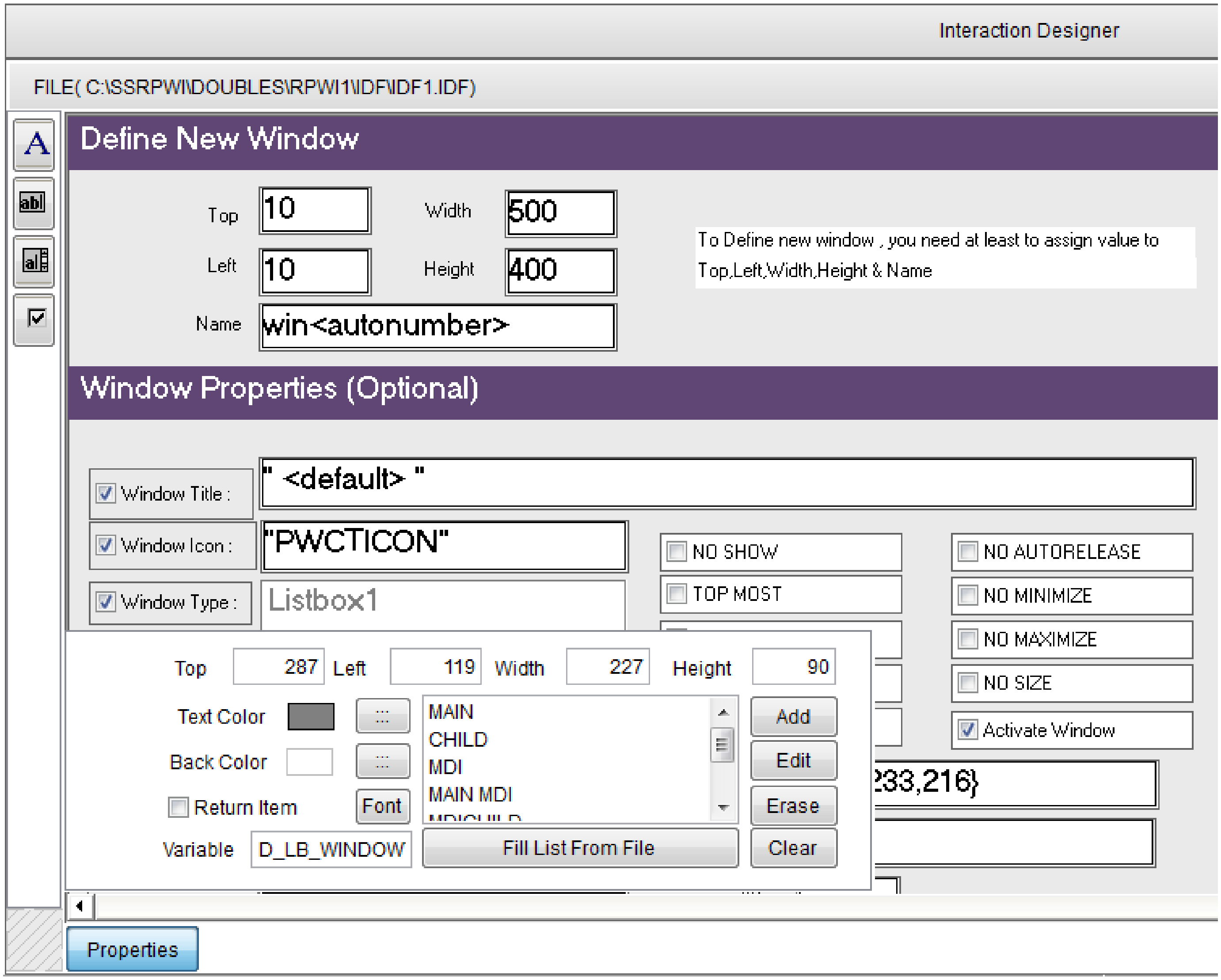

Figure 55: Interaction Designer

## (2) Transporter Designer (Component Designer)

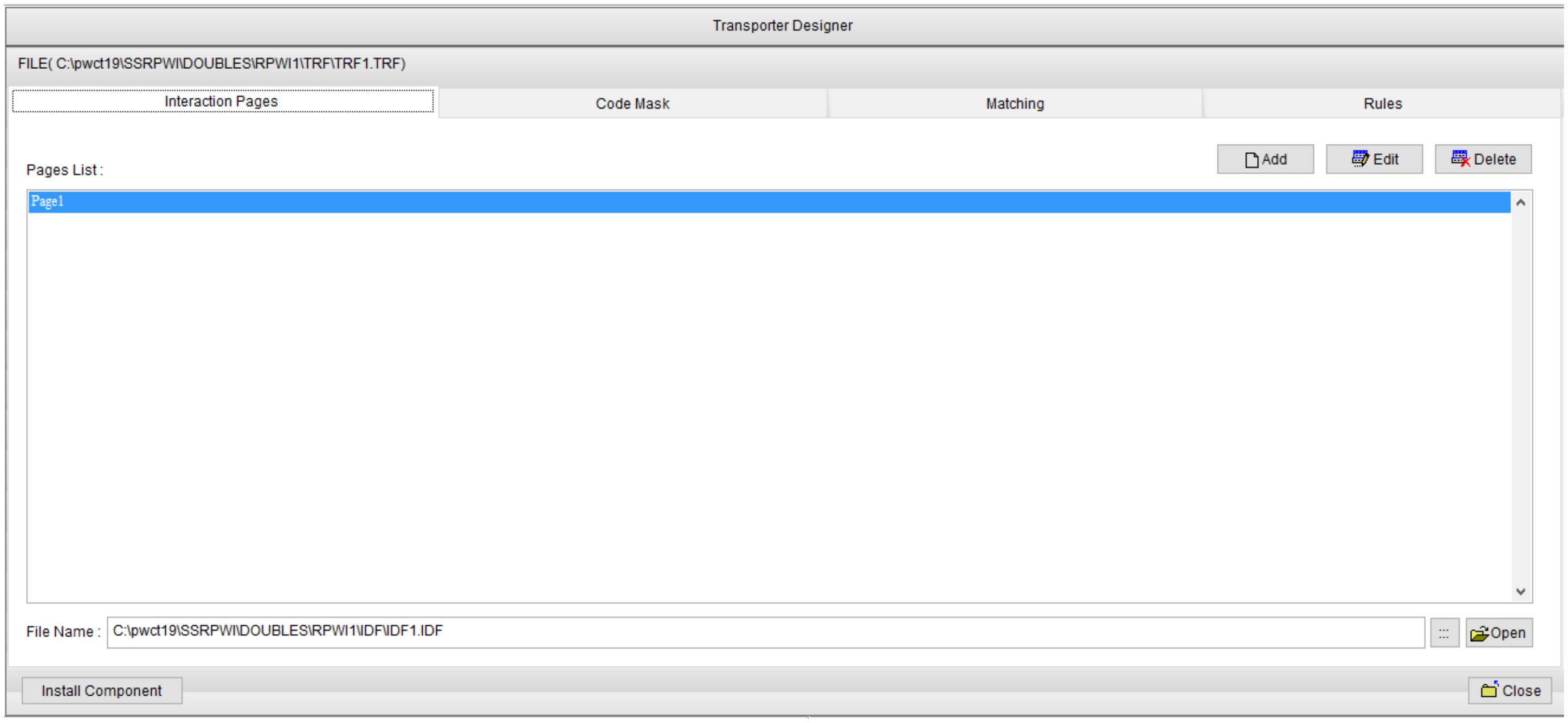

Figure 56: Transporter/Component Designer – Interaction Pages

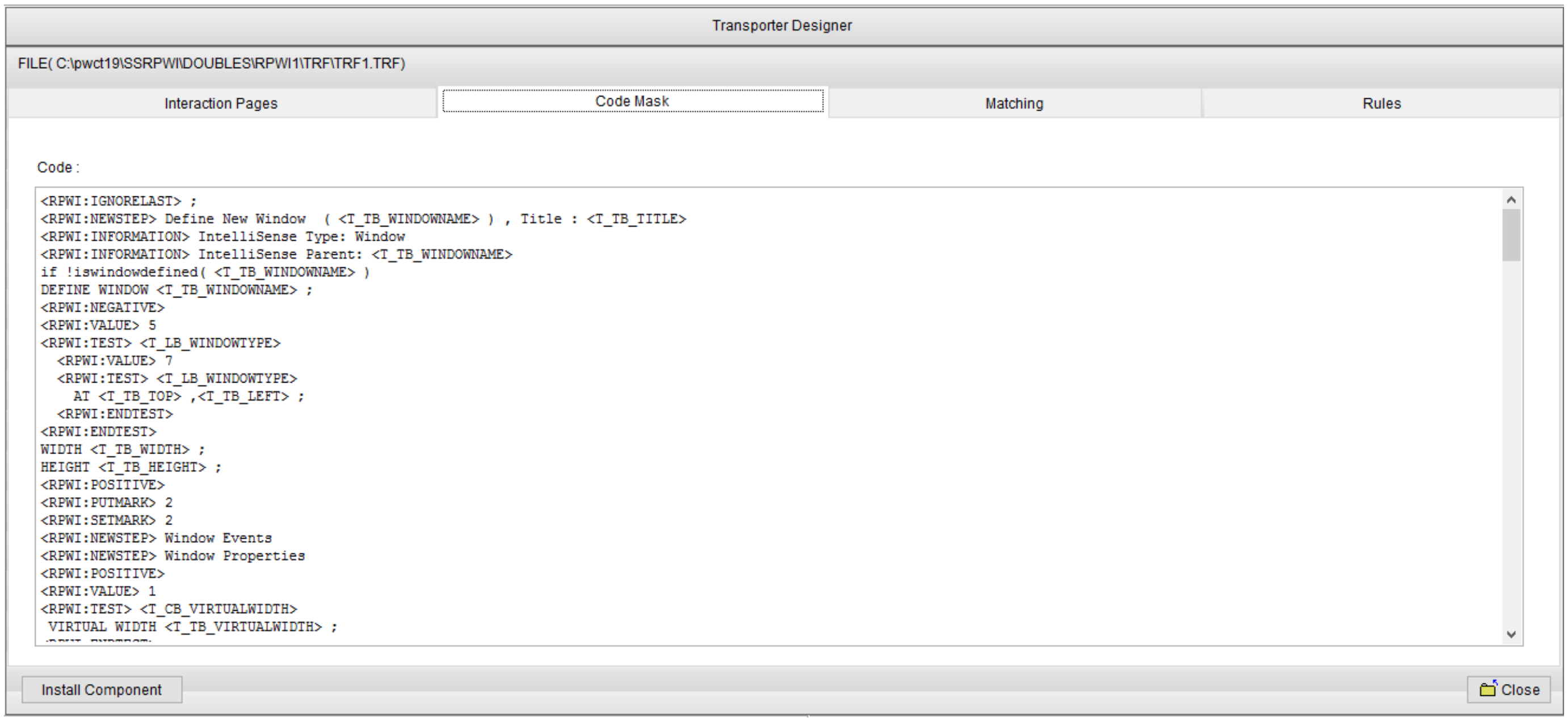

Figure 57: Transporter/Component Designer – Code Mask (Script)

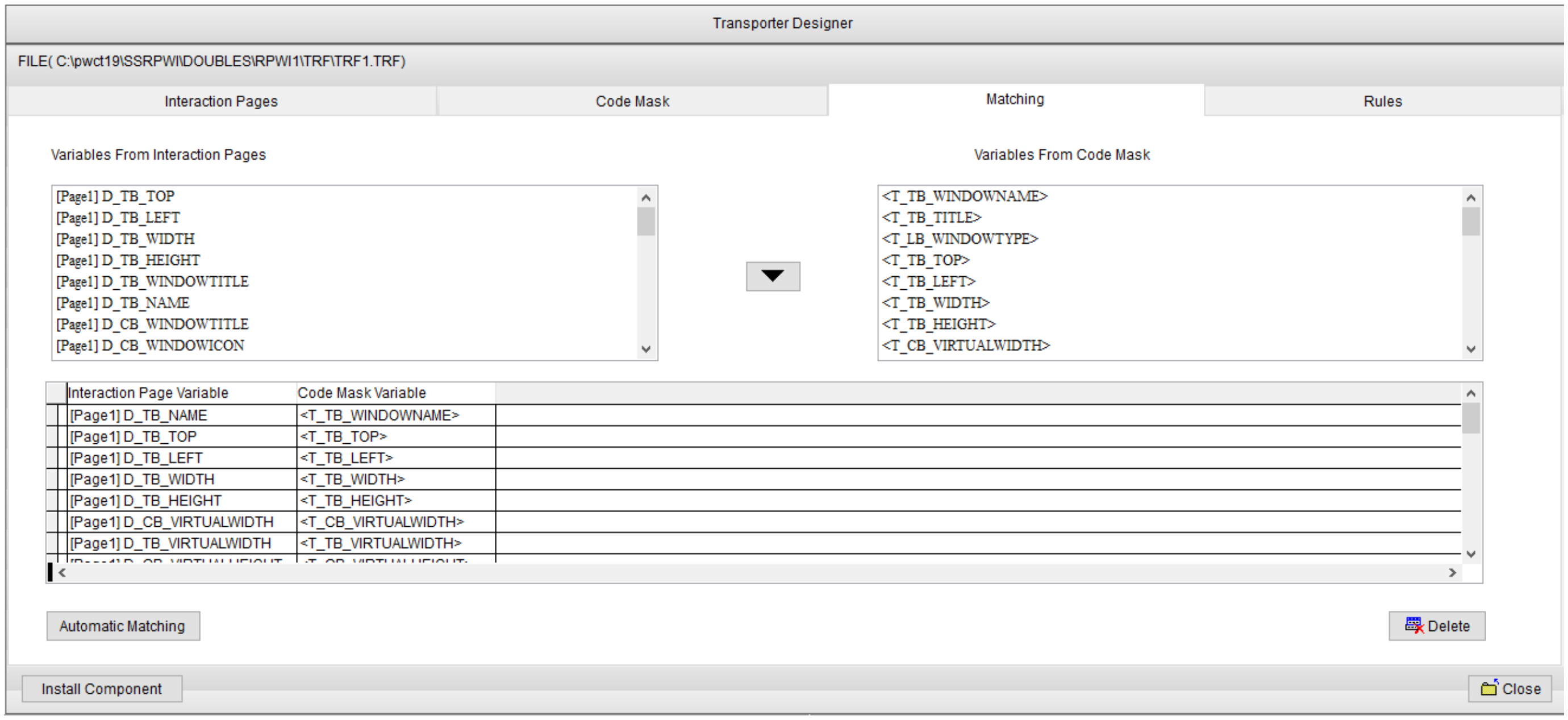

Figure 58: Transporter/Component Designer – Matching

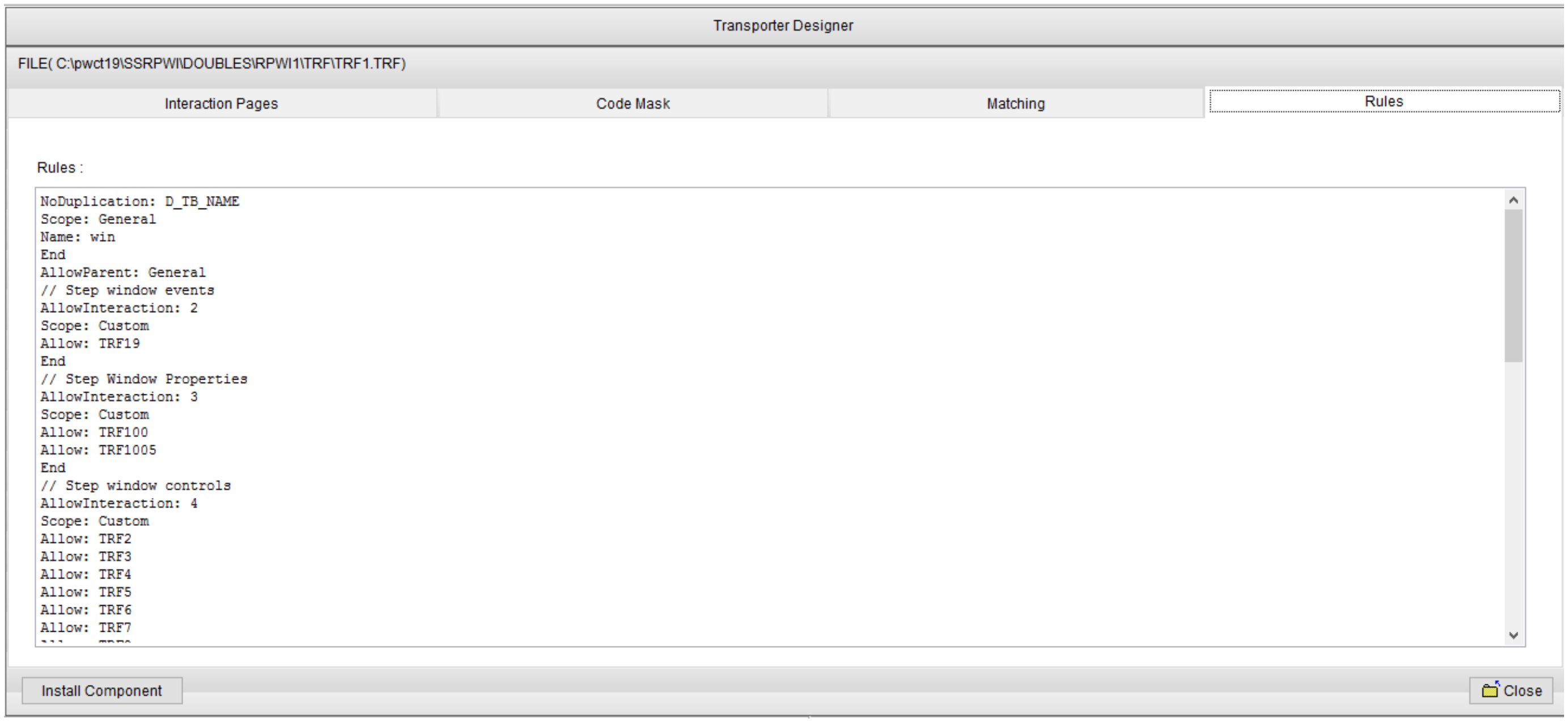

Figure 59: Transporter/Component Designer – Rules